\documentclass[twocolumn]{aastex631}

\usepackage{savesym}
\savesymbol{tablenum}
\usepackage{siunitx}
\usepackage{lipsum}

\newcommand{\microns}{\si{\micro \meter}}

\received{March 7, 2025}
\revised{May 19, 2025}

\submitjournal{AJ}

\begin{document}

\title{A Panchromatic Characterization of the Evening and Morning Atmosphere of WASP-107~b: Composition and Cloud Variations, and Insight into the Effect of Stellar Contamination}

\correspondingauthor{Matthew M. Murphy}
\email{mmmurphy@arizona.edu}

\author[0000-0002-8517-8857]{Matthew M. Murphy}
\affiliation{Steward Observatory, 933 North Cherry Avenue, Tucson, AZ 85721, USA}

\author[0000-0002-9539-4203]{Thomas G. Beatty}
\affiliation{Department of Astronomy, University of Wisconsin--Madison, Madison, WI 53703, USA}
\author[0000-0001-8291-6490]{Everett Schlawin}
\affiliation{Steward Observatory, 933 North Cherry Avenue, Tucson, AZ 85721, USA}
\author[0000-0003-4177-2149]{Taylor J. Bell}
\affiliation{Bay Area Environmental Research Institute, NASA's Ames Research Center, Moffett Field, CA 94035, USA}
\affiliation{Space Science and Astrobiology Division, NASA's Ames Research Center, Moffett Field, CA 94035, USA}
\affiliation{AURA for the European Space Agency (ESA), Space Telescope Science Institute, 3700 San Martin Drive, Baltimore, MD 21218, USA}
\author[0000-0002-3328-1203]{Michael Radica}
\altaffiliation{NSERC Postdoctoral Fellow}
\affiliation{Department of Astronomy \& Astrophysics, University of Chicago, 5640 South Ellis Avenue, Chicago, IL 60637, USA}
\affiliation{Institut Trottier de Recherche sur les Exoplanètes and Département de Physique, Université de Montréal, 1375 Avenue Thérèse-Lavoie-Roux, Montréal, QC, H2V 0B3, Canada}
\author[0000-0002-2984-3250]{Thomas D. Kennedy}
\affiliation{Department of Astronomy and Astrophysics, University of Michigan, Ann Arbor, MI, 48109, USA}
\author[0000-0001-6086-4175]{Nishil Mehta}
\affiliation{Laboratoire Lagrange, Observatoire de la Côte d’Azur, CNRS, Université Côte d’Azur, Nice, France}
\author[0000-0003-0156-4564]{Luis Welbanks}
\affiliation{School of Earth and Space Exploration, Arizona State University, Tempe, AZ, USA}
\author[0000-0002-2338-476X]{Michael R. Line}
\affiliation{School of Earth and Space Exploration, Arizona State University, Tempe, AZ, USA}
\author[0000-0001-9521-6258]{Vivien Parmentier}
\affiliation{Laboratoire Lagrange, Observatoire de la Côte d’Azur, CNRS, Université Côte d’Azur, Nice, France}
\author[0000-0002-8963-8056]{Thomas P. Greene}
\affiliation{Space Science and Astrobiology Division, NASA's Ames Research Center, Moffett Field, CA 94035, USA}
\author[0000-0003-1622-1302]{Sagnick Mukherjee}
\affiliation{Department of Astronomy and Astrophysics, University of California, Santa Cruz, CA, USA}
\affiliation{Department of Physics and Astronomy, Johns Hopkins University, Baltimore, MD, USA \\ }
\author[0000-0002-9843-4354]{Jonathan J. Fortney}
\affiliation{Department of Astronomy and Astrophysics, University of California, Santa Cruz, CA, USA}
\author[0000-0003-3290-6758]{Kazumasa Ohno}
\affil{Division of Science, National Astronomical Observatory of Japan, 2-12-1 Osawa, Mitaka-shi 1818588 Tokyo, Japan}
\author[0000-0002-3295-1279]{Lindsey Wiser}
\affiliation{School of Earth and Space Exploration, Arizona State University, Tempe, AZ, USA}
\author[0000-0002-3034-8505]{Kenneth Arnold}
\affiliation{Department of Astronomy, University of Wisconsin--Madison, Madison, WI 53703, USA}
\author[0000-0003-3963-9672]{Emily Rauscher}
\affiliation{Department of Astronomy and Astrophysics, University of Michigan, Ann Arbor, MI, 48109, USA}
\author[0000-0001-8745-2613]{Isaac R. Edelman}
\affiliation{Bay Area Environmental Research Institute, NASA's Ames Research Center, Moffett Field, CA 94035, USA}
\author[0000-0002-7893-6170]{Marcia J. Rieke}
\affiliation{Steward Observatory, 933 North Cherry Avenue, Tucson, AZ 85721, USA}

\begin{abstract}
Limb-resolved transmission spectroscopy has the potential to transform our understanding of exoplanetary atmospheres. By separately measuring the transmission spectra of the evening and morning limbs, these atmospheric regions can be individually characterized, shedding light into the global distribution and transport of key atmospheric properties from transit observations alone. 
In this work, we follow up the recent detection of limb asymmetry on the exoplanet WASP-107~b \citep{murphy24} by reanalyzing literature observations of WASP-107~b using all of JWST's science intruments (NIRISS, NIRCam, NIRSpec, and MIRI) to measure its limb transmission spectra from $\sim$1--12~\microns{}. We confirm the evening--morning temperature difference inferred previously and find that it is qualitatively consistent with predictions from global circulation models. We find evidence for evening--morning variation in SO$_2$ and CO$_2$ abundance, and significant cloud coverage only on WASP-107~b's morning limb. We find that the NIRISS and NIRSpec observations are potentially contaminated by occulted starspots, which we leverage to investigate stellar contamination's impact on limb asymmetry measurements. We find that starspot crossings can significantly bias the inferred evening and morning transmission spectra depending on when they occur during the transit, and develop a simple correction model which successfully brings these instruments' spectra into agreement with the uncontaminated observations. 
\end{abstract}

\section{Introduction} \label{sec:intro}

Exoplanetary science has entered a new era in which detailed characterization of exoplanetary atmospheres is readily possible. Observations using the James Webb Space Telescope (JWST) have enabled transmission and emission spectroscopy at both higher precision and wider continuous wavelength coverage than previously possible. Since JWST's launch, humanity's ability to characterize an exoplanet has shifted from the mere detection of singular molecular species in the atmosphere to the measurement of the abundance of most major carbon- and oxygen-bearing molecules \citep[e.g.,][]{welbanks24, beatty24_gj3470}. These observations have made new and powerful constraints on the circulation patterns \citep[e.g.,][]{kempton2023_gj1214b}, interior heating processes \citep[e.g.,][]{welbanks24, sing2024_wasp107b}, and nature of clouds and hazes \citep[e.g.,][]{grant23_wasp17b, dyrek2023_wasp107b, schlawin24_wasp69b, bell2024_wasp43b} that shape exoplanetary atmospheres. 

Besides expanding the scope and precision of prevalent techniques, JWST has enabled a new mode to probe exoplanet atmospheres. This new mode --- limb-resolved transmission spectroscopy --- probes spatial atmospheric heterogeneity by individually measuring the transmission spectrum of each planetary limb, rather than their combination as measured by traditional transmission spectroscopy. Several specific techniques have been developed to do this, including modeling the occulting planetary disk as independent hemispheres \citep{espinoza2021_catwoman, jones2022_catwoman}, modeling the disk's shape using a Fourier series \citep{grant2023_transmissionstrings}, and fitting specific sections of the transit ingress and egress separately \citep{espinoza24_wasp39b}. By separating each limb's transmission spectrum, the evening and morning terminator regions can be separately characterized. As a result, heterogeneity in fundamental atmospheric properties between terminators, and by proxy the day and nightsides, often predicted by 3-dimensional atmospheric models \citep[e.g.][]{kataria2016_gcmgrid, powell2019_HJlimbasymmetries, zamyatina24_wasp96bGCMs_limbasym} can be studied. 

Limb-resolved transmission spectroscopy with JWST is not the only technique that can deliver insight into the heterogeneity of transiting exoplanet atmospheres. For over a decade, phase curve observations have been used with great success to globally map the hemisphere-integrated distribution of atmospheric temperature and clouds, primarily for hot Jupiters \citep[e.g.][]{knutson2012_hd189phasecurve, beatty2019_phasecurves, may2022_phasecurvestudy, murphy2023_wasp43phasecurves}. Also, high-resolution cross-correlation spectroscopy, typically from the ground, can separate the spectral signatures of the eastern and western hemispheres of transiting ultra-hot Jupiters due to their different rotational Doppler shifts, which has been used to infer heterogeneous wind speeds and molecular absorption \citep[e.g.][]{bourrier2020_wasp121basymmetry, hoeijmakers2020_mascara2basymmetry, ehrenreich2020_wasp76ironasymmetry, kesseli2021_wasp76ironasymmetry, gandhi2022_w76bhighresretrieval, wardenier2023_w76bGCM, maguire2024_wasp76limbasymmetries}. Limb-resolved transmission spectroscopy is complementary to these techniques and expands their discovery space. Phase curves and ground-based spectroscopy are generally limited to planets on very close-in orbits, typically with orbital periods of one-to-two days or less. On the other hand, limb-resolved transmission spectroscopy can reach planets with multi-day periods and thus greatly expand the set of planets and physical conditions that can be probed. Also, limb-resolved transmission spectroscopy involves only a transit observation, meaning it requires significantly less time investment than a phase curve. Altogether, limb-resolved transmission spectroscopy with JWST is a powerful new tool with the potential to significantly expand the community's ability to study exoplanet atmospheres. 

So far, limb-resolved transmission spectroscopy with JWST has been used to detect limb asymmetry on two exoplanets: WASP-107~b \citep{murphy24} and WASP-39~b \citep{espinoza24_wasp39b}. WASP-107~b is a $T_{\rm eq}\approx~$750~K, P = 5.72~day inflated Neptune \citep{anderson2017_wasp107b, piaulet2021_wasp107b, welbanks24}, on which \cite{murphy24} inferred a $\sim$180~K temperature gradient between its evening and morning limbs from a single JWST/NIRCam F322W2 transit observation. WASP-39~b is a $T_{\rm eq}\approx~$1100~K, P = 4.05~day hot Jupiter \citep{mancini2018_wasp39b}, on which \cite{espinoza24_wasp39b} inferred a $\sim$177~K evening--morning temperature gradient from a single JWST/NIRSpec PRISM transit observation. 

The similarity between WASP-107~b's and WASP-39~b's evening--morning temperature gradients is surprising given the large difference between their equilibrium temperatures. General circulation models \citep[e.g.][]{kataria2016_gcmgrid, roth2024_HJmodels} and microphysical cloud formation models \citep[e.g.][]{powell2019_HJlimbasymmetries, powell2024twoDcloudmodels} typically predict that the limbs homogenize toward cooler temperatures, and that planets below $\sim$1000~K should have relatively homogeneous atmospheres. Therefore, the cooler WASP-107~b would generally not be expected to have comparable, or even stronger, heterogeneity than the hotter WASP-39~b. WASP-107~b has been extensively studied in the literature, which has revealed a number of other interesting properties. WASP-107~b appears to be on a near-polar orbit \citep{dai2017_wasp107b, rubenzahl2021_wasp107b}, indicative of an active dynamical history. As mentioned, WASP-107~b's atmosphere is highly inflated to the point of being a Jupiter-sized planet despite being Neptune-mass \citep{anderson2017_wasp107b, piaulet2021_wasp107b}, which is likely driven by intense tidal heating in the planet's interior \citep{welbanks24, sing2024_wasp107b} and related to its dynamical history. Fortunately, this inflated atmosphere makes WASP-107~b an excellent target to characterize in transmission, particularly using limb-resolved transmission spectroscopy. In addition, WASP-107~b is currently experiencing significant atmospheric loss, evidenced by an immense ($\approx$ 7~$\times$ the planet's optical radius) outflowing exosphere of helium \citep{spake2018_wasp107b, kirk2020_wasp107b, spake2021_wasp107b}. Altogether, these raise the question of whether WASP-107~b is an outlier in the warm planet population, or whether limb asymmetry may actually be common below $\sim$1000~K. 

In this work, we follow up on the work of \cite{murphy24} with additional observations of WASP-107~b to confirm the presence of limb asymmetry and further characterize its evening and morning limbs.
We also perform an initial exploration of the impact that stellar contamination, specifically of the form of occulted starspots, has on limb-resolved transmission spectroscopy and measuring limb asymmetry with JWST. This is motivated by apparent spot crossings in two of the observations of WASP-107~b we analyze. 
Therefore, we aim to develop a simple model for the effect of occulted starspots on measuring limb asymmetry. 

\section{Observations} \label{sec:observations}

We reanalyzed five transit observations of WASP-107b from JWST. These include two observations with JWST's Near-Infrared Camera (NIRCam) using its long-wave grism with both the F322W2 and F444W filters, one with its Mid-Infrared Instrument (MIRI) in Low-resolution Spectroscopy (LRS) mode, one with its Near Infrared Imager and Slitless Spectrograph (NIRISS) in Single-Object Slitless Spectroscopy (SOSS) mode, and finally one observation with its Near Infrared Spectrograph (NIRSpec) with its G395H disperser. The details of these observations and the limb-combined transmission spectra derived therefrom have each been previously reported \citep[][and Krishnamurthy et al. submitted.]{dyrek2023_wasp107b, welbanks24, sing2024_wasp107b}. We summarize the key details here. 

The NIRCam observations were taken on UT 2023 January 14 (for the visit using F322W2) and UT 2023 July 04 (F444W) as part of the MANATEE program \citep[JWST-GTO-1185, PI: T. Greene;][]{schlawin2018_manateepaper}. The visit using F322W2 yielded a spectrophotometric time series at 20.22~second cadence, with 1293 integrations in total, sampling $\sim$2.4 -- 4.0~\microns{}. The visit using F444W took 625 integrations at 41.7~second cadence, sampling $\sim$3.9 -- 5.0~\microns{}. The relative cadences of these two visits were designed to achieve the same signal in the two different wavelength regimes. 
The MIRI observation was taken on UT 2023 January 19 as part of JWST-GTO-1280 (PI: P.O. Lagage), exactly one full orbital period of WASP-107~b after the F322W2 visit. This MIRI visit took 4045 integrations at 6.5~second cadence, sampling wavelengths $\sim$5 -- 14~\microns.
The NIRISS observation was taken on UT 2023 June 11 as part of JWST-GTO-1201 (PI: D. Lafrenière). This NIRISS visit took 678 integrations at approximately 11.5 second cadence, sampling wavelengths $\sim$0.85 -- 2.7~\microns{} in its first spectral order. In this work, we neglect NIRISS SOSS' higher spectral orders due to their lower signal-to-noise ratios and limited additional wavelength coverage. 
The \mbox{NIRSpec} observation was taken on UT 2023 June 23 as part of JWST-GTO-1224 (PI: S. Birkmann). This NIRSpec visit took 1230 integrations at approximately 18~second cadence, discontinuously sampling wavelengths $\sim$2.7--5.2~\microns{} across two detectors (NRS1 and NRS2). 
These five observations coincidentally captured two distinct epochs in time: The F322W2 and MIRI observations probed consecutive transits (5.72 days apart) in January of 2023 while the NIRISS, NIRSpec, and F444W observations probed transits that were each two periods (11.44 days apart) after one another in the summer of 2023. We also list the exact dates in chronological order in the Appendix Table~\ref{apxtab:obs_dates}. For comparison, the rotation period of the star WASP-107 is estimated to be approximately 17 days \citep{dai2017_wasp107b, mocnik2017_wasp107b, piaulet2021_wasp107b}, relevant for our discussion of apparent spot crossings in some of these data in Section~\ref{sec:stellarcontam}. 
All together, these JWST observations continuously sampled WASP-107b's transmission spectrum from $\sim$0.85 -- 14~\microns{}, uniquely enabling a panchromatic investigation of limb asymmetry on WASP-107~b.

\section{Data Analysis} \label{sec:dataanalysis}

\subsection{Data Reduction} \label{subsec:datareduction}

\subsubsection{JWST/NIRCam} \label{subsubsec:reduction_NIRCamMIRI}
For the NIRCam F322W2 data, we use the same reduction with the \texttt{tshirt} \citep{TSHIRTPIPELINE} pipeline that was described and presented in \citet{murphy24}. We summarize the key details here. We used JWST pipeline version 1.8.4 and JWST CRDS context \texttt{jwst\_1039.pmap}. We ran the default JWST pipeline Stage 1 reduction except we replaced the \texttt{refpix} step with a Row-by-row, odd/even by amplifier (ROEBA) correction \citep{schlawin20_1fnoise} using background pixels for improved 1/$f$ noise reduction. We used the reference pixels of the bottom four rows for the odd/even direction, and pixels from X=1846--2043 along the rows. We raised the default jump step threshold to 6-$\sigma$. We performed column-by-column background subtraction using a linear fit of pixels Y=5 to Y=24 and Y=44 to Y=65. We applied covariance-weighted extraction \citep{schlawin20_1fnoise} using a rectangular aperture with a full width of 10 pixels between Y=29 and Y=39. All pixel positions are given in zero-based counting. 

Similarly, we used the same \texttt{tshirt} reduction for NIRCam F444W as described and presented in \citet{welbanks24}, which we summarize here. We used JWST pipeline version 1.10.2 and JWST CRDS context \texttt{jwst\_1093.pmap} \citep{bushouse2023jwst}. We ran the initial \texttt{dq\_init}, \texttt{saturation} and \texttt{superbias} steps of the \texttt{jwst} pipeline and again replaced the \texttt{refpix} step with a ROEBA correction \citep{schlawin20_1fnoise}. We used pixels 4 through 637 for row-by-row 1/f subtractions of the third and fourth amplifier (i.e. pixels 1538 through 2044) and pixels 512 through 637 for the second amplifier (i.e. pixels 637 through 1024) on a group-by-group basis. We proceeded with the rest of the \texttt{jwst} stage 1 pipeline, again raising the jump step threshold in the \texttt{jwst} pipeline to 6-$\sigma$. For the lightcurve extraction, we again used a column-by-column background subtraction with a linear robust line fit with 3-$\sigma$ clipping for a background pixel region from Y=5 to Y=21 and Y=41 to Y=64. We did covariance-weighted extraction \citep{schlawin20_1fnoise} using a rectangular aperture with a full width of 10 pixels between Y=26 and Y=36. 

Since the measurement of limb asymmetry from a light curve may be highly sensitive to systematic timing errors \citep{vonparis16_LA, line16_nonunifclouds, espinoza2021_catwoman, murphy24_timingdegen}, we took extra caution when calculating the time value of each \mbox{NIRCam} light curve point since the NIRCam detector is read out along the dispersion direction. The \texttt{INT\_TIMES} time-stamp and \texttt{int\_mid\_BJD\_TDB} column provided in the data corresponds to the last column read out. To account for this, we uniformly subtracted a 0.67~second offset, half of the read delay across the detector, from the timestamps in each channel of each NIRCam dataset so that our time arrays correspond to the middle of the array read-out. 

\subsubsection{JWST/MIRI} \label{subsubsec:reduction_MIRI}
We used the \texttt{Eureka!} pipeline \citep{EUREKAPIPELINE} reduction for MIRI LRS also described and presented in \citet{welbanks24}, which we summarize here. We use version 0.9 of \texttt{Eureka!} with CRDS context `1097' and JWST pipeline version 1.10.2. The \texttt{Eureka!} control and parameter files we used are available for download at \url{https://zenodo.org/records/10780449}. We generally follow the default Stage 1 processing, but turn on the first-frame and last-frame steps (which discard these frames) and increase the jump rejection threshold to 7-$\sigma$. We perform two iterations of 5-$\sigma$ clipping along the time axis for background pixels and one iteration of 5-$\sigma$ clipping along the spatial axis, as well as integration-level background subtracting using Y=11 to Y=61. We perform optimal spectral extraction of pixels within 4 pixels of the spectra trace determined using a median integration. The MIRI time series exhibited a strong systematic trend at the beginning of the observation, typical of MIRI observations \citep[e.g.,][]{kempton2023_gj1214b, bell2023_w80}, so we clipped the first 400 integrations ($\sim$40~minutes). 

\texttt{tshirt} and \texttt{Eureka!} have been shown to produce consistent results from the same data \citep{bell2023_w80, murphy24, welbanks24}, but \texttt{tshirt} performs a group-by-group correction step for 1/$f$ noise in NIRCam data \citep{schlawin20_1fnoise} using sky pixels whereas \texttt{Eureka!} uses reference pixels. On the other hand, \texttt{Eureka!} is better tuned for MIRI reduction than \texttt{tshirt} \citep[see e.g.][]{bell2024_wasp43b}. This is why we choose \texttt{tshirt} for NIRCam and \texttt{Eureka!} for MIRI.

\subsubsection{JWST/NIRISS} \label{subsubsec:reduction_NIRISS}
We reduced the NIRISS/SOSS \citep{doyon_near_2023, albert_near_2023} data using the publicly available \texttt{exoTEDRF} pipeline \citep{feinstein_early_2023, radica_awesome_2023, exotedrfJOSS} pipeline, which has been widely used to analyze both NIRISS \citep{coulombe_broadband_2023, lim_atmospheric_2023, fournier-tondreau_near_2024, gressier_jwst_2024, radica_promise_2024, radica_muted_2024}, and recently, \mbox{NIRSpec} (\citealp{benneke_jwst_2024, schmidt_comprehensive_2025}; Ahrer et al.~submitted) observations. These observations of WASP-107~b were originally analyzed and published in Krishnamurthy et al.~(submitted), and we re-use the same stellar spectra for this analysis --- though we note that we re-model the light curves specifically for this work, as noted in Section~\ref{subsec:datafitting}. We provide a brief overview of the reduction process here, but refer the reader to Krishnamurthy et al.~(submitted) for a more in-depth description.

We perform the standard \texttt{exoTEDRF} stage 1 calibrations \citep[e.g.,][]{radica_muted_2024, piaulet_jwst_2024} including superbias and non-linearity corrections, a time-domain cosmic ray flagging routine, and ramp fitting. We additionally correct the 1/$f$ noise before ramp fitting (i.e., at the group-level) using the \texttt{scale-achromatic} method \citep{radica_awesome_2023}. Our stage 2 calibrations include flat field correction and bad pixel interpolation, for which we use a 7-$\sigma$ threshold for both the temporal and spatial filtering. We subtract the background by scaling the standard STScI background model independently on either side of the background ``step'' \citep[e.g.,][]{lim_atmospheric_2023, fournier-tondreau_near_2024}. Finally, we extract the 2D stellar spectra with a simple box aperture of 32 pixels as the transit depth bias resulting from the order self-contamination is expected to be $<$2\,ppm below 2.6\,µm and $<$10\,ppm from 2.6--2.85\,µm \citep{radica_applesoss_2022, dareau-bernier_atoca_2022}.

Finally, we perform one extra calibration which was not part of the Krishnamurthy et al.~(submitted) workflow. SOSS has the longest frame time of any of the JWST time series modes (5.494\,s) --- meaning that the longest wavelengths are read $\sim$5.5\,s after the shortest. Since assessing the signatures of limb asymmetries in light curves requires precise knowledge of the transit timing \citep{murphy24_timingdegen} systematic biases in the light curve time stamps can, in turn, confound the ability to detect signatures of asymmetries. To this end, we use the \texttt{refine\_soss\_timestamps} routine within \texttt{exoTEDRF} to calculate timestamps for each individual detector column (i.e., wavelength), taking into account the precise time it is read relative to mid-integration --- thereby ensuring the most reliable possible timing information for our light curves.  

\subsubsection{JWST/NIRSpec} \label{subsubsec:reduction_NIRSpec}
Finally, we re-reduced the NIRSpec G395H data using \texttt{tshirt}.
We used a similar raw data reduction and lightcurve extraction as in \citet{schlawin24_gj1214b}.
As with the NIRCam analysis, we use the \texttt{jwst} pipeline for the Stage 1 (i.e. ramps-to-slopes) calculation with a modification to the reference pixel correction.
We use JWST version 1.13.4, CRDS context \texttt{jwst\_1276.pmap} and CRDS version 11.17.5.
We initially run the \texttt{dq\_init}, \texttt{saturation} and \texttt{superbias} steps with default parameters.
We replaced the reference pixel correction with a column-by-column ROEBA correction along the fast-read direction \citep{schlawin20_1fnoise}, using a Gaussian process kernel to interpolate an approximately 1/f noise signal across the source aperture and between columns with \texttt{celerite2} and a Savistky-Golay filter.
We select the background pixels as all pixels below 1.5 DN/s from the MAST rate file automatically generated with JWST version 1.14.0 and CRDS context \texttt{jwst\_1253.pmap}.
After mitigating for 1/f noise with this column-by-column subtraction on a group-by-group basis, we return to the \texttt{jwst} pipeline linearity, dark current, jump and ramp fitting steps using default parameters except for the jump step threshold of 10-$\sigma$.

We next extracted spectroscopic lightcurves from the ramp-fit data products.
We initially fit a Gaussian to a reference image on a column-by-column basis and use a 3rd order polynomial fit to the Gaussian centroids to create a trace and fix this trace for all integrations.
We then subtract the background on a column-by-column basis using all active pixels in a column that are more than 6 pixels from the source centroid, rounded to the nearest whole pixel.
We then sum the source pixels using co-variance weighted extraction \citep{schlawin20_1fnoise} with an aperture full width of 10 pixels, rounded to the nearest whole pixel.
We collect all time stamps using the \texttt{INT\_TIMES} extension with the \texttt{int\_mid\_BJD\_TDB} column.

\subsubsection{Data Binning} \label{subsubsec:reduction_binning}

Following \cite{murphy24}, we binned the \mbox{NIRCam} F322W2 spectra into 30 equally-sized 0.05~\microns{} full-width (R$\sim$64) channels from 2.475 to 3.925~\microns{}. We similarly binned the F444W spectra into 20 equally-sized 0.055~\microns{} (R$\sim$81) channels from 3.927 to 4.971~\microns{}, and the NIRISS SOSS data into 40 equally-sized 0.05~\microns{} (R$\sim$35) channels from 0.875 to 2.8~\microns{}. We binned the MIRI LRS data into 37 total bins that were 0.1665~\microns{} wide from 5--8~\microns{} and 0.25~\microns{} wide onward. These bins for MIRI were designed to maintain a spectral resolution of R$\sim$40 across the entire bandpass, despite using uniform bin sizes. Then, since the NIRSpec data are fully encompassed by the NIRCam F322W2 and F444W bandpasses, we binned the NIRSpec data into the same channels as the NIRCam data where they overlapped (2.775 to 3.674~\microns{} for NRS1, 3.927 to 4.971~\microns{} for NRS2) to enable a direct comparison between the two instruments. These wavelength bins for each instrument were chosen to balance sufficient spectral sampling with minimized light curve scatter. 

\subsection{Light Curve Modeling} \label{subsec:datafitting}

We modeled the light curve of each $i$th channel as the product of an astrophysical model, consisting of just a transit model $T_i(\theta, t)$, and a systematics model $S_i (\theta, t)$. For the transit model, we used the package \texttt{catwoman} \citep{espinoza2021_catwoman,jones2022_catwoman}, which enables simultaneously fitting for the evening and morning planet-star radius ratio $R_{\rm p} / R_{\rm \star}$, with a quadratic limb darkening treatment. For the systematics models we used visit-long linear ramps of the form
\begin{equation}
    S_i (t) = m_i \left(t - t_{median} \right) + b_i,
\end{equation}
where $m$ is the slope and $b$ is the intercept. Each light curve model is then $T_i \times S_i$. To determine the best-fit model parameters $\theta$, we employed Bayesian inference via the Markov Chain Monte Carlo method (MCMC) using the package \texttt{emcee} \citep{emcee}, treating each channel independently. We fixed most orbital parameters (time of conjunction $t_c$, orbital period $P$, semi-major axis $a/R_\star$, inclination $i$, eccentricity $e$, and argument of periastron $\omega$) to the values previously constrained by \citet{murphy24} from a simultaneous fit of nearly all available transit (JWST, TESS, Spitzer, SOAR) and radial velocity observations, which we have copied in Table~\ref{tab:orbitalparams} for convenience. Then, we fit for the evening and morning planet-star radius ratios $R_{\rm p,eve}/R_{\rm \star}$ and $R_{\rm p,morn}/R_{\rm \star}$, quadratic limb darkening coefficients $u_1$ and $u_2$, the slope $m$, intercept $b$, and a relative flux uncertainty multiplier $\sigma$. For the limb darkening coefficients $u_1$ and $u_2$, we applied Gaussian priors set to the values and 1$\sigma$ uncertainties predicted by an Kurucz/ATLAS9 stellar model \citep{castelli2004newgridsatlas9model} of WASP-107 calculated using ExoCTK \citep{exoctk}. Note that we fit directly for these coefficients, rather than reparametrizing them. We did not enforce priors (i.e., we use unbounded uniform priors) on the other fitting parameters. The multiplier term $\sigma$ is a single value multiplied to all individual relative flux uncertainties in an individual time series, meant to ensure these uncertainties match the intrinsic light curve scatter. We used 21 walkers in the MCMC, set to 3 times the number of free parameters. We ran each channel's MCMC sampling for 5000 steps after an initial 1000 step burn-in, which was over 60 times each parameter's autocorrelation time and thus sufficient for each walker to converge. 

\begin{table}[h!]
    \centering
    \caption{Orbital parameters of WASP-107b derived by \citet{murphy24} and fixed in each of our fits.}
    \begin{tabular}{c|c}
        Parameter    &  Value \\ \hline 
        Time of conjunction, t$_c$ (day)       &  2459958.747244 $\pm$ 8.2e-6 \\
        Orbital Period, $P$ (day)              &  5.7214872 $\pm$ 3.e-7 \\
        Semi-major axis, $a$ ($R_\star$)       &  18.05 $\pm$ 0.1 \\
        Inclination, $i$ (deg)                 &  89.57 $\pm$ 0.03 \\
        Eccentricity, $e$                      &  0.05 $\pm$ 0.01 \\
        Argument of Periapsis, $\omega$ (deg)  & -2.3 $\pm$ 6.1
    \end{tabular}
    \label{tab:orbitalparams}
\end{table}

We achieved good fits to each light curve with reduced $\chi^2$ values generally ranging from 0.996 to 1.005. 
Galleries of the residual time series for each instrument are shown in Figures~\ref{apxfig:residualgalleryF322W2}--\ref{apxfig:residualgalleryNIRSpec} of the Appendix. The mean residuals were all consistent with zero and
the standard deviation of residuals ranged from 587--877~ppm for NIRCam F322W2, 426--997~ppm for NIRCam F444W, 1209--7300~ppm for MIRI, 217--985~ppm for NIRISS, and 387--886~ppm for NIRSpec. 
To evaluate the robustness of our evening and morning transmission spectra against assumptions made during the fitting process, we ran several parameter variation tests which are described in Appendix Section~\ref{apx:paramvartests}. In summary, we verified that our results were robust to the uncertainties in each fixed orbital parameter. 

\section{Results} \label{subsec:ALfits_results}

\subsection{Evening and Morning Transmission Spectra}

\begin{figure*}[th!]
    \centering
    \includegraphics[width=\textwidth]{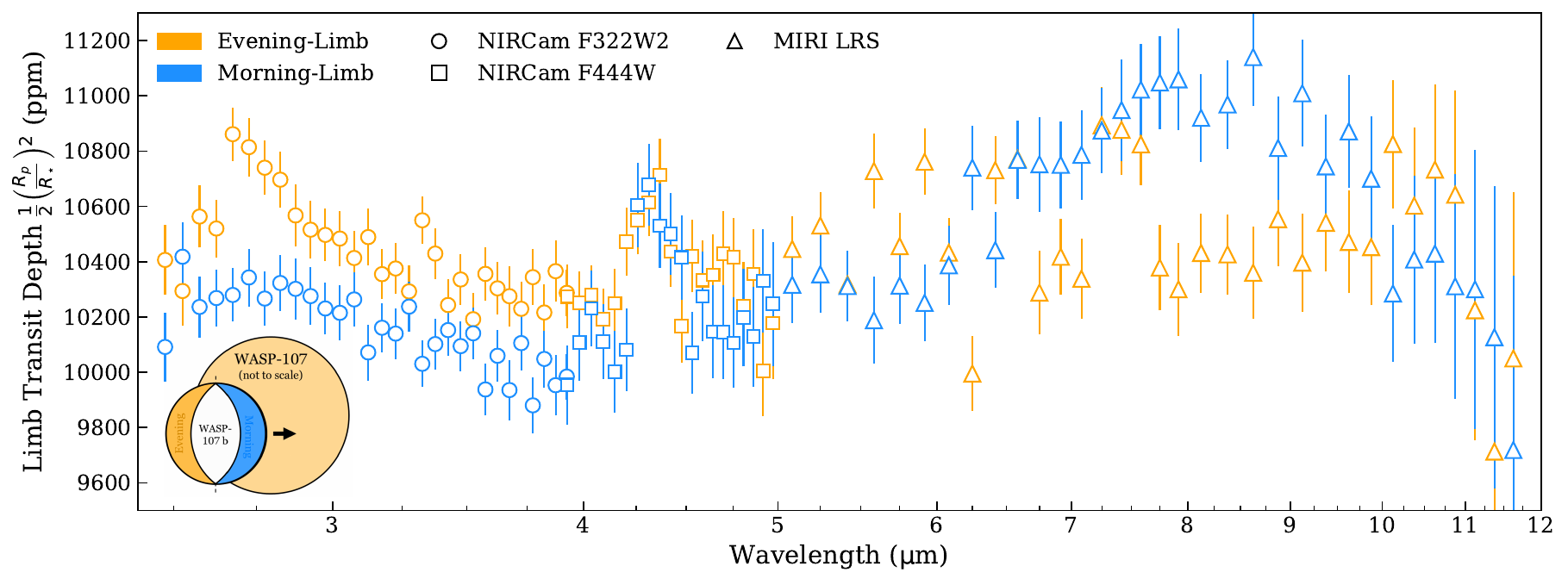}
    \caption{The evening and morning limb transmission spectra of WASP-107b as observed by JWST NIRCam F322W2 (circles), NIRCam F444W (squares), and MIRI/LRS (triangles). The evening limb spectrum is shown in orange, and the morning limb spectrum in blue. The diagram in the lower left illustrates the orientation of each limb relative to WASP-107~b's transit trajectory, but note that the bodies in the diagram are not to scale, they are not aligned to the true orbital geometry, and the exact shape of the colored slices do not accurately represent the true shape and scale of the limbs as probed in transmission. The NIRCam F322W2 portion of the spectrum shown here matches that first presented by \citet{murphy24}. In addition, while not shown, the combination of our morning and evening spectra are consistent with the panchromatic limb-combined spectrum of these instruments presented in \cite{welbanks24}.}
    \label{fig:spectra_dataonly}
\end{figure*} 

Our best-fit evening and morning limb transmission spectra of WASP-107b are shown in Figure~\ref{fig:spectra_dataonly}. We only show the results from NIRCam and MIRI here as we find evidence that the NIRISS and NIRSpec spectra are contaminated by starspot crossings during those transits, discussed further in Section~\ref{sec:stellarcontam}. The best-fit NIRISS and NIRSpec spectra, including an attempt to correct them for this stellar contamination, are subsequently shown and compared to the NIRCam spectra in Figure~\ref{fig:contam_spectra}. We find the same vertical offset between the NIRCam and MIRI spectra seen in the limb-combined spectrum by \cite{welbanks24}, most likely caused by the different linearity properties of each detector. We apply a manual 285~ppm upward offset to the MIRI spectra as determined through an atmospheric retrieval by \cite{welbanks24}. All together, we find that the best-fit evening and morning transmission spectra are consistent between NIRCam and MIRI. The NIRCam F322W2 and F444W spectra overlap by one channel, in which both the morning and evening depths agree nearly exactly. The overall offset between the evening and morning spectra is also visually consistent going between F322W2 and F444W, though this is modulated by an absorption feature due to SO$_2$ (discussed further later) at the start of the F444W bandpass. When including the offset to the MIRI spectra, the F444W and MIRI spectra appear to align consistently as well. While there are no overlapping channels between F444W and MIRI, the absolute evening and morning limb depths are similar. There appears to be significant noise at the red-ends of both the F444W and MIRI spectra though, where the limb depths become relatively scattered and have larger uncertainties. These were also channels where the light curves themselves had the most intrinsic scatter, particularly beyond 11~\microns{}, so we caution against over-interpreting those individual points. Potential correlated noise in the MIRI data may be contributing here as well, which we discuss further in Section~\ref{subsec:stellarcontam_lcnoise}. We compare these observed spectra to model transmission spectra later in Section~\ref{subsec:compare2models}.

We see several interesting features in the spectra shown in Figure~\ref{fig:spectra_dataonly}. First, the overall offset between the evening and morning spectra, initially seen by \cite{murphy24} in just the F322W2 data, is generally consistent across the entire spectrum. There are some important exceptions due to molecular features, which we return to. \cite{murphy24} found that this offset is primarily driven by a temperature gradient between the two terminators, which is supported by the consistency of the magnitude of this offset between observations, which we return to in Section~\ref{subsec:compare2models}. The F322W2 spectra are negligibly changed from that of \cite{murphy24}, displaying strong features around 2.7~\microns{}, shown to be caused by water and CO$_2$, and 3.3~\microns{}, caused by methane \citep{welbanks24}. The F444W morning limb spectrum displays a notable feature at 4~\microns{}, shown to be due to SO$_2$ \citep{welbanks24}, but interestingly the same feature does not appear in the evening limb spectrum. However, the evening limb spectrum does exhibit known SO$_2$ features at 7.5~\microns{} and between 8--10~\microns{}. This hints at a potential difference in the SO$_2$ abundances between the evening and morning terminators, which we explore further in Section~\ref{subsec:SO2variations}. Both limbs exhibit the strong 4.3~\microns{} CO$_2$ feature in the F444W data, but with different relative amplitudes. This may hint at a difference in the CO$_2$ abundances as well, which we explore further in Section~\ref{subsec:CO2variations}. Finally, we see a large, broad feature from approximately 6--11~\microns{} in the morning limb spectrum which does not appear in the evening limb spectrum. This feature is reminiscent of the cloud transmission features inferred from the combined-limb MIRI spectrum by both \cite{dyrek2023_wasp107b} and \cite{welbanks24}, and its prevalence on just the morning limb may indicate an inhomogeneous cloud distribution. We discuss this further in Section~\ref{subsec:clouds}

\subsection{The Impact of Stellar Contamination on Measuring Limb Asymmetry}\label{sec:stellarcontam}


\subsubsection{NIRISS and NIRSpec Results} \label{subsec:stellarcontam_niriss_and_nirspec_obs}

\begin{figure*}[ht!]
    \centering
    \includegraphics[width=\textwidth]{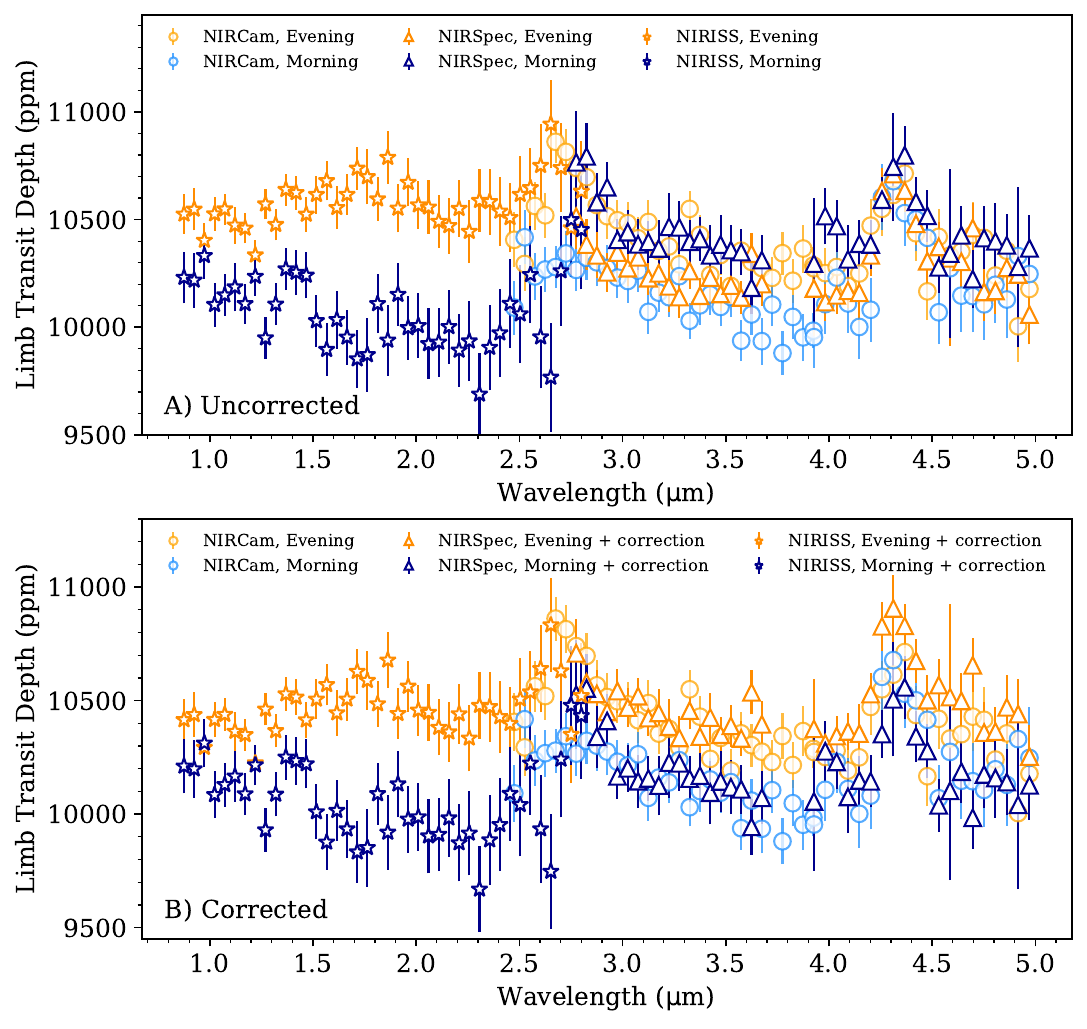}
    \caption{Comparing the evening (orange) and morning (blue) limb transmission spectra observed with JWST/NIRISS SOSS (stars) and NIRSpec G395H (triangles) to those with JWST/NIRCam F322W2 and F444W (both circles). Panel~A shows the nominal results for each instrument. Panel~B shows the result of applying the correction we develop for contamination from starspot crossings to the NIRISS and NIRSpec spectra, discussed in Section~\ref{subsec:stellarcontam_corrections}. These corrections are +110~ppm and +20~ppm uniform offsets to the NIRISS evening and morning limb spectra, respectively, and -195~ppm and +240~ppm offsets to the NIRSpec evening and morning limb spectra, respectively. 
    }
    \label{fig:contam_spectra}
\end{figure*}

As mentioned in the previous section, we find potential evidence that stellar contamination has affected the NIRISS and NIRSpec limb tranmission spectra. Panel~A of Figure~\ref{fig:contam_spectra} shows the best-fit NIRISS and NIRSpec limb spectra, with NIRISS represented by the star-shaped points and NIRSpec/G395H as the triangle points, compared to the NIRCam spectra (both F322W2 and F444W) from Figure~\ref{fig:spectra_dataonly} shown as the circular points. Consistent with NIRCam and MIRI, the NIRISS limb spectra exhibit limb asymmetry where the evening limb is generally larger. In the several channels where NIRISS and NIRCam overlap, the morning limb spectra agree well, although the NIRISS spectrum exhibits large scatter toward the red-edge of its bandpass. In NIRISS' evening limb spectrum, the relative amplitude of the 2.7~\microns{} feature is similar to that of NIRCam, but appears to be offset to higher absolute transit depths than NIRCam. The NIRSpec spectra, on the other hand, appear far more discrepant. The NIRSpec spectra are morphologically similar to the NIRCam spectra, evident by the similar slopes from $\sim$3.0--3.5~\microns{}, the 4~\microns{} feature in the morning limb spectrum, and the large 4.3~\microns{} features in each limb's spectrum. However, there appears to be a significant relative offset, such that the polarity of the limb asymmetry in the NIRSpec spectra is opposite that of NIRCam. The NIRSpec and NIRISS spectra also overlap by roughly two channels, and exhibit the same polarity difference. 

Ignoring the discrepancies relative to NIRCam and one another for now, there are several interesting potential features in the NIRISS and NIRSpec limb spectra. The NIRISS morning limb spectrum appears to have a strong blueward slope from $\sim$2.4~\microns{} to the blue end of the bandpass. If real, this slope could be indicative of the presence of some cloud or haze, potentially the same aerosol that causes the broad absorption seen in MIRI. On the other hand, the NIRISS evening limb spectrum does not appear to have such a strong blueward slope, potentially supporting the idea that WASP-107~b's evening limb is relatively clear compared to its morning limb (discussed further in Section~\ref{subsec:clouds}). For NIRSpec, similar to NIRCam F444W, the morning limb spectrum exhibits a strong feature at 4~\microns{}, most likely caused by SO$_2$, whereas this region is relatively flat in the evening limb spectrum. Neglecting their offsets, the relative amplitudes of this 4~\microns{} feature between the evening and morning limb spectra are consistent with that of NIRCam, and potentially support the idea that WASP-107~b's morning limb has higher SO$_2$ abundance. 

\subsubsection{Potential Evidence for Spot Crossings} \label{subsec:stellarcontam_crossings}

In order to investigate potential light curve features, like spot crossings, in each observation, we derived the broadband (i.e., band-integrated) light curve for each instrument/mode's observation following the same reduction methods previously described (Section~\ref{subsec:datareduction}). We turned to the broadband light curves since they have generally lower photometric scatter than in-band spectroscopic light curves, and marginalize over any noise that may only be present in one or a few individual channels. We then fit each broadband light curve individually, following the same procedure we used to fit the spectroscopic light curves (Section~\ref{subsec:datafitting}). We note that the best-fit limb depths were consistent with the spectroscopic limb depths in each case. 

\begin{figure*}
    \centering
    \includegraphics[width=\textwidth]{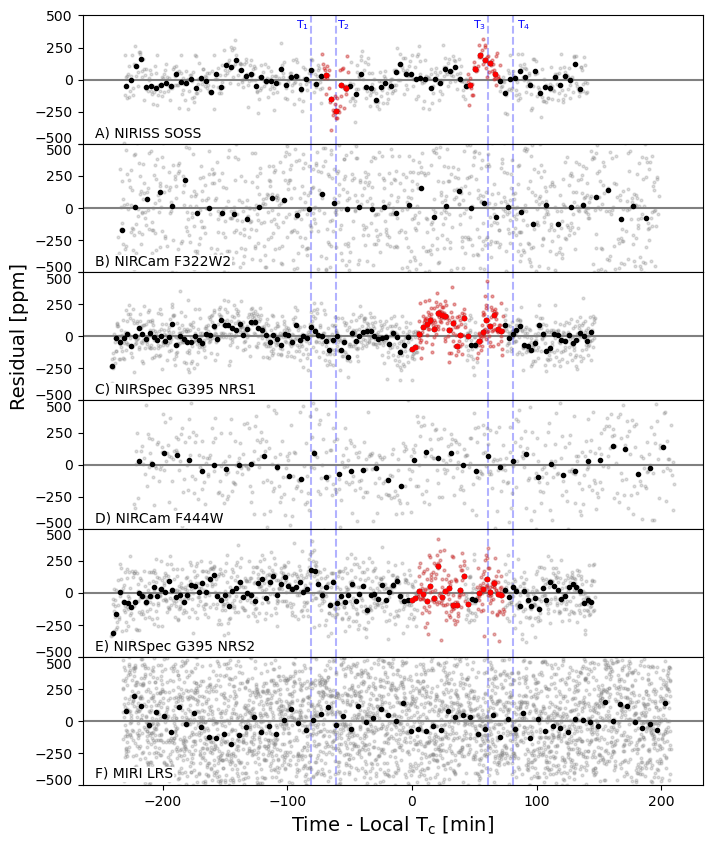}
    \caption{Data-model residuals from fitting the broadband transit light curves of each instrument's observation of WASP-107~b. Each time series is aligned relative to its local time of conjunction using the ephemeris of \cite{murphy24}. We binned the SOSS, F322W2, NRS1, F444W, NRS2, and MIRI light curves into 2.2, 3.4, 0.9, 7.0, 0.9, and 0.7~minute bins, respectively, so that all time series shown have similar scatter of $\sim$75~ppm. Binned data are in black, and the unbinned data are in gray. The vertical blue dashed lines mark the four approximate transit contact points for reference. We see potential starspot crossings, marked in red, during the NIRISS SOSS visit (Panel~A) and the NIRSpec G395H visit (Panels~C \& E, simultaneous observations in two bands). We do not see evidence for spot crossings during either of the NIRCam or MIRI visits.
    }
    \label{fig:broadbandresidualsgallery}
\end{figure*}

Figure~\ref{fig:broadbandresidualsgallery} shows the time series of residuals between each instrument/mode's broadband light curves and the corresponding best-fit model. We order the time series according to wavelength. There appear to be several large bump-like features during the NIRISS (Panel~A) and NIRSpec (Panels B \& E) visits, highlighted in red, that are potentially starspot crossings. We identified these by eye by their similar shape and duration to previous starspot crossings observed in this system by \cite{mocnik2017_wasp107b}, \cite{dai2017_wasp107b}, and \cite{sing2024_wasp107b}. We identify two potential spot crossings during the NIRISS visit: one coinciding with the end of transit ingress near contact point T$_2$, and another at the beginning of egress near contact point T$_3$. There also appear to be two spot crossings during the NIRSpec visit: one near the middle of transit, and another also coinciding with T$_3$. We note that the mid-transit spot in the NIRSpec visit was also identified by \cite{sing2024_wasp107b} in their analysis of the data, but not the egress spot. These are most visible in the NRS1 time series (Panel~C) due to its reduced scatter relative to NRS2 (Panel~E). On the other hand, we see no spot crossings or bump-like features during either of the NIRCam or MIRI visits. We do note that the photometric precision of the NIRCam and MIRI light curves are relatively lower than that of NIRISS and NIRSpec though, which may hide low-amplitude crossings or features. 

\subsubsection{Potential Correlated Noise} \label{subsec:stellarcontam_lcnoise}

Another possible explanation for the various features seen in the residual time series shown in Figure~\ref{fig:broadbandresidualsgallery} could be correlated light curve noise. JWST is still early in its total science lifetime, especially if it indeed operates for the anticipated 20~years until its propellant is expended \citep{Rigby2023}, so the full behavior and correction of systematic errors in each JWST instrument are still being determined. There have been several instances already where systematic noise, sometimes of an undetermined source, has affected the results of JWST transmission spectroscopy \citep[e.g.,][]{alderson2024_toi836b_t0systematicerrors, cartermay2024_w39datasynth}.

\begin{figure}
    \centering
    \includegraphics[width=\columnwidth]{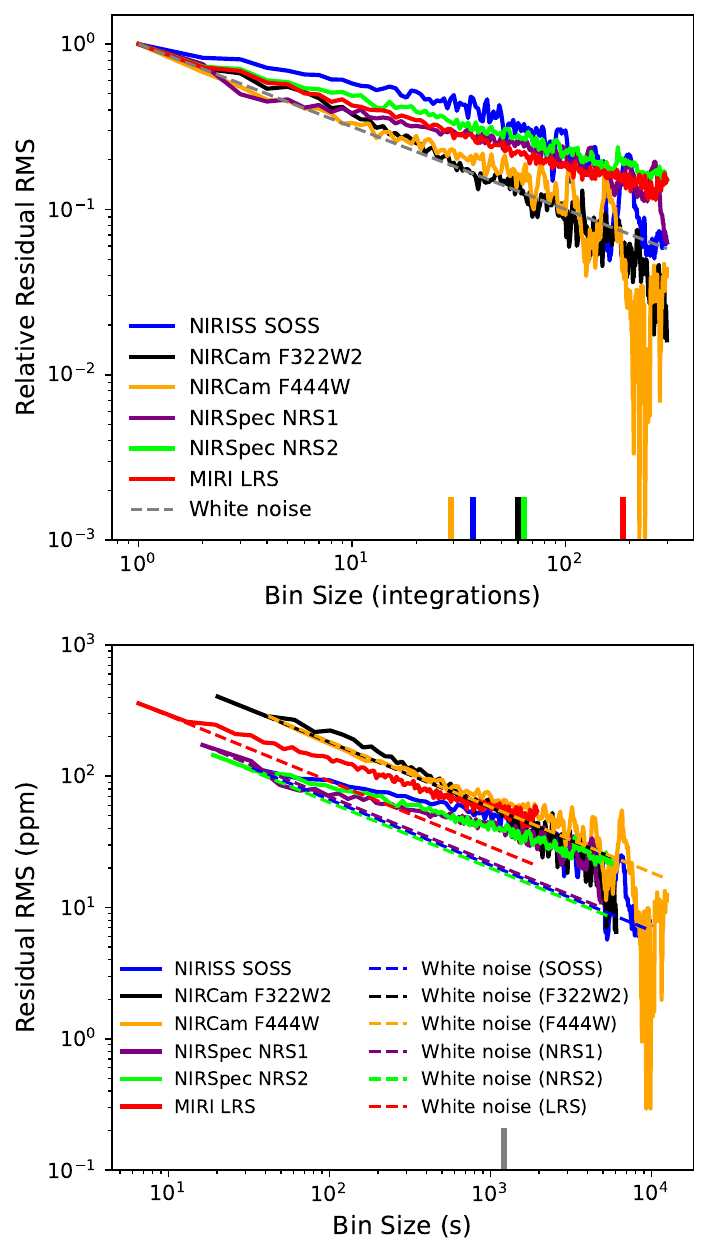}
    \caption{Results of time-averaging tests on each of our broadband fits (Figure~\ref{fig:broadbandresidualsgallery}). Top: Plotting the root-mean-square (RMS) of the data-model residuals, relative to the unbinned case, as a function of the number of integrations per bin $N$. The expected behavior of pure white noise is indicated by the dashed gray line, which falls off as $1/\sqrt{N}$. Bottom: Plotting the raw RMS as a function of the actual temporal size of each bin. The expected white noise behaviors, different for each instrument, are plotted as dashed colored lines. Results for NIRISS are shown in blue, NIRCam F322W2 in black, NIRCam F444W in orange, NIRSpec NRS1 in purple, NIRSpec NRS2 in lime, and MIRI LRS in red. At the bottom, the small vertical lines indicate the length of transit ingress/egress for reference, either in units of each instrument's integration time (top) or the actual time (bottom).}
    \label{fig:broadbandAVplots}
\end{figure}

To explore whether correlated noise is still present in the light curves, we conducted time-averaging tests, as introduced by \cite{pont2006}, on each of the broadband fits, measuring the root-mean-square of data-model residuals as a function of temporal bin size. Figure~\ref{fig:broadbandAVplots} shows the results of these tests for each instrument/mode, showing both the relative (top) and absolute (bottom) RMS trends for the fairest comparison between instruments. When the residual RMS deviates from the expected behavior for white noise, which falls off with the square root of the number of integrations per bin, this is evidence for additional correlated or red noise in the data. We find that both NIRCam observations, shown as the black and orange lines, are fully consistent with white noise behavior and exhibit no evidence for unmitigated correlated noise. On the other hand, the NIRISS (blue), NIRSpec (purple and lime), and MIRI (red) results deviate from the white noise expectation, both in the relative and absolute cases, suggesting that residual red noise may be present in these data. 

If correlated red noise in the NIRISS and NIRSpec data is the source of the various bumps seen in the broadband residuals (Figure~\ref{fig:broadbandresidualsgallery}), then this noise may also be causing the difference in these instruments' limb spectra relative to NIRCam. The MIRI limb spectra do not appear significantly inconsistent with NIRCam (Figure~\ref{fig:spectra_dataonly}), though such comparison is less precise without overlapping wavelength bins. In either case, red noise may at least be causing some of the scatter seen in the MIRI limb spectra (e.g., shown later in comparison to the models in Figure~\ref{fig:spectra_withmodels}), particularly at its shortest wavelengths. We cannot fully rule out this red noise being the cause of the discrepancies we seen between the limb spectra. However, given the similar durations of each spot-like bump in the residual time series and the fact that spot crossings have been previously observed several times for this system \citep{mocnik2017_wasp107b, dai2017_wasp107b}, spots are also a likely explanation. Further, even if spots are not the cause in our case, understanding what effect spot crossings, or more generally any bump-like feature in the light curve, would have on observed limb spectra has yet to be explored in the literature, and is necessary to investigate. For these reasons, in the rest of this work we assume spot crossings to be the cause of these features in the broadband residuals, and the primary contributors of the offsets we observed in the NIRISS and NIRSpec limb spectra (Figure~\ref{fig:contam_spectra}). As we will show later on, the corrections we derive based on this assumption work well to bring these data into agreement with \mbox{NIRCam}, and may still be applicable even if correlated noise is the true root of this issue. 

\subsubsection{Modeling The Effect of Occulted Spots on Limb-resolved Spectra} \label{subsec:stellarcontam_spotmodels}

To investigate what impact these spot crossings may be having on the observed limb transmission spectra (Figure~\ref{fig:contam_spectra}, Panel~A), we performed a series of injection-recovery tests. We describe these tests in detail in the Section~\ref{apxsec:starspots} of the Appendix. In short, using \texttt{starry} \citep{luger19_starry_main, luger21_starry_extra1, luger21_starry_extra2}, we simulated the surface intensity map of WASP-107 in both a spot-free case and a case with one spot, basing the model spot's properties on the observed crossings in the NIRISS and NIRSpec data. Then, using \texttt{catwoman}, we simulated and fit transit observations of a hypothetical WASP-107~b across each stellar disk for both a control case in which WASP-107~b has uniform limbs, and a test case in which WASP-107~b has limb asymmetry to evaluate how well the true limb depths can be recovered. The result of these tests is a model of the bias on the measured limb depths induced by a spot crossing, which can be used to correct for this bias in our real data. 

\subsubsection{Correcting for Spots in the NIRISS and NIRSpec Observations} \label{subsec:stellarcontam_corrections}

Using our model for the effect of spot crossings on the observed limb transmission spectra derived in Section~\ref{apxsec:starspots} of the Appendix, we can correct the nominal NIRISS and NIRSpec limb transmission spectra for the crossings during each visit. For each individual spot, we use the difference between the retrieved evening and morning limb transit depths and the true, injected value in the control case (Panel~E of Figure~\ref{fig:spoteffects}) as an additive correction factor. Comparing the time series in Figure~\ref{fig:broadbandresidualsgallery} to the mapping between time (during transit) and spot position in Panel~C of Figure~\ref{fig:spoteffects}, the two spots during the NIRISS visit are located at longitudes of approximately -60$^\circ$ and +60$^\circ$ on the stellar disk, and the two spots during the NIRSpec visit at approximately +20$^\circ$ and +60$^\circ$. Due to the spot at -60$^\circ$ in the NIRISS data, our tests suggest the observed evening limb transit depth is increased by 230~ppm and the morning limb depth is decreased by 150~ppm. Due to the spot at +60$^\circ$ in both data sets, the evening limb depth is decreased by 120~ppm and the morning limb depth is increased by 170~ppm. Then, due to the spot at +20$^\circ$ in the NIRSpec data, the evening limb depth is decreased by 75~ppm and the morning limb depth is increased by 70~ppm. Each of these values accounts for the fact that the mere presence of a spot biases the observed depth down by approximately 47~ppm. Adding these offsets together, assuming for simplicity that they are not chromatic, the NIRISS evening limb spectrum is offset by +110~ppm and the NIRISS morning limb spectrum by +20~ppm, while the NIRSpec evening limb spectrum is offset by -195~ppm and the NIRSpec morning limb spectrum by +240~ppm. It is important to note that, in reality, starspots have wavelength-dependent contrasts. Nevertheless, as discussed in the next paragraph we find that this simplified correction still works quite well, and fully measuring and accounting for the chromatic spot contrast is beyond the scope of this work. 

We correct for these spot-induced offsets by subtracting them from the nominal best-fit spectra (Panel~A of Figure~\ref{fig:contam_spectra}). The resulting spot-corrected NIRISS and NIRSpec limb spectra are shown in Panel~B of Figure~\ref{fig:contam_spectra}, again compared to the spot-free NIRCam spectra. Though this correction method is simple and makes a number of assumptions, we find that it brings the NIRISS and NIRSpec limb spectra into relatively good agreement with the NIRCam spectra. The evening and morning depths between the corrected NIRISS and NIRCam spectra in the 2.7~\microns{} feature overlap well, though there are a couple of points of high scatter at the red-edge of the NIRISS spectrum. The overall evening--morning offset between the corrected NIRSpec spectra agrees nearly exactly with what is seen in NIRCam, except for some points of high scatter at NIRSpec's blue edge. Altogether, this shows that while spots can have a significant impact on the observed limb transmission spectra, they can be corrected for. Further modeling on this effect, particularly into its chromaticity, would be beneficial to the community going forward. 


\subsection{Comparison to Literature Model Spectra} \label{subsec:compare2models}

\begin{figure*}[th!]
    \centering
    \includegraphics[width=\textwidth]{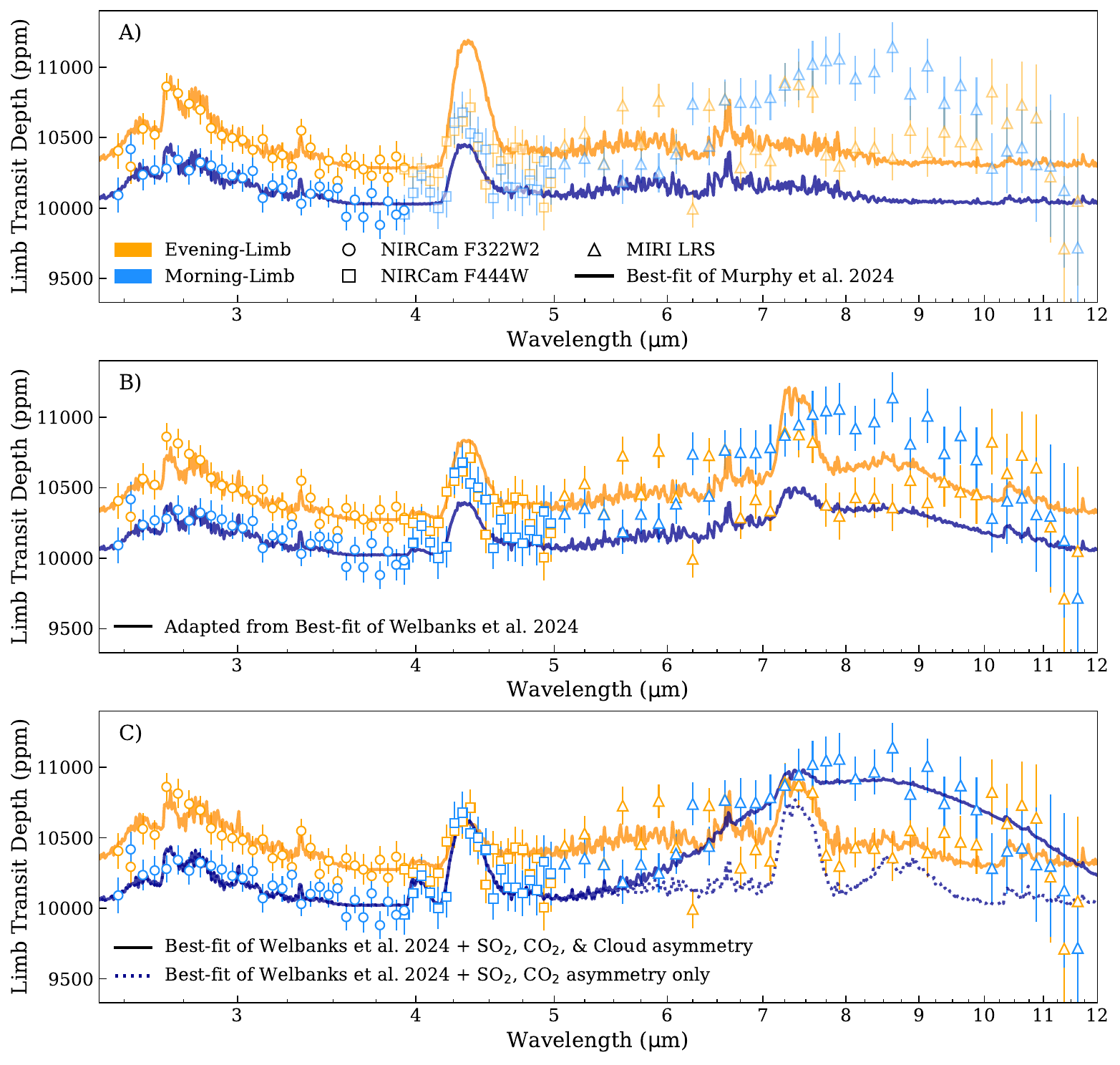}
    \caption{Comparing various model scenarios to our observed limb spectra of WASP-107~b. Panel~A shows the best-fit models of \cite{murphy24}, whom initially inferred an evening--morning temperature difference based on the NIRCam F322W2 spectra, extended to our wider wavelength range. We reduced the opacity of the F444W and MIRI points to emphasize that these data were not used in \cite{murphy24}. Panel~B shows the best-fit models of \cite{welbanks24}, recomputed for the differing evening and morning limb temperatures. We find evidence for evening--morning variation in the abundances of SO$_2$ and CO$_2$ as well as heterogeneous cloud distribution. The solid line in Panel~C shows the result of including these asymmetries, while the dotted line includes just the molecular asymmetries.}
    \label{fig:spectra_withmodels}
\end{figure*} 

\citet{murphy24} performed initial fits to just the F322W2 evening and morning spectra, fixing the majority of atmospheric parameters to values based on \citet{dyrek2023_wasp107b} and \citet{welbanks24} and then grid-fitting for the temperature and gray cloud opacity on each limb. From this, \cite{murphy24} inferred temperatures of approximately 790~K and 610~K on the evening and morning limbs (a difference of $\sim$180~K), respectively, and found tentative but inconclusive evidence for a small difference in gray cloud opacity as well. We compare our panchromatic spectra to the best-fit models of \citet{murphy24}, extended in wavelength range, in Panel~A of Figure~\ref{fig:spectra_withmodels}. We find that while these models fit the F322W2 portion of the data quite well, there are several shortcomings at longer wavelengths largely because F322W2 alone has insufficient constraining power for other molecules. These models did not include SO$_2$ and naturally do not fit the SO$_2$ features seen around 4~\microns{} and 7.5--10~\microns{}. Also, these models assume the equilibrium abundance of CO$_2$ at the corresponding temperatures, which we find overpredicts the size of the 4.5~\microns{} CO$_2$ feature on the evening limb, and underpredicts it on the morning limb. Finally, with no cloud opacity besides a gray cloud in the model, these models cannot fit the large feature from 6--10~$\mathrm{\mu}$m on the morning limb. Besides these shortcomings with individual spectral features, the overall offset between the model evening and morning limb spectra is consistent with that seen in the observed spectra, outside of strong molecular features. This supports the evening--morning temperature difference initially measured by \cite{murphy24}.

\citet{welbanks24} performed various retrievals on the combined F322W2, F444W, MIRI/LRS, as well as an HST/WFC3 combined-limb spectrum. From these, \citet{welbanks24} precisely measured the average abundances of the major molecules in WASP-107b's atmosphere including H$_2$O, CO$_2$, CH$_4$, and SO$_2$. Also, \citet{welbanks24} found that the 6--11~\microns{} cloud feature was best fit by a microphysics-agnostic Gaussian-shaped feature after finding that models with typically-considered cloud species, including silicates and salts, could not match the shape of the spectrum. We recomputed the best-fit models from \cite{welbanks24} at the morning and evening limb temperatures inferred by \cite{murphy24}, keeping the abundances of H$_2$O, CO$_2$, CH$_4$, SO$_2$, and NH$_3$ fixed to the \texttt{Aurora} retrieval values (see Extended Data Table 2 of \cite{welbanks24}). These are shown in Panel~B of Figure~\ref{fig:spectra_withmodels}. In comparison to our evening and morning spectra, these models improve considerably upon those of \cite{murphy24} outside of the F322W2 bandpass, and are generally able to fit the majority of the spectrum well. This further supports the $\sim$180~K evening-morning temperature difference from \cite{murphy24}. We still see several discrepancies though. The evening SO$_2$ feature at 4~\microns{} is well fit, but this feature is underpredicted on the morning limb. The same is true for the CO$_2$ feature at 4.5~\microns{}. The evening limb spectrum redward of $\sim$6.5~\microns{} is generally overpredicted by a nearly uniform amount, likely due to the added cloud opacity at these wavelengths in the model. On the other hand, the morning limb spectrum in MIRI is underpredicted, and the broad observed feature at these wavelengths is not well fit. 
%


From this comparison to the adapted models of \cite{welbanks24}, we find supporting evidence for potential chemical variations in SO$_2$ and CO$_2$ as well as heterogeneous cloud distribution on top of the underlying evening--morning temperature gradient inferred by \cite{murphy24}. We explore each of these three possibilities in detail in the following Sections~\ref{subsec:SO2variations} -- \ref{subsec:clouds}. As a preview of the results, the solid line Panel~C of Figure~\ref{fig:spectra_withmodels} shows the result of including these asymmetries into the model of \cite{welbanks24}, yielding a good and self-consistent fit of both the evening and morning-limb spectra from $\sim$2.5--11~\microns{}. For reference of the individual effect of this potential cloud in the morning-limb spectrum, the blue dotted line shows the morning-limb model without including it. All together, the NIRCam and MIRI limb spectra are fit well considering the evening-morning temperature difference from \cite{murphy24} in addition to these compositional and cloud heterogeneities.

\subsection{Evidence for Morning-Evening SO$_2$ Variations} \label{subsec:SO2variations}

By separating the transmission spectra of the evening and morning terminators, we have a unique opportunity to probe spatial variations in the abundance of the photochemical product SO$_2$ in an exoplanet's atmosphere. There have already been several estimates of the average SO$_2$ abundance in WASP-107~b's atmosphere in the literature, each derived from the limb-combined transmission spectrum. Using just MIRI, which probes two strong SO$_2$ features, \cite{dyrek2023_wasp107b} infer log$_{10}$ SO$_2$ volume mixing ratios of $\text{log}_{10}\left(\text{SO}_2\right)$ = -5.03$^{+0.33}_{-0.18}$ and -6.72$^{+0.30}_{-0.23}$ from two different retrieval frameworks: \texttt{ARCiS} and \texttt{pRT}, respectively. Using NIRCam in addition to MIRI, and thus probing three strong SO$_2$ features, \cite{welbanks24} infer intermediate values of $\text{log}_{10}\left(\text{SO}_2\right)$ = -5.7 $\pm$ 0.4 from a free retrieval with \texttt{Aurora} and $\text{log}_{10}\left(\text{SO}_2\right)$ = -5.2 $\pm$ 0.2 from grid-based retrieval with \texttt{CHIMERA}. As seen in Figure~\ref{fig:spectra_withmodels} though, these updated SO$_2$ abundances from \cite{welbanks24} do not seem to well fit the SO$_2$ absorption features in individual limb transmission spectra. \cite{sing2024_wasp107b} also estimated the SO$_2$ abundance using just the NIRSpec observations, which cover just the single $\sim$4.0~\microns{} feature. They find relatively larger values of $\text{log}_{10}\left(\text{SO}_2\right)$ = -5.06$^{+0.14}_{-0.15}$ from a free retrieval with \texttt{ATMO} and $\text{log}_{10}\left(\text{SO}_2\right)$ = -4.59$^{+0.19}_{-0.18}$ from a free retrieval with \texttt{NEMESIS}.

\begin{figure*}
    \centering
    \includegraphics[width=\textwidth]{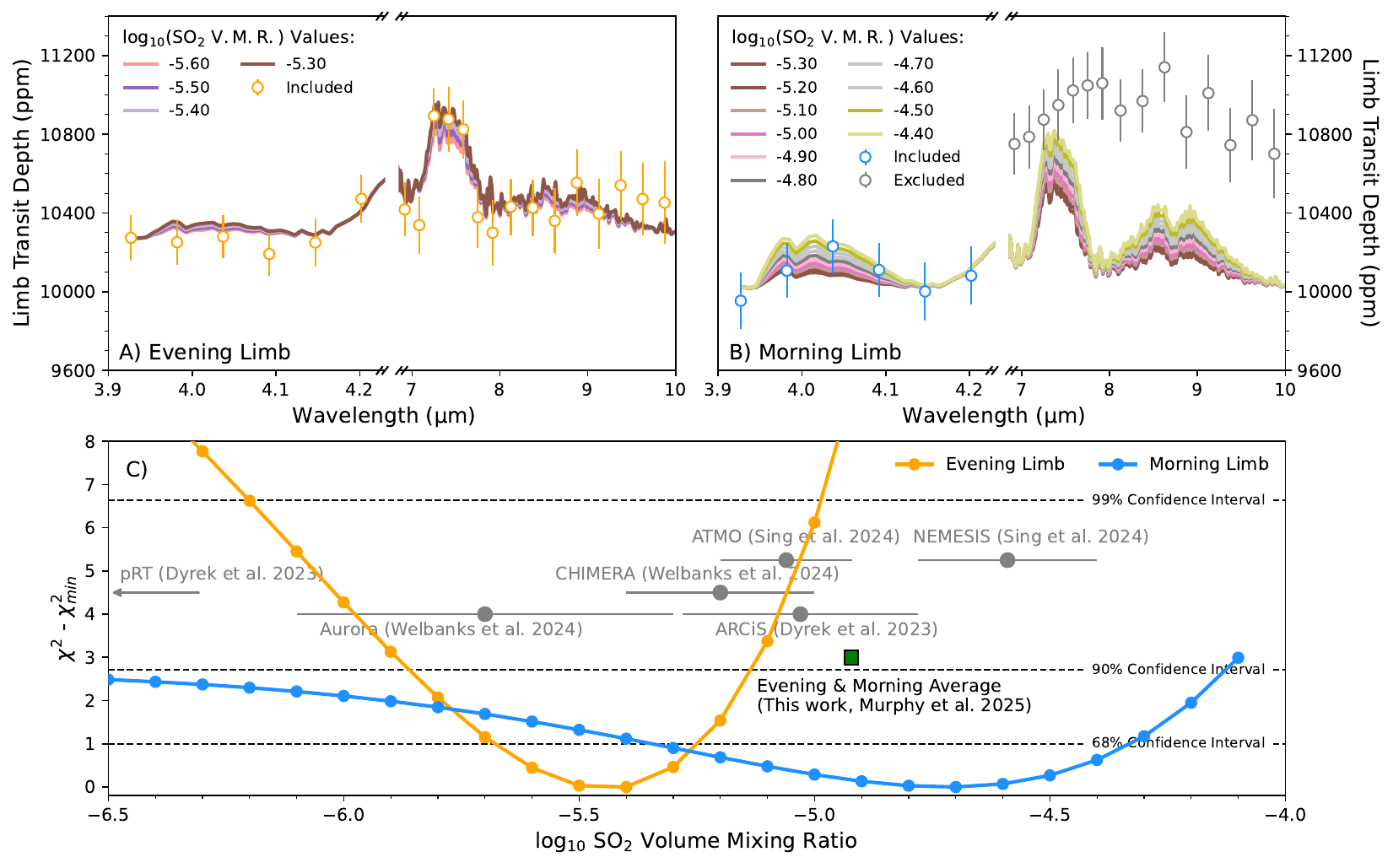}
    \caption{Investigating the abundances of SO$_2$ on WASP-107~b's evening and morning limbs. We performed a $\chi^2$ comparison between our observed limb spectra and model spectra with varying SO$_2$ abundance. Panel~A shows the evening limb spectrum from NIRCam F444W and MIRI LRS truncated to points within prominent SO$_2$ features, with all models consistent within the 68$\%$ confidence level. Panel~B shows the same for the morning limb spectrum, with the points excluded from this calculation due to strong overlapping cloud absorption in MIRI plotted in gray. Panel~C shows the full results of these tests, plotting the calculated $\chi^2$ value, relative to the minimum, as a function of the input SO$_2$ abundance for the evening (orange) and morning (blue) limb cases. The corresponding confidence intervals based on \cite{practicalstatsforastrobook} are plotted as horizontal lines. For reference, we also show SO$_2$ abundances previously inferred from the limb-combined spectrum in the literature.}
    \label{fig:so2biggallery}
\end{figure*}

To test for and quantify a potential evening--morning variation in SO$_2$ abundance, we generated a grid of model spectra for each limb with varying SO$_2$ abundances. We set the limb temperatures based on \cite{murphy24} and all other abundances based on \cite{welbanks24} as in Panel~B of Figure~\ref{fig:spectra_withmodels}, then we varied the volume mixing ratio from $\text{log}_{10}\left(\text{SO}_2\right)$ = -6.5 to -4.0 in steps of 0.1. We then calculated the $\chi^2$ value between these model spectra and the observed spectra, focusing only on points at wavelengths of significant SO$_2$ absorption. For the evening limb, this included 6 channels from $\sim$3.9 to 4.2~\microns{} to cover the feature in the NIRCam/F444W bandpass and 15 channels from $\sim$7 to 10~\microns{} to cover the several features in the MIRI/LRS bandpass, as indicated in Panel~A of Figure~\ref{fig:so2biggallery}. For the morning limb, we include the same 6 NIRCam/F444W channels but exclude the MIRI/LRS channels due to the apparently dominant overlapping cloud absorption at those wavelengths. These chosen channels are indicated in Panel~B of Figure~\ref{fig:so2biggallery}. We used the $\chi^2$ value to evaluate the goodness-of-fit of each model following the method of \cite{practicalstatsforastrobook}, where the confidence intervals at which a given model can be rejected are defined relative to the minimum $\chi^2$ value for a given number of degrees of freedom. In this case, there is one degree of freedom (the SO$_2$ abundance) and the 68\%, 90\%, and 99\% confidence intervals are at relative $\chi^2$ values of 1, 2.71, and 6.63, respectively.

Figure~\ref{fig:so2biggallery} shows the results of this $\chi^2$ test. Panel~A shows the observed evening limb spectrum, focused just on the relevant channels between NIRCam and MIRI, compared to the models that fall below the 68\% confidence interval. Panel~B shows the same for the morning limb spectrum. Panel~C shows the relative $\chi^2$ curves as a function of the input SO$_2$ abundance for each limb, compared to the literature average abundance values previously mentioned. These tests indeed show a higher volume mixing ratio of SO$_2$ on WASP-107~b's morning limb than on its evening limb, at least at the pressures probed in transmission. For the evening limb, we find the minimum $\chi^2$ at a volume mixing ratio of $\text{log}_{10}\left(\text{SO}_2\right)$ = -5.4, and values between $\text{log}_{10}\left(\text{SO}_2\right)$ = -5.7 and -5.3 cannot be rejected beyond 68\% confidence. The morning limb distribution is less tightly constrained, likely due to the fewer data points used in the calculation, but prefers significantly higher amounts of SO$_2$. The distributions negligibly overlap below the 68\% confidence interval. The morning limb's minimum $\chi^2$ is at a value of $\text{log}_{10}\left(\text{SO}_2\right)$ = -4.7, and values between approximately $\text{log}_{10}\left(\text{SO}_2\right)$ = -5.3 and -4.3 cannot be rejected. These suggest that WASP-107~b's SO$_2$ abundance may be 5$\times$ higher at its cooler morning terminator than at the hotter evening terminator.

To compare our limb-resolved abundances to the limb-combined retrieved values in the literature, we calculated the arithmetic mean of the minimum $\chi^2$ values for each limb. This yields an average SO$_2$ volume mixing ratio of $\text{log}_{10}\left(\text{SO}_2\right)$ = -4.92. As shown in Panel~C of Figure~\ref{fig:so2biggallery}, this value is slightly larger than the literature values except for the \texttt{NEMESIS} retrieval of \cite{sing2024_wasp107b}, but is relatively consistent with those of \texttt{ARCiS} by \cite{dyrek2023_wasp107b}, \texttt{CHIMERA} by \cite{welbanks24}, and \texttt{ATMO} by \cite{sing2024_wasp107b}. This demonstrates that retrievals on the limb-combined transmission spectrum can accurately retrieve the average abundance of a molecule. 

Since they overlap in wavelength, the NIRSpec G395H limb transmission spectra could provide an independent check of this SO$_2$ asymmetry seen by NIRCam. However, we did not include them in the above tests due to the spot contamination and a likely degeneracy between this contamination, our attempted correction (Section~\ref{subsec:stellarcontam_corrections}), and the modelled SO$_2$ abundance. Nevertheless, we can use the corrected NIRSpec limb spectra as a qualitative check. As can be seen in Panel~B of Figure~\ref{fig:contam_spectra}, the amplitudes of the 4.0~\microns{} SO$_2$ feature on each limb are very consistent between NIRSpec (corrected) and NIRCam. This supports the existance of a heterogeneous SO$_2$ distribution on WASP-107~b.

\subsection{Evidence for Morning-Evening CO$_2$ Variations} \label{subsec:CO2variations}

\begin{figure*}[ht!]
    \centering
    \includegraphics[width=\textwidth]{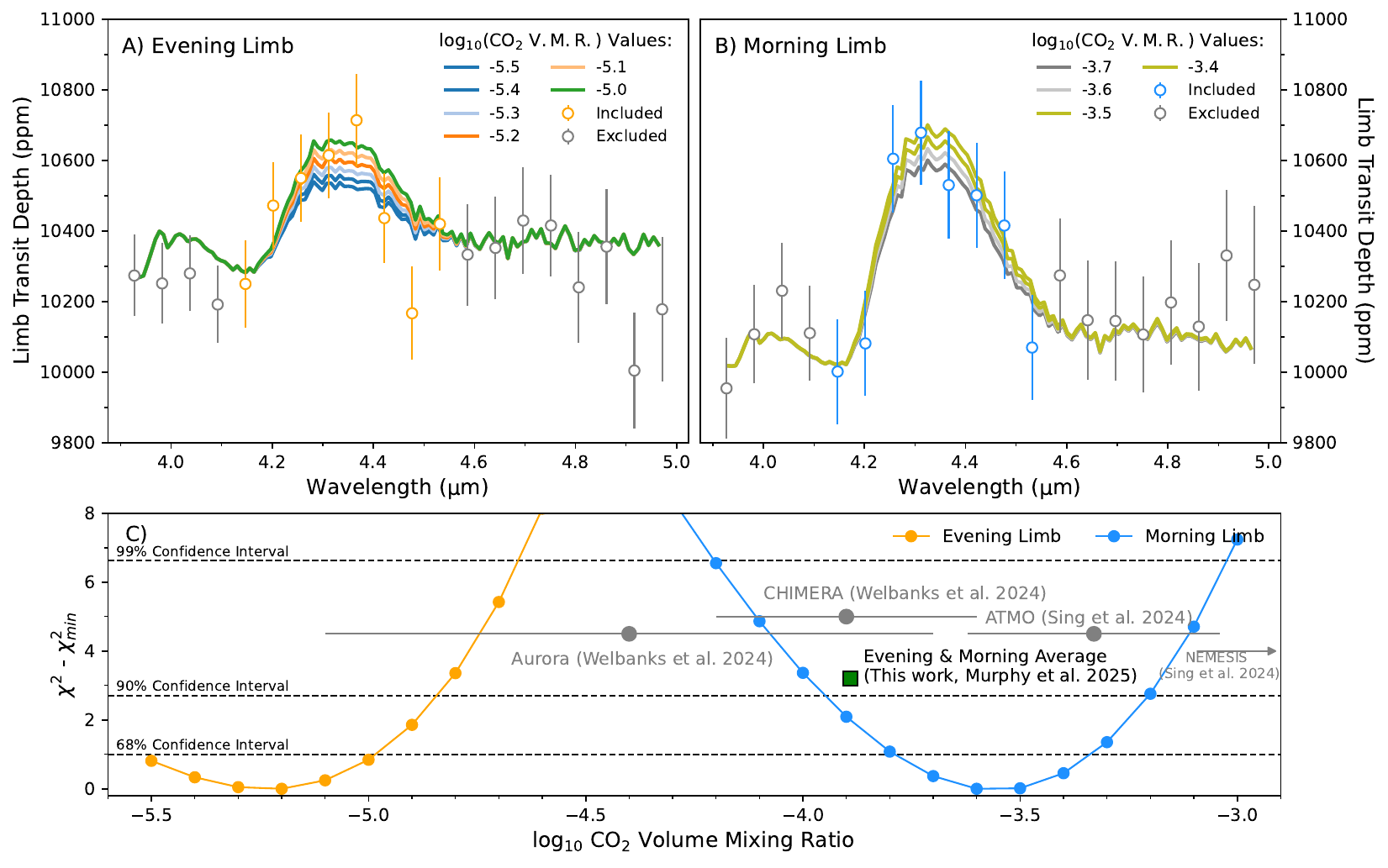}
    \caption{Investigating the abundances of CO$_2$ on WASP-107~b's evening and morning limbs. We performed a $\chi^2$ comparison between our observed limb spectra and model spectra with varying CO$_2$ abundance, focusing on the $\sim$4.5~\microns{} feature in NIRCam/F444W. Panel~A shows the evening limb spectrum with all models consistent within the 68$\%$ confidence level, and Panel~B shows the same for the morning limb spectrum. Panel~C shows the full results of these tests, plotting the $\chi^2$ value, relative to the minimum, as a function of the input CO$_2$ abundance for both the evening (orange) and morning (blue) limb cases. The corresponding confidence intervals based on \cite{practicalstatsforastrobook}, are plotted as horizontal lines. For reference, we also show CO$_2$ abundances previously inferred for WASP-107~b from the limb-combined spectrum in the literature.}
    \label{fig:CO2figure}
\end{figure*} 

Similar to SO$_2$, we also observe an asymmetry in the amplitude of the $\sim$4.35~$\mu$m CO$_2$ feature between WASP-107~b's evening and morning limbs, as highlighted in Figure~\ref{fig:spectra_withmodels}. The limb depths within this feature reach the same peak value, but since the morning limb's surrounding baseline is lower this means the morning limb feature is larger, and suggests that more CO$_2$ is present on the morning-limb as well. In the literature, \cite{dyrek2023_wasp107b} were unable to reliably estimate the average CO$_2$ abundance due to its insignificant contribution to the transmission spectrum in the MIRI bandpass. From the NIRCam data, \cite{welbanks24} inferred values of $\text{log}_{10}\left(\text{CO}_2\right)$ = -4.4 $\pm$ 0.7 with \texttt{Aurora} and $\text{log}_{10}\left(\text{CO}_2\right)$ = -3.9 $\pm$ 0.3 using \texttt{CHIMERA}. From the NIRSpec data, \cite{sing2024_wasp107b} inferred larger values of $\text{log}_{10}\left(\text{CO}_2\right)$ = -3.33$^{+0.29}_{-0.25}$ with \texttt{ATMO} and $\text{log}_{10}\left(\text{CO}_2\right)$ = -2.62$^{+0.36}_{-0.34}$ with \texttt{NEMESIS}. 

To further investigate this asymmetry, we repeated the same $\chi^2$ comparison test that we did for SO$_2$ in the previous section. We generated model transmission spectra at the temperature of each limb, varying the volume mixing ratios from $\text{log}_{10}\left(\text{CO}_2\right)$ = -5.5 to -3.5 in steps of 0.1. Then, we calculated the $\chi^2$ value between these models and the observations, restricting this calculation only to points within the $\sim$4.35~$\mu$m CO$_2$ feature. These included 8 channels between $\sim$4.2 and 4.55~\microns{}, indicated by the colored points in Panels~A and B of Figure~\ref{fig:CO2figure} which were the same for each limb. 

Figure~\ref{fig:CO2figure} shows the results of this CO$_2$ $\chi^2$ test. As before, Panel~A shows the observed evening limb spectrum compared to the models within the 68\% confidence interval, Panel~B shows the same for the morning limb spectrum, and Panel~C shows the relative $\chi^2$ curves as a function of CO$_2$ volume mixing ratio on each limb. In Panel~C, the average abundances inferred by \cite{welbanks24} and \cite{sing2024_wasp107b} are shown for comparison. Similar to SO$_2$, these tests suggest that there is a higher volume mixing ratio of CO$_2$ on WASP-107~b's cooler morning limb than on its evening limb. The difference in CO$_2$ abundance appears more extreme, and of higher statistical significance than for SO$_2$. For the evening limb, we find the minimum $\chi^2$ at a volume mixing ratio of $\text{log}_{10}\left(\text{CO}_2\right)$ = -5.2, with values less than -5.5 or greater than -5.0 disfavored at over 68\% confidence. For the morning limb, we find the minimum $\chi^2$ at $\text{log}_{10}\left(\text{CO}_2\right)$ = -3.6, with values less than -3.8 or greater than -3.3 disfavored. These distributions do not overlap even within the 99\% confidence interval, and suggest that CO$_2$ may be up to 40$\times$ more abundant on the morning limb than on the evening limb. 

The average of our $\chi^2$-minimizing evening and morning limb abundances is $\text{log}_{10}\left(\text{CO}_2\right)$ = -3.9. This is nearly exactly the value inferred by the \texttt{CHIMERA} grid-based retrievals of \cite{welbanks24} on the limb-combined spectrum, and within the 1-$\sigma$ uncertainty of their \texttt{Aurora} retrievals. This value is lower than that of both \cite{sing2024_wasp107b} retrievals. The consistency with at least the \texttt{CHIMERA} retrievals again demonstrates that retrievals on a limb-combined transmission spectrum can well retrieve the average abundance of a molecule. Further, this and the SO$_2$ average value from the previous section show that the average abundance value is dominated by the value on the limb where the molecule is most abundant. In the case of WASP-107~b, both were the morning limb. 

As in the previous section, we can use the corrected NIRSpec limb spectra as a qualitative check of this finding based on NIRCam. Interestingly, the relative strengths of the CO$_2$ feature on each limb appear to be slightly different between NIRSpec (corrected) and NIRCam, with NIRSpec having a slightly higher amplitude evening-limb feature and a smaller amplitude morning-limb feature than NIRCam. This would potentially suggest a smaller asymmetry in CO$_2$ abundance than the NIRCam data imply, assuming no temporal variation. However, the differences between the feature amplitudes are smaller than the corresponding uncertainties on those transit depths. Given this potential discrepancy, our inferred CO$_2$ asymmetry should be treated as an upper limit. It is also worth highlighting that the limb-combined CO$_2$ abundances inferred by \cite{sing2024_wasp107b} from NIRSpec are generally higher than those by \cite{welbanks24} from NIRCam, indicating that temporal or, more likely, systematic differences between the two instruments may also be at play here. Additional follow-up observations would be useful to verify the true level of asymmetry. 

In these tests (and the SO$_2$ tests in the previous section), we assume that the cloud opacity is the same on each limb in order to isolate the effect of changing the molecular abundance. However, it is possible that cloud opacity could also be contributing to the difference in feature amplitudes, especially given that the peaks of each limbs' CO$_2$ feature appear to reach the same absolute transit depth. This potential degeneracy is already partially explored by the models in Panel~A of Figure~\ref{fig:spectra_withmodels}, in which the CO$_2$ abundance is set to the equilibrium value at each limb's temperature, as determined by \texttt{CHIMERA}, but there is a difference in gray cloud opacity, with log$_{10} \left( \kappa_{cld} / cm^2~g^{-1} \right)$ = -1.78 on the morning limb and -1.58 on the evening limb \citep{murphy24}. As seen in Panel~A though, this simple scenario does not explain the data and, as mentioned, the equilibrium abundance of CO$_2$ vastly overpredicts the observed feature amplitude. To reduce the CO$_2$ feature's amplitude to the observed level relative to the surrounding wavelengths, the cloud opacity there would have to significantly increase. However, as discussed further in the next section, this seems inconsistent with the apparent lack of a cloud absorption feature at MIRI wavelengths. Follow-up work with more detailed modeling that encapsulates the combined and competing effects of temperature, molecular abundance, and cloud opacity -- including potentially non-gray clouds -- is needed to better understand this potential asymmetry in CO$_2$ abundance. 

\subsection{Evidence for Uniform H$_2$O Abundance} \label{subsec:H2Ovariations}

Motivated by the evidence for limb-to-limb variations in SO$_2$ and CO$_2$ abundance found in the previous sections, we used the same techniques to investigate potential variation in H$_2$O as well. There was no obvious hint of H$_2$O variation from the model comparisons in Figure~\ref{fig:spectra_withmodels}, and both chemical and 3D models generally predict that H$_2$O should be homogeneously distributed throughout the atmosphere \citep{fortney2020, zamyatina24_wasp96bGCMs_limbasym}. Under the assumption that H$_2$O should be homogeneous, repeating this test therefore also serves as a validity check for the tests in Sections~\ref{subsec:SO2variations} and \ref{subsec:CO2variations}. Further, this additional test may reveal the contribution of degenerate factors, such as clouds, to the SO$_2$ and CO$_2$ variations we infer, as discussed in the previous section.

\begin{figure*}
    \centering
    \includegraphics[width=\textwidth]{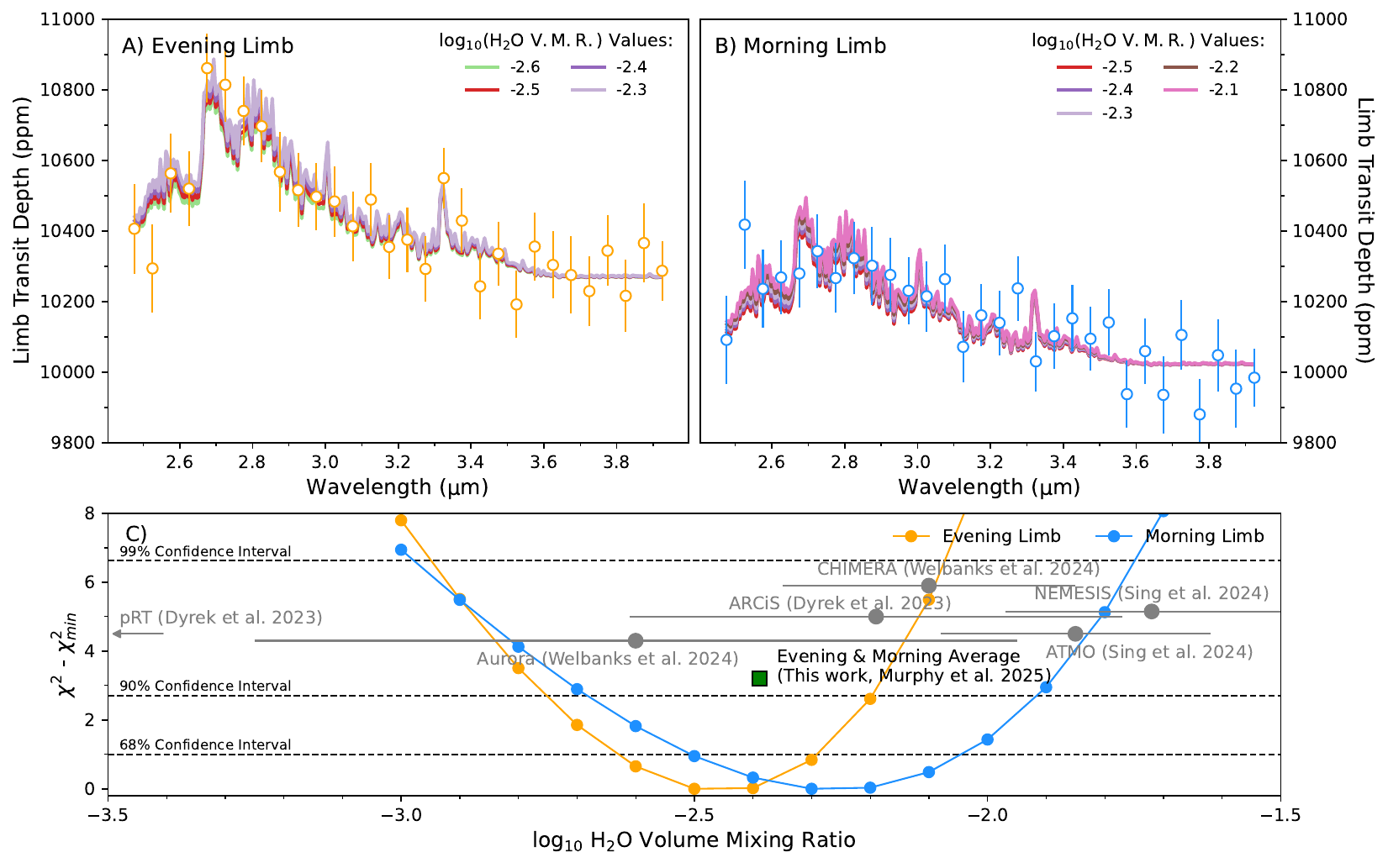}
    \caption{Investigating the abundances of H$_2$O on WASP-107~b's evening and morning limbs. As in Sections~\ref{subsec:SO2variations} and \ref{subsec:CO2variations}, we performed a $\chi^2$ comparison between our observed limb spectra and model spectra with varying H$_2$O abundance, focusing on NIRCam/F322W2. Panel~A shows the evening limb spectrum with all models consistent within the 68$\%$ confidence level, and Panel~B shows the same for the morning limb spectrum. Panel~C shows the full results of these tests, plotting the $\chi^2$ value, relative to the minimum, as a function of the input H$_2$O abundance for both the evening (orange) and morning (blue) limb cases. The corresponding confidence intervals based on \cite{practicalstatsforastrobook}, are plotted as horizontal lines. For reference, we also show H$_2$O abundances previously inferred for WASP-107~b from the limb-combined spectrum in the literature. 
    }
    \label{fig:H2Ofigure}
\end{figure*}

We repeated the $\chi^2$ comparison test using the same technique as in the previous sections, varying the volume mixing ratio of water from $\text{log}_{10}\left(\mathrm{H_2O}\right)$ = -3.5 to -1.5 in steps of 0.1. This range encompasses the various retrieved values from the limb-combined spectra by \cite{dyrek2023_wasp107b}, \cite{welbanks24}, and \cite{sing2024_wasp107b}. Then, we calculated the $\chi^2$ value between these models and each limb's spectrum, focusing on the NIRCam/F322W2 spectrum due to the strong and broad H$_2$O feature at $\sim$2.7~$\mu$m. Figure~\ref{fig:H2Ofigure} shows the results, again compared to literature retrieved values. Unlike for SO$_2$ (Figure~\ref{fig:so2biggallery}) and CO$_2$ (Figure~\ref{fig:CO2figure}), we find no evidence for a variation in the H$_2$O abundance between WASP-107~b's evening and morning limbs. The minimum $\chi^2$ is achieved at $log_{10}\left(\mathrm{H_2O}\right)$ = -2.5 for the evening limb and $log_{10}\left(\mathrm{H_2O}\right)$ = -2.3 for the morning limb, and the $\chi^2$ curves overlap well within the 68\% confidence interval. This test therefore suggests WASP-107~b has a homogeneous H$_2$O distribution, consistent with theoretical expectations \citep{fortney2020, zamyatina24_wasp96bGCMs_limbasym}. Further, since the potential contribution of clouds to the spectrum is likely very similar between F322W2 and F444W \citep{dyrek2023_wasp107b, welbanks24} this result suggests that any potential contamination from clouds to the apparent asymmetry in SO$_2$ and CO$_2$ is likely small.

\subsection{Evidence for Morning-limb Clouds} \label{subsec:clouds}

\cite{dyrek2023_wasp107b} and \cite{welbanks24} both inferred the presence of clouds in WASP-107~b's atmosphere via strong, broad absorption in the MIRI LRS bandpass. We see the same broad absorption feature from $\sim$6 to 11~\microns{} in the morning-limb spectrum. The individual contribution of this putative cloud can be seen by comparing the solid blue model in Panel~C of Figure~\ref{fig:spectra_withmodels}, which includes the cloud following the parametrization of \cite{welbanks24}, and the dotted line, which is the exact same model except without this cloud absorption. This broad absorption cannot be explained by any known gaseous opacity sources \citep{welbanks24}. The morning limb SO$_2$ feature is also enhanced in Panel~C, as discussed in Section~\ref{subsec:SO2variations}, which slightly affects the SO$_2$ feature around 7.5~\microns{} but this is overshadowed by the cloud feature and cannot explain the broad feature by itself. Conversely, we do not see evidence that such absorption is present on the evening limb. This can be seen comparing Panels~B and C of Figure~\ref{fig:spectra_withmodels}. The evening-limb model in Panel~C does not have the parametric cloud added, which we see fits the observed spectrum quite well. However, the model in Panel~B does have this cloud added using the parameters from \cite{welbanks24}, which yields a model that overpredicts the transit depth across these wavelengths. 

The presence of this broad feature in only the morning limb spectrum suggests WASP-107~b may have a heterogeneous cloud distribution on WASP-107~b; clouds may be present only on WASP-107~b's morning limb, and likely nightside as well. If true, this is likely a consequence of the temperature difference between the limbs, which may help narrow down the exact type of condensate responsible for this feature. We explore this further in the following sections. It is important to note though that, as discussed in Section~\ref{subsec:stellarcontam_lcnoise}, the MIRI data does exhibits some evidence for residual correlated noise which may contribute to some of this apparent asymmetry. However, it seems unlikely that correlated noise by itself, especially in the absence of large-scale features in the light curve (as we see for NIRISS and NIRSpec, Section~\ref{sec:stellarcontam}) could drive such a spectral feature, particularly when it is also observed in the limb-combined spectrum of independent reductions \citep{dyrek2023_wasp107b, welbanks24}.

\section{Comparison to 3D Circulation Models} \label{sec:GCMcomparisons}

To further interpret WASP-107~b's evening-morning differences that we infer from our limb transmission spectra, we compare our results to general circulation models (GCMs). We run new GCMs of WASP-107~b using two different frameworks: RM-GCM and SPARC/MITgcm; described below. 

\subsection{GCM Setup} \label{subsec:GCMsetup}

The RM-GCM was adapted from the University of Reading's Intermediate General Circulation Model \citep{hoskins1975} for hot Jupiter atmospheres by \cite{rauscher2010}, and has since been coupled to a plane-parallel radiative transfer code using double-gray \citep{rauscher2012} and more recently picket-fence \citep{malsky2024} treatments. The model solves the primitive equations in spectral space horizontally, and on a logarithmically-spaced grid of $\sigma$ (P/P$_{\rm surf}$) vertically. In the models presented here, opacities were generated using the picket-fence method with two thermal and three starlight channels \citep{parmentier2014, parmentier2015}, as implemented in the RM-GCM by \citet{malsky2024}. For the cloudy model, we applied the method of \citet{roman2019}, with the updates described in \citet{roman2021} and \citet{malsky2024}. Clouds are presumed to be present wherever local conditions are cooler than a species' condensation curve. Where clouds are present, solid abundance is set as one tenth of the bulk atmospheric abundance of the limiting constituent of each species. This one tenth limit is imposed primarily for numerical stability in the radiative transfer calculations, and is required due to the inclusion of silicate cloud species. Cloud formation simulations typically predict the fraction of potential cloud-forming mass that condenses to be close to one, but this can be decreased by a factor of several with different choices for poorly constrained microphysical parameters \citep[e.g.,][]{powell2024twoDcloudmodels}. Particle sizes are prescribed as a function of pressure, fixed to 0.1 $\mu$m at pressures less than 10 mbar, and increasing linearly with pressure in the deeper atmosphere. Together, the mass abundance and particle size yield the optical depth per bar, the single-scattering albedo, and the asymmetry parameter, which are then applied in combination with the gas-phase radiative transfer. We include cloud opacity from KCl, Cr, SiO$_2$, Mg$_2$SiO$_4$, VO, Ca$_2$SiO$_4$, CaTiO$_3$, and Al$_2$O$_3$, but the strongest contributors to radiative feedback are SiO$_2$ and Mg$_2$SiO$_4$. To treat the super-solar metallicity of WASP-107b, the picket-fence coefficients derived for solar-composition atmospheres by \citet{parmentier2015} were applied, with pressure- and temperature-dependent Rosseland mean opacities using 10~x solar metallicity from \citet{freedman2014}. We assumed a solar carbon-to-oxygen ratio, and an internal heat flux corresponding to 400 K. RM-GCM version 5.0, used for the models presented here, is available online \citep{RM-GCM}.

The SPARC/MITgcm model \citep{showman19_gcm} couples the MITgcm, a General Circulation Model maintained at the Massachusetts Institute of Technology \citep{adcroft_mitgcm} with the plane-parallel radiative transfer code of \cite{marley1999} and solves the primitive equations on a cube-sphere grid. SPARC/MITgcm has previously been widely applied to modelling the atmospheric dynamics of giant exoplanets \citep[e.g.][]{showman2009_gcms, parmentier2013_gcms, kataria2015_w43gcms, kataria2016_gcmgrid, parmentier2016_cloudygcms, steinrueck2019_gcms, parmentier2021_hJphasecurves}. We ran both a clear and cloudy atmosphere case. We treated clouds in the form of active tracers \citep{parmentier2013_gcmtracers} following the method of \cite{parmentier2021_hJphasecurves} and \cite{tanshowman21b}, considering the molecules MnS and Na$_2$S. These tracer molecule clouds are added in the form of two tracer equations representing condensable vapor, and are simultaneously integrated. In other words, these clouds are created and destroyed according to the local temperature and pressure following the condensation curve of each species. The tracers also move in the atmosphere according to the atmospheric circulation. These tracer-based clouds are also radiatively active (i.e., radiative feedback from the cloud is accounted for). All cloud species are initialized with the solar abundance scaled to the metallicity of the planet. To best match the expected bulk physical properties of WASP-107~b, we set the model to have a metallicity of 10~x the solar value, a sub-solar carbon-to-oxygen ratio of 0.25, and an internal heat flux corresonding to a temperature of 350~K \citep{welbanks24}. 

\subsection{GCM Results} \label{subsec:GCMresults}

\begin{figure*}[ht!]
    \centering
    \includegraphics[width=\textwidth]{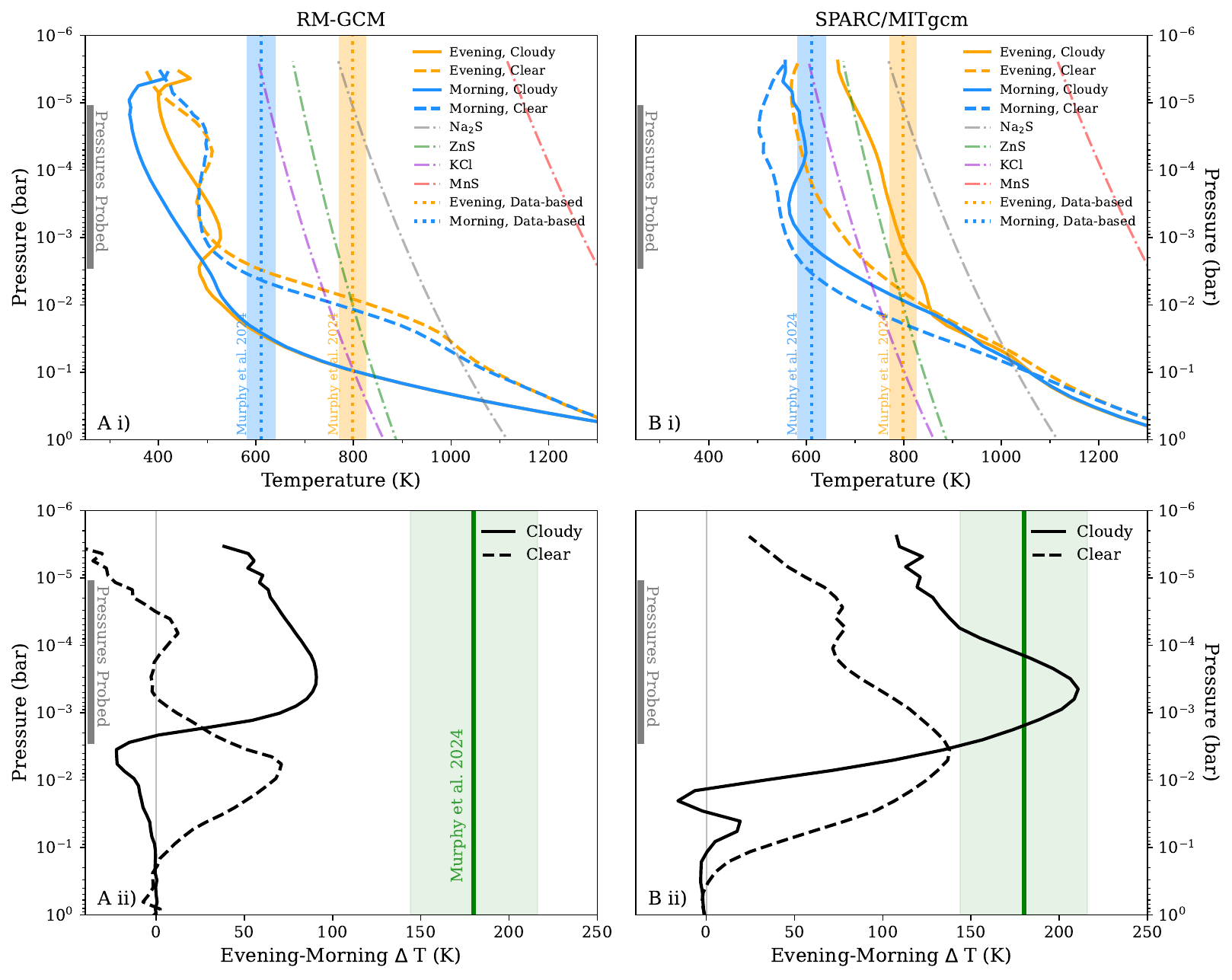}
    \caption{Results from our two GCM families' runs of WASP-107~b. The left column (panels Ai and Aii) show the results from RM-GCM and the right column (panels Bi and Bii) from SPARC/MITgcm. The top panels (Ai and Bi) show the evening (orange) and morning (blue) terminator temperature-pressure profiles in each model for both a cloudy (solid) and clear (dashed) atmosphere case. For reference, the vertical dotted lines and shaded regions are the evening and morning limb temperatures and uncertainties determined from the observed limb spectra by \cite{murphy24}. We also plot the approximate condensation curves for relevant species (Na$_2$S, ZnS, KCl, MnS) following the equations of \cite{morley2012cloudcondensationcurves} for 10x solar metallicity, but note these are independent of the GCM. The bottom panels (Aii and Bii) show the evening--morning temperature difference as a function of pressure, compared to that inferred by \cite{murphy24}. For reference, in each panel, the gray bar indicates the approximate range of pressures that our observations probe based on \cite{dyrek2023_wasp107b}.}
    \label{fig:GCM_tempprofiles}
\end{figure*}

Once each GCM converged, we extracted the latitude-averaged temperature-pressure profile at the evening and morning terminators (longitudes of $\pm$90$^\circ$), averaging together the profiles from longitudes within $\pm$10$^\circ$. The results are shown in Figure~\ref{fig:GCM_tempprofiles}, with the RM-GCM results shown in the left-hand column (Panels~Ai \& Aii) and the SPARC/MITgcm results in the right-hand column (Panels~Bi \& Bii). The top panels (Ai \& Bi) show the temperature-pressure profiles for each limb, with the cloudy case shown as solid lines and the clear case as dashed lines. The bottom panels (Aii \& Bii) show the evening-morning temperature difference as a function of pressure, calculated as the difference between the profiles shown in the top panels. For reference, we indicate the approximate range of pressures probed by our observations based on the contribution function determined by \cite{dyrek2023_wasp107b}, as well as the evening and morning limb temperatures and 1-$\sigma$ uncertainties determined by \cite{murphy24} as the vertical orange and blue shaded regions, respectively.

We find that the two GCMs predict fairly different temperature pressure profiles. As can be seen comparing Panels Ai and Bi of Figure~\ref{fig:GCM_tempprofiles}, the RM-GCM temperature profiles lie at cooler temperatures than that of SPARC/MITgcm by several hundred Kelvin. That the GCM models differ in the upper atmosphere is not unexpected. The two models use different prescriptions for radiative forcing, with SPARC/MITgcm using correlated-k radiative transfer with a sub-solar carbon-to-oxygen ratio of 0.25 and RM-GCM using picket-fence radiative transfer with a solar carbon-to-oxygen ratio. Differences in the degree and implementation of numerical dissipation used in the models may also be playing a role. The terminators are especially sensitive to model differences since radiation there is strongly inhomogeneous and temperature gradients are large. The larger difference between the two cloudy models can be ascribed to the different cloud species, cloud distributions, and particle sizes employed by the two models. The inclusion of clouds in the RM-GCM generally cools each limb relative to the cloud-free case most likely because the clouds are dominated by silicates which form deeper in the atmosphere and are more reflective, yielding a relatively low greenhouse effect and high albedo to cool the atmosphere. On the other hand, the MnS and Na$_2$S clouds in SPARC/MITgcm form higher in the atmosphere and contribute a stronger greenhouse effect, heating the atmosphere instead of cooling it. Treating clouds in GCMs requires simplified parameterizations for numerical feasibility, and a range of reasonable assumptions can produce significant differences in results. The differences between models from the two different codes in some sense represent the level of uncertainty intrinsic in modeling a complex atmosphere, where many different physical and numerical choices are required.

The RM-GCM predicts very homogeneous terminators in the clear atmosphere case, with an evening-morning temperature difference of only $\sim$10~K at the pressures probed in transmission. However, the inclusion of clouds drives a stronger heterogeneity, raising the limb temperature difference to a maximum of approximately 100~K. Although these profiles qualitatively agree with the results of \cite{murphy24} and this work in that the evening limb is hotter than the morning limb, the evening-morning difference in the RM-GCM is relatively smaller and the absolute temperatures of each limb are much cooler. Similar to the discussion above, part of this discrepancy may be due to the fact that the picket-fence radiative transfer method used by the RM-GCM is not as physically realistic as that of SPARC/MITgcm at the low pressures probed in transmission. 

The SPARC/MITgcm, on the other hand, predicts a strong temperature heterogeneity in both the clear and cloudy atmosphere case, both also with a hotter evening limb. The peak temperature difference is approximately 125~K for the clear atmosphere, slightly below the pressures probed, and 210~K for the cloudy atmosphere, within the pressures probed. This cloudy model evening-morning temperature difference agrees extremely well with the 180 $\pm$ 36~K difference inferred by \cite{murphy24}, as indicated in green in Panel~Bii of Figure~\ref{fig:GCM_tempprofiles}. In addition, as seen in Panel~Bi, the cloudy evening and morning limb temperature profiles from SPARC/MITgcm are very consistent with the limb temperatures inferred from the observed limb spectra by \cite{murphy24}. All together, these cloudy SPARC/MITgcm results agree very well with our observations, in terms of both the absolute temperature on each limb and the relative temperature gradient between them. Due to this agreement and the fact that SPARC/MITgcm's radiative transfer method is more physically realistic at the relevant pressures, we consider it our ``nominal" GCM model going forward.


\begin{figure*}[ht!]
    \centering
    \includegraphics[width=\textwidth]{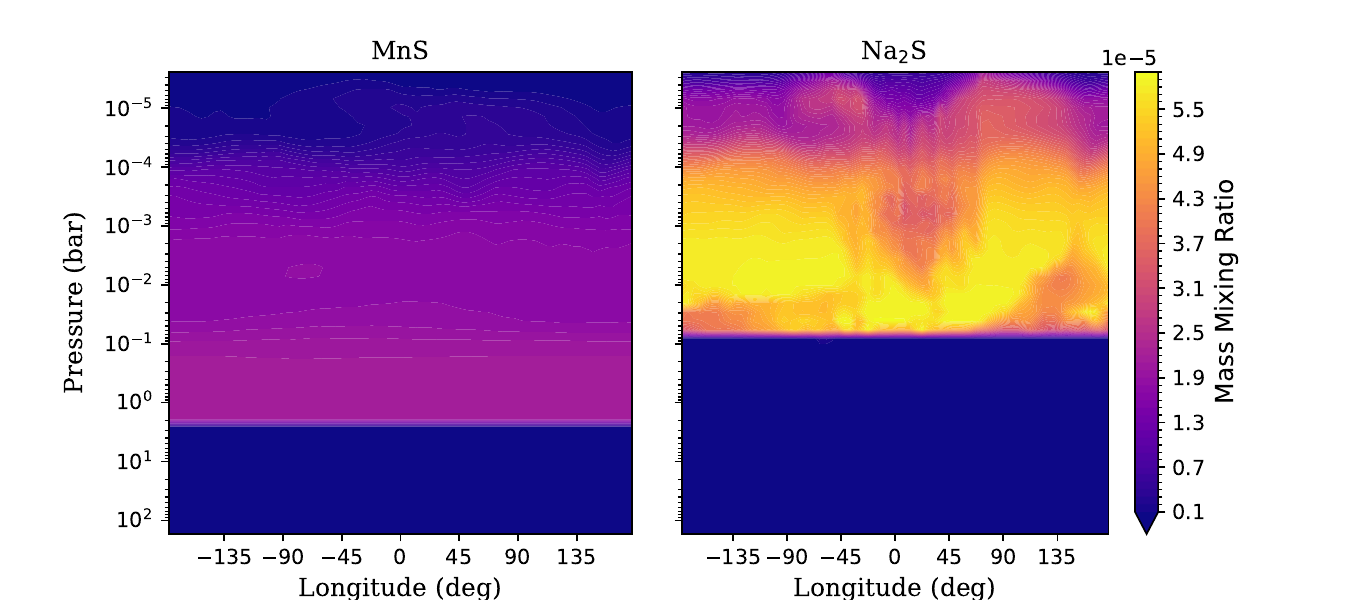}
    \caption{Spatial distributions of MnS (left) and Na$_2$S (right) clouds as a function of longitude, averaged over all latitudes, and pressure in our cloudy SPARC/MITgcm model. The color mapping represents the mass mixing ratio of the cloud particle, with the same scaling in each panel. }
    \label{fig:sparc_clouddist}
\end{figure*}

Both GCMs predict a larger evening--morning temperature difference at the pressures probed in transmission in their cloudy case than in their clear case. This shows the important role that cloud opacity plays in driving atmospheric heterogeneity, especially since both models treated clouds differently and considered different condensable species. The presence of clouds is also qualitatively consistent with the retrieval results of \cite{dyrek2023_wasp107b} and \cite{welbanks24}, and the potential cloud heterogeneity we infer (Section~\ref{subsec:clouds}). To explore this further, Figure~\ref{fig:sparc_clouddist} shows the vertical-longitudinal distribution of clouds formed in our nominal cloudy SPARC/MITgcm model, in terms of their mass mixing ratio. These distributions are averaged over latitude. For reference, we also show the full thermal and circulation maps at various pressure levels in the Appendix Figure~\ref{apxfig:sparc_temperaturewindmaps}. We find that MnS condenses homogeneously in the deep atmosphere, with a cloud base around a few bars, and is well below the expected photosphere \citep[around 10$^{-3}$ - 10$^{-4}$~bar;][]{dyrek2023_wasp107b}. At these deep pressures, internal heat dominates and is globally homogeneous with no large-scale circulation structures, yielding this homogeneous cloud distribution. On the other hand, Na$_2$S condenses higher up in the atmosphere with a cloud base around 6 $\times$ 10$^{-2}$~bar but the majority of cloud mass above it. Therefore, in this model the Na$_2$S cloud most significantly affects the atmosphere observed in transmission.

Unlike the MnS cloud, the Na$_2$S cloud has a heterogeneous distribution. In particular, there is a hole in the cloud on the dayside (longitude of 0~deg) that extends below the expected photosphere. The cloud mass is not zero on the dayside, but is significantly lower than elsewhere in the atmosphere. This relative clearness extends into the evening limb as well, though not entirely, with the cloud mass becoming significant again near the evening terminator around pressures of 5 $\times$ $10^{-3}$~bar. The peak cloud mass is relatively uniformly distributed across the nightside, however there is a region at low pressures near the antistellar point (longitude of -180~deg) where the cloud dissipates near its base. From the antistellar point through the morning limb, though, the cloud mass is at its most significant and is uniformly distributed in longitude all the way through the morning limb until before the dayside again. Na$_2$S has significant feedback on the temperature of the atmosphere, and this cloud-break on the dayside and half of the evening limb is likely what drives the divergence between this model's morning and evening limb T-P profiles.
Altogether, this Na$_2$S cloud has a spatially heterogeneous distribution and is most significantly concentrated around the morning limb, with a relatively clearer (but not entirely cloud-free) dayside and evening limb. This is very consistent with the potential cloud heterogeneity we see in the limb spectra (Section~\ref{subsec:clouds}). We discuss this further in the next section. 

The observations presented herein span two epochs separated, at most, by 194 days or approximately 33 orbits of WASP-107~b. In light of this long baseline, we also looked for temporal atmospheric variability in our GCMs to assess whether any variability should be expected between our observing epochs. Neither cloudy GCM varied significantly in time though. In both models, each limb's temperature at any pressure did not vary by more than approximately 10~K over the last 200 days of the simulation.


\section{Discussion} \label{sec:discussion}

\subsection{Potential Explanations for WASP-107~b's Heterogeneous Cloud} \label{subsec:discussion_clouds}

In Section~\ref{subsec:clouds}, we found evidence of a heterogeneous cloud distribution on WASP-107~b: a large cloud-like \citep{dyrek2023_wasp107b, welbanks24} absorption feature present only in the morning limb spectrum, suggesting that clouds may be located on WASP-107~b's morning terminator but not its evening terminator. These morning clouds would likely be present on the planet's nightside as well, and may originally condense there before being circulated through the morning limb. 

\cite{dyrek2023_wasp107b} originally inferred the presence of clouds in WASP-107~b's atmosphere based on the MIRI limb-combined spectrum, which exhibited a similar spectral shape to what we isolate in the morning limb spectrum. They found that a mix of silicates, particularly SiO$_2$, best fit their spectrum. \cite{welbanks24}, in contrast, argued that such species may not be physically feasible at the relatively cool temperatures in WASP-107~b's upper atmosphere and that, although species like SiO$_2$, Mg$_2$SiO$_4$, and MgSiO$_3$ have strong spectroscopic features in the MIRI LRS bandpass, they could not explain the panchromatic limb-combined transmission spectrum. At WASP-107~b's relatively cool temperature, silicate clouds should be horizontally homogeneous throughout the atmosphere. This was indeed true in our cloudy RM-GCM model, which includes these species. The homogeneity of these clouds through the atmosphere is therefore inconsistent with the spectral heterogeneity we observe in the limb-resolved spectra (Section~\ref{subsec:clouds}). Further, these silicates are highly reflective and led to a relatively small evening--morning temperature gradient in the RM-GCM which is also in conflict with the limb-resolved spectra. 

Unlike the silicate clouds, Na$_2$S clouds in our cloudy SPARC/MITgcm model yielded significant spatial heterogeneity, with a cloudier morning limb, and a large evening--morning temperature gradient consistent with the limb-resolved spectra. Therefore, Na$_2$S clouds may be a viable explanation for WASP-107~b's limb asymmetry. Similarly, though ZnS and KCl are not included in the SPARC/MITgcm model, their condensation curves lie between the evening and morning temperatures as inferred by \cite{murphy24} and by the GCM's T-P profiles, as shown in Figure~\ref{fig:GCM_tempprofiles}. This suggests that ZnS and KCl could condense on WASP-107~b's morning limb but not its evening limb. If present, clouds of these species should also be heterogeneously distributed, potentially more strongly than Na$_2$S due to their lower condensation temperatures. Therefore, clouds made up of one of or a mixture of Na$_2$S, ZnS, and KCl are consistent with WASP-107~b's limb asymmetry. 

One concern for this Na$_2$S, ZnS, or KCl scenario is that these species do not have strong absorption features, neither narrow or broad, in the MIRI LRS bandpass. As with the silicates, \cite{welbanks24} also showed that WASP-107~b's panchromatic limb-combined spectrum could not be well fit by any of these species individually. \cite{dyrek2023_wasp107b} similarly found that these species provided a poorer fit to their data than the silicate clouds. In fact, this lack of any good fits from a specific species is what led \cite{welbanks24} to develop a parametric description for this cloud, which we also used in this work. 

A potential resolution may be the way these condensates are mixed. Recent work by \cite{kiefer2024_mixedclouds} showed that, for the same cloud composition (fixed to the best-fit of \citealt{dyrek2023_wasp107b}), simply changing the method used to mix the individual components can significantly change the resulting transmission spectrum. Methods like the Landau-Lifshitz-Looyeng effective medium theory can result in a broad and relatively amorphous cloud absorption feature whereas other methods like the System of Spheres Approximation can result in a feature with distinct peaks \citep[e.g. see Figure~14 in][]{kiefer2024_mixedclouds}. However, this is likely only relevant when mixing species which each have strong absorption features at the relevant wavelengths to begin with, so it is unclear whether this could be a viable solution for Na$_2$S, ZnS, and KCl. Further modeling and consideration of this complexity in composition and cloud structure is necessary to better understand the nature of WASP-107~b's clouds. 

Observations of WASP-107~b in eclipse will also be important for solving this problem. Eclipse spectroscopy, particularly using JWST MIRI/LRS to probe these same wavelengths, would be able to definitely confirm whether silicate clouds persist on the dayside or the dayside is indeed relatively clear with some sulfide clouds as our limb spectra and nominal GCM suggest. The eclipse observations using JWST NIRSpec currently being executed for JWST-GTO-1201 (PI: D. Lafrenière) will be useful as well, but due to \mbox{NIRSpec's} wavelength range these observations may not be able to constrain the cloud properties as well as MIRI would.

\subsection{Linking chemical variations to atmospheric circulation} \label{subsec:discussion_variations}

Evening-morning chemical composition variations are intrinsically linked to atmospheric dynamics. Recent 2D \citep{tsai2023_so2transport} and 3D \citep{zamyatina24_wasp96bGCMs_limbasym} circulation models have found that efficient circulation can act to homogenize the global distribution of specific molecules compared to the potentially heterogeneous distribution under equilibrium chemistry alone, with some dependence on local temperatures \citep{tsai2023_so2transport} and whether the molecule is vertically quenched in the lower or upper atmosphere \citep{zamyatina24_wasp96bGCMs_limbasym}. For instance, \cite{zamyatina24_wasp96bGCMs_limbasym} predict that the distribution of CH$_4$ should be fairly homogeneous between the limbs under the action of circulation whereas CO$_2$, which quenches at relatively lower pressures, is less affected by circulation and can still have a heterogeneous distribution determined primarily by equilibrium chemistry. In fact, both \cite{zamyatina24_wasp96bGCMs_limbasym} and \cite{tsai2023_so2production} predict that CO$_2$ should be more abundant on a planet's cooler morning limb, which is precisely what we observe on WASP-107~b (Section~\ref{subsec:CO2variations}). However, \cite{tsai2023_so2production} predict an abundance difference of only a few to 10$\times$ difference in CO$_2$, which is slightly smaller than the $\sim$40$\times$ difference we inferred for WASP-107~b, though we note that their models were tuned for the different exoplanet WASP-39~b. \cite{zamyatina24_wasp96bGCMs_limbasym} predicted a similarly smaller difference of only a few to 10$\times$ difference, in terms of the mole fraction, though similarly their models were tuned for the exoplanet WASP-96~b. It is possible that degeneracy with a cloudiness gradient, and potentially temperature too, as mentioned in Section~\ref{subsec:CO2variations} not captured in our simple tests could lead to a slight overestimation of the evening--morning difference on WASP-107~b. We did also test whether our chosen spectral resolution could be biasing the inferred difference by repeating the test described in Section~\ref{subsec:CO2variations} at higher resolution, but found the same result.

Similar to CO$_2$, SO$_2$ may also quench in the upper atmosphere since it is a photochemical product. However, \cite{zamyatina24_wasp96bGCMs_limbasym} did not include photochemistry or SO$_2$ in their models. \cite{tsai2023_so2production}'s 1D photochemical models for the hotter exoplanet WASP-39~b predict that, in the absence of mixing, SO$_2$ should also be more abundant on the cooler morning limb, provided it is not too cold that SO$_2$ production is shut off \citep{tsai2023_so2transport}. Even if the morning limb temperature is too cold for local SO$_2$ production, as in \cite{tsai2023_so2transport}, horizontal mixing transports SO$_2$ away from the evening limb where it can build up on the nightside due to the lack of photodissociation. In this case, the morning limb SO$_2$ abundance can be built up to a comparable or larger value that on the evening limb. This is consistent with the SO$_2$ heterogeneity we observe on WASP-107~b.

\subsection{Potential Impact of Unocculted Starspots on Limb-resolved Transmission Spectroscopy} \label{subsec:discussion_TLSE}

In Section~\ref{sec:stellarcontam}, we explored how the difference in observed limb transmission spectra between instruments/visits can be explained by starspot crossings during certain visits, and demonstrated that these can be corrected for. Here, we speculate on what impact unocculted starspots, as well as unocculted faculae, would have on limb transmission spectra. Given the several spot crossings observed during these JWST visits and in several other previous observations of WASP-107~b \citep{mocnik2017_wasp107b}, it is not unlikely that heterogeneities are present on WASP-107's surface during a given observation even if a spot crossing is not observed. Based on the models of \cite{rackham18_TLSEmstars} and \cite{rackham19_TLSEfgkstars}, unocculted spots/faculae can induce strong slopes, broad molecular features, and generally chromatic offsets in the observed limb-combined transmission spectrum relative to a spot-free scenario through the transit light source effect (TLSE), which has been seen in several JWST transit observations \citep[e.g.][]{moran2023_observedTLSE, may2023_observedTLSE, radica_promise_2024, fournier-tondreau_near_2024}. However, whether the same effects would be had on the limb-resolved transmission spectra is an open question that requires additional modeling. The same effects are indeed likely had on the two limb spectra together, which should be considered when combining multiple instruments or visits. We are more concerned, though, with whether the TLSE affects the relative difference between the limb spectra. Except in the edge case where an unocculted heterogeneity rotates out of view during the transit, and thus affects one period of transit more than another, it seems unlikely that unocculted heterogeneities would impact one limb's spectrum differently than the other. Rather, we would expect the two limb spectra to be biased uniformly, and thus the TLSE would not create any false positive or negative limb asymmetry. However, further in-depth modeling of the TLSE in this context is necessary to verify this speculation. 

\subsection{Comparison to WASP-39~b Limb Transmission Spectra} \label{subsec:discussion_w39comparison}

\begin{figure}
    \centering
    \includegraphics[width=1\columnwidth]{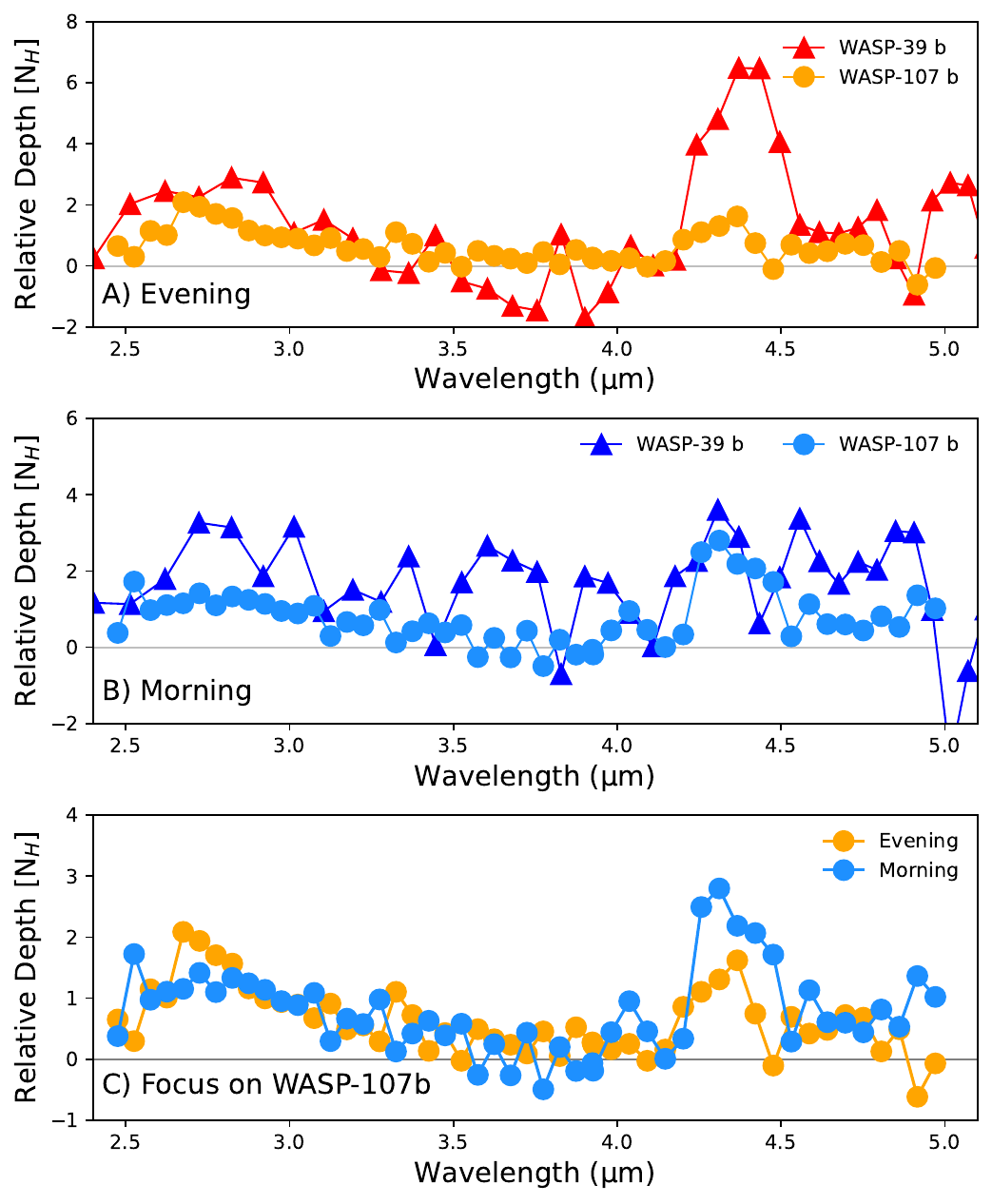}
    \caption{Comparing the relative limb transit depths of WASP-107~b to WASP-39~b \citep{espinoza24_wasp39b}. Panel~A shows the evening limb spectra for each planet, relative to the depth at 4.15~\microns{} and scaled to each planet's scale height. Panel~B shows the same for the morning limb spectra, and Panel~C shows the relative evening and morning limb spectra just for WASP-107~b.}
    \label{fig:w39_w107_comparison}
\end{figure}

Limb asymmetry was also recently detected on WASP-39~b \citep{espinoza24_wasp39b}. Here, we compare the published evening and morning limb spectra of WASP-39~b to WASP-107~b. For WASP-39~b, we use the \texttt{catwoman (MM)} spectra from \cite{espinoza24_wasp39b} whose methods align most closely to ours for WASP-107~b. To make the comparison fair in relative terms, we scale the transit depths by the corresponding planet's scale height, and compare relative to a zero-point defined by the depth at 4.15~\microns{}. Assuming a mean molecular weight of $\mu$ = 2.3, we calculate scale heights of approximately 1017~km and 778~km for WASP~107~b's evening and morning limbs, using T$_{\mathrm{eve}}$ = 798~K and T$_{\mathrm{morn}}$ = 611~K from \cite{murphy24}; and 861~km and 716~km for WASP-39~b's evening and morning limbs, using T$_{\mathrm{eve}}$ = 1068~K and T$_{\mathrm{morn}}$ = 889~K from \cite{espinoza24_wasp39b}. These correspond to evening and morning transit depths of 634~ppm and 485~ppm for WASP-107~b and 361~ppm and 300~ppm for WASP-39~b. 
We note that the WASP-39~b spectra are derived from $\sim$2--5~\microns{} NIRSpec/PRISM spectra which overlap with both the NIRISS/SOSS and NIRCam spectra for WASP-107~b. However, given the issue with stellar contamination for NIRISS/SOSS discussed in Section~\ref{sec:stellarcontam}, to be conservative we restrict this comparison to our WASP-107~b NIRCam spectra from $\sim$2.5--5~\microns{}. Figure~\ref{fig:w39_w107_comparison} shows this comparison, where Panel~A compares the evening-limb spectra for each planet, Panel~B compares the morning-limb spectra for each planet, and Panel~C compares the evening and morning-limb spectra of just WASP-107~b for reference. 

From Panels~A and B of Figure~\ref{fig:w39_w107_comparison}, we see that the relative limb transmission depths are similar between WASP-107~b and WASP-39~b. The amplitude of the absorption feature around 2.7~\microns{} and the redward pseudocontinuum are 1-2~scale heights. Shown in Panel~A, the relative evening limb depths are quite similar at all wavelengths between the two planets. The notable exception is the evening limb CO$_2$ feature at 4.4~\microns{}, which is of much higher amplitude ($\sim$6~scale heights) for WASP-39~b than WASP-107~b ($\sim$2~scale heights). \cite{espinoza24_wasp39b} compared their limb spectra to a number of models, including from GCMs, forward models, and retrievals that considered various physical scenarios (e.g., chemical disequilibrium and clouds versus cloud-free atmosphere), but none of their models could reproduce the large amplitude of this feature. 
On the other hand, the relative morning limb depths for WASP-107~b are slightly smaller than those of WASP-39~b, except in the CO$_2$ feature where they have approximately the same relative amplitude. From Panel~C, we see that the WASP-107~b's relative transit depths for each limb are nearly identical at all wavelengths, particularly between $\sim$2.8--4.1~\microns{}, except at 4.4~\microns{} where the morning limb feature is slightly stronger.

One leading hypothesis put forward by \cite{espinoza24_wasp39b} for WASP-39~b's oversized evening limb CO$_2$ feature is that WASP-39~b's evening limb is clear while the morning limb is relatively cloudy. Based on our MIRI limb spectra (Section~\ref{subsec:clouds}) we inferred a similar scenario on WASP-107~b. It may be the case that WASP-107~b's evening limb is not entirely cloud free, especially at its lower temperature than WASP-39~b's evening limb ($\sim$800~K versus 1100~K), reducing the amplitude of its CO$_2$ feature relative to that of WASP-39~b. Differences between the two planets' bulk parameters (e.g., surface gravity, metallicity, C/O ratio) are likely contributing as well. Additional modeling of these two planets, as well as investigations into potential limb asymmetry on other planets -- especially at intermediate temperatures -- will be needed to fully understand the differences between these two planets, and what they generally imply about atmospheric circulation, chemistry, and cloud formation.

\section{Conclusions} \label{sec:conclusions}

We reanalyzed panchromatic literature transit observations of WASP-107~b using JWST/NIRISS, NIRCam, NIRSpec, and MIRI to measure WASP-107~b's evening and morning limb transmission spectra from $\sim$1 to 12~\microns{}. Our primary result is the combined evening and morning limb transmission spectra from NIRCam and MIRI, shown in Figures~\ref{fig:spectra_dataonly} and \ref{fig:spectra_withmodels}. Our limb spectra are consistent with the NIRCam F322W2-only results from \cite{murphy24} and, further, we find:
\begin{itemize}
    \item The $\sim$180~K evening-morning temperature difference inferred by \cite{murphy24} using the JWST/NIRCam F322W2 limb spectra can well explain the full panchromatic limb spectra. We ran new clear and cloudy GCMs of WASP-107~b, and our cloudy SPARC/MITgcm model matched this result very well in terms of both the individual temperature of each limb and the evening--morning temperature difference.
    \item Evidence for evening--morning variation in the abundance of SO$_2$, with relatively more SO$_2$ on WASP-107~b's morning limb. Our comparison to forward models finds volume mixing ratios of approximately $\text{log}_{10}\left(\text{SO}_2\right)$ = -4.7 on the morning limb and $\text{log}_{10}\left(\text{SO}_2\right)$ = \mbox{-5.4} on the evening limb, which is approximately a difference of 5$\times$. The average of these abundances is consistent with literature limb-averaged abundances determined through retrievals on the limb-combined spectrum \citep{dyrek2023_wasp107b, welbanks24}.
    \item Evidence for evening--morning variation in the abundance of CO$_2$, also with relatively more CO$_2$ on WASP-107~b's morning limb. Our forward model comparison finds volume mixing ratios of approximately $\text{log}_{10}\left(\text{CO}_2\right)$ = -3.6 on the morning limb and $\text{log}_{10}\left(\text{CO}_2\right)$ = -5.2 on the evening limb, nearly a difference of 40$\times$. The average of these is also consistent with literature limb-averaged abundances determined through retrievals using the limb-combined spectrum \citep{welbanks24}.
    \item Evidence for heterogeneous cloud coverage on WASP-107~b. We observe the broad transmission feature caused by clouds previously detected by \cite{dyrek2023_wasp107b} and \cite{welbanks24}, and find that it is only present in the morning limb transmission spectrum. This is consistent with our cloud SPARC/MITgcm model which predicts a similar heterogeneous Na$_2$S cloud, though it is unclear whether Na$_2$S could explain the observed spectral feature. Though not present in our model, the condensation curves of ZnS and KCl also lie between our model's evening and morning temperature-pressure profiles, and thus may also be responsible for this heterogeneity.  
\end{itemize}
We found evidence for starspot crossings during the NIRISS and NIRSpec transits which appear to significantly impact the observed limb spectra. Conversely, we saw no evidence for spot crossings during the NIRCam or MIRI visits, though the MIRI visit does exhibit some unmitigated red noise. Using simple starspot models combined with simulated transit observations and injection-recovery tests, we developed a model for how these spots affect the observed limb asymmetry as a function of the spot's location in the light curve. We used this model to attempt to correct for the spots' affect on the NIRISS and NIRSpec spectra, and found that it can qualitatively bring them into consistency with the spot-free NIRCam observations. 

Follow-up modeling work, particularly atmospheric retrievals, would be useful to verify the evening--morning chemical abundance variations we infer, and explore potential degeneracies with temperature, cloud properties, and disequilibrium effects in more detail. Similarly, emission spectroscopy observations of WASP-107~b would be useful to characterize its dayside atmosphere and verify these potential heterogeneities. We recommend eclipse observations using MIRI LRS which would be highly sensitive to the dayside cloud properties, and thus able to directly verify the cloud heterogeneity we infer in transmission. 

\section*{Acknowledgments}
The authors would like to acknowledge and thank Dr. Dominic Samra for a helpful discussion that aided this manuscript, and \cite{kipping2025exoplaneteerscallingplotsallan} for clarifying the naming of ``time-averaging tests" used in this manuscript.
M.M.M., E.S., and M.J.R. acknowledge funding from the NASA Goddard Spaceflight Center via NASA contract NAS5-02105.
T.J.B. and T.P.G.~acknowledge funding support from the NASA Next Generation Space Telescope Flight Investigations program (now JWST) via WBS 411672.07.04.01.02, and T.P.G acknowledges also funding support from WBS 411672.07.05.05.03.02 in the same program (NIRCam).
M.R.L. acknowledges support from STScI award HST-AR-16139. 
SM acknowledges the Templeton TEX cross-training fellowship for supporting this work. JJF acknowledges the support of JWST Theory Grant JWST-AR-02232-001-A.
K.O. acknowledges support from the JSPS KAKENHI Grant Number JP23K19072.   
M.R.\ acknowledges support from the Natural Sciences and Engineering Research Council of Canada, the Canadian Space Agency, and the fonds de recherche du Québec - nature et technologie. 
This work benefited from the 2023 and 2024 Exoplanet Summer Program in the Other Worlds Laboratory (OWL) at the University of California, Santa Cruz, a program funded by the Heising-Simons Foundation and NASA.
This research made use of the open source Python package \texttt{exoctk}, the Exoplanet Characterization Toolkit \citep{exoctk}.

%

\vspace{5mm}
\facilities{JWST (NIRISS, NIRCam, NIRSpec, MIRI)}


\software{
    \texttt{ExoCTK} \citep{exoctk},
    \texttt{ExoTiC-LD} \citep{exoticld},
    \texttt{numpy} \citep{numpy},
    \texttt{matplotlib} \citep{matplotlib},
    \texttt{catwoman} \citep{espinoza2021_catwoman,jones2022_catwoman},
    \texttt{starry} \citep{luger19_starry_main, luger21_starry_extra1, luger21_starry_extra2}
          }



\clearpage

\appendix
\restartappendixnumbering

\section{Observation Details} \label{apx:obsdetails}
\begin{table}[h!]
    \centering
    \caption{Dates of each transit observation of WASP-107~b using JWST}
    \begin{tabular}{c|c|c|c|c}
         Instrument & Mode/Disperser & Date (UT) & Program & Reference  \\ \hline 
         NIRCam & F322W2 & 2023 Jan 14 & JWST-GTO-1185 & \cite{welbanks24}, \cite{murphy24}, this work. \\
         MIRI   & LRS    & 2023 Jan 19 & JWST-GTO-1280 & \cite{dyrek2023_wasp107b}, \cite{welbanks24}, this work. \\
         NIRISS & SOSS   & 2023 Jun 11 & JWST-GTO-1201 & Krishnamurthy et al.~(submitted), this work. \\
         NIRSpec & G395H & 2023 Jun 23 & JWST-GTO-1224 & \cite{sing2024_wasp107b}, this work. \\
         NIRCam & F444W  & 2023 Jul 04 & JWST-GTO-1185 & \cite{welbanks24}, this work. 
    \end{tabular}
    \label{apxtab:obs_dates}
\end{table}

\section{Light Curve Residual Galleries} \label{apx:residualgalleries}

\begin{figure*}[h!]
    \centering
    \includegraphics[width=0.9\textwidth]{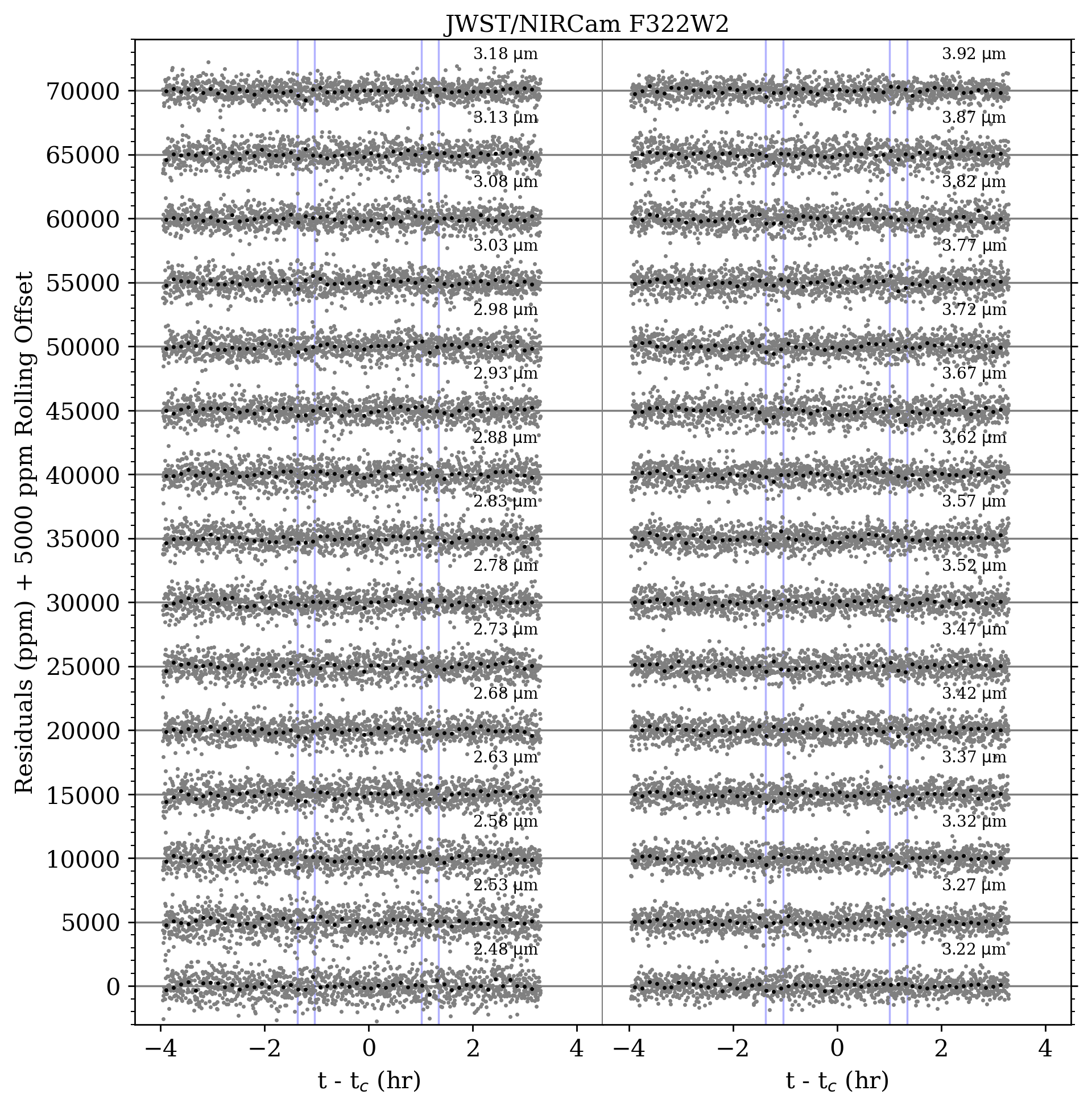}
    \caption{Gallery of light curve residuals from our fits to the JWST/NIRCam F322W2 spectroscopic light curves. Gray points are the light curve at native cadence, black points are binned to $\sim$500~s cadence chosen to be consistent across all instruments (Figures~\ref{apxfig:residualgalleryF444W}--\ref{apxfig:residualgalleryNIRSpec}). Each channel's residual time series is vertically offset for visual separation. The vertical blue lines demarcate the approximate four transit contacts to indicate ingress and egress.}
    \label{apxfig:residualgalleryF322W2}
\end{figure*} \clearpage

\begin{figure*}[h!]
    \centering
    \includegraphics[width=0.9\textwidth]{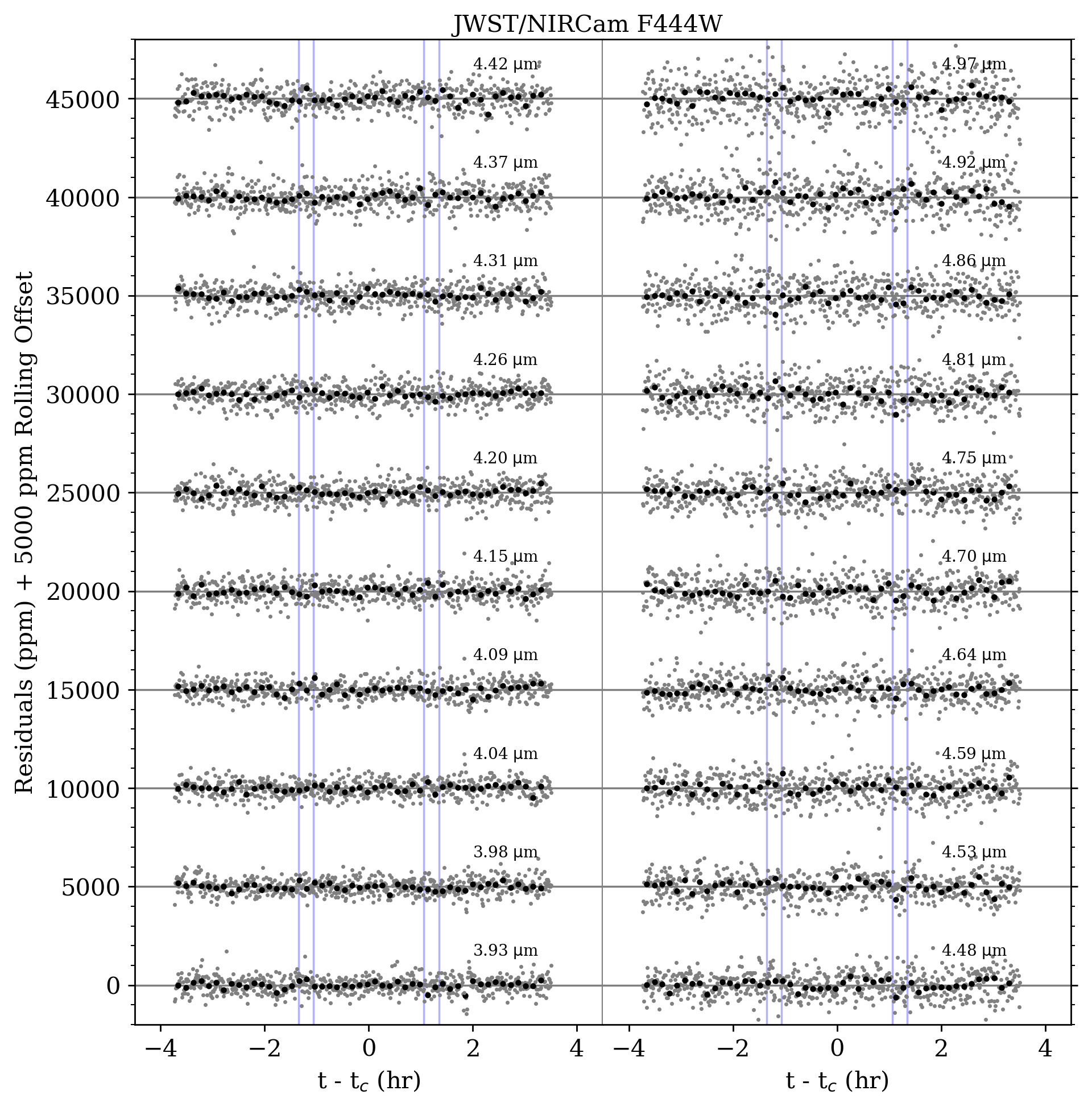}
    \caption{Gallery of light curve residuals from our fits to the JWST/NIRCam F444W spectroscopic light curves. Gray points are the light curve at native cadence, black points are binned to $\sim$500~s cadence chosen to be consistent across all instruments (Figures~\ref{apxfig:residualgalleryF322W2}--\ref{apxfig:residualgalleryNIRSpec}). Each channel's residual time series is vertically offset for visual separation. The vertical blue lines demarcate the approximate four transit contacts to indicate ingress and egress.}
    \label{apxfig:residualgalleryF444W}
\end{figure*}    \clearpage

\begin{figure*}[h!]
    \centering
    \includegraphics[width=0.9\textwidth]{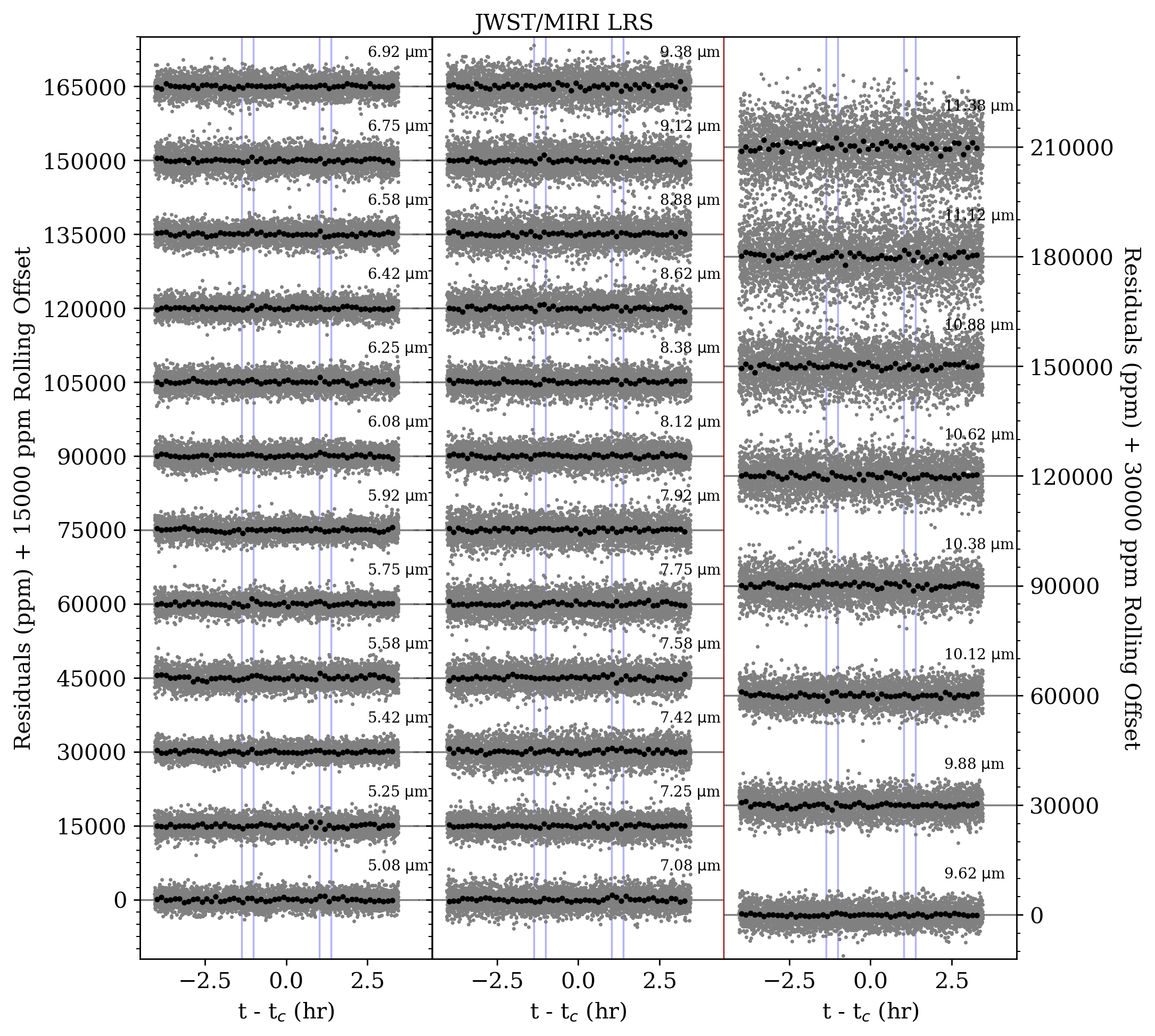}
    \caption{Gallery of light curve residuals from our fits to the JWST/MIRI LRS spectroscopic light curves. Gray points are the light curve at native cadence, black points are binned to $\sim$500~s cadence chosen to be consistent across all instruments (Figures~\ref{apxfig:residualgalleryF322W2}--\ref{apxfig:residualgalleryNIRSpec}). Each channel's residual time series is vertically offset for visual separation. Note that the left and middle panels share an offset, and the right panel has a different offset. The vertical blue lines demarcate the approximate four transit contacts to indicate ingress and egress.}
    \label{apxfig:residualgalleryMIRI}
\end{figure*}  \clearpage  

\begin{figure*}
    \centering
    \includegraphics[width=0.9\textwidth]{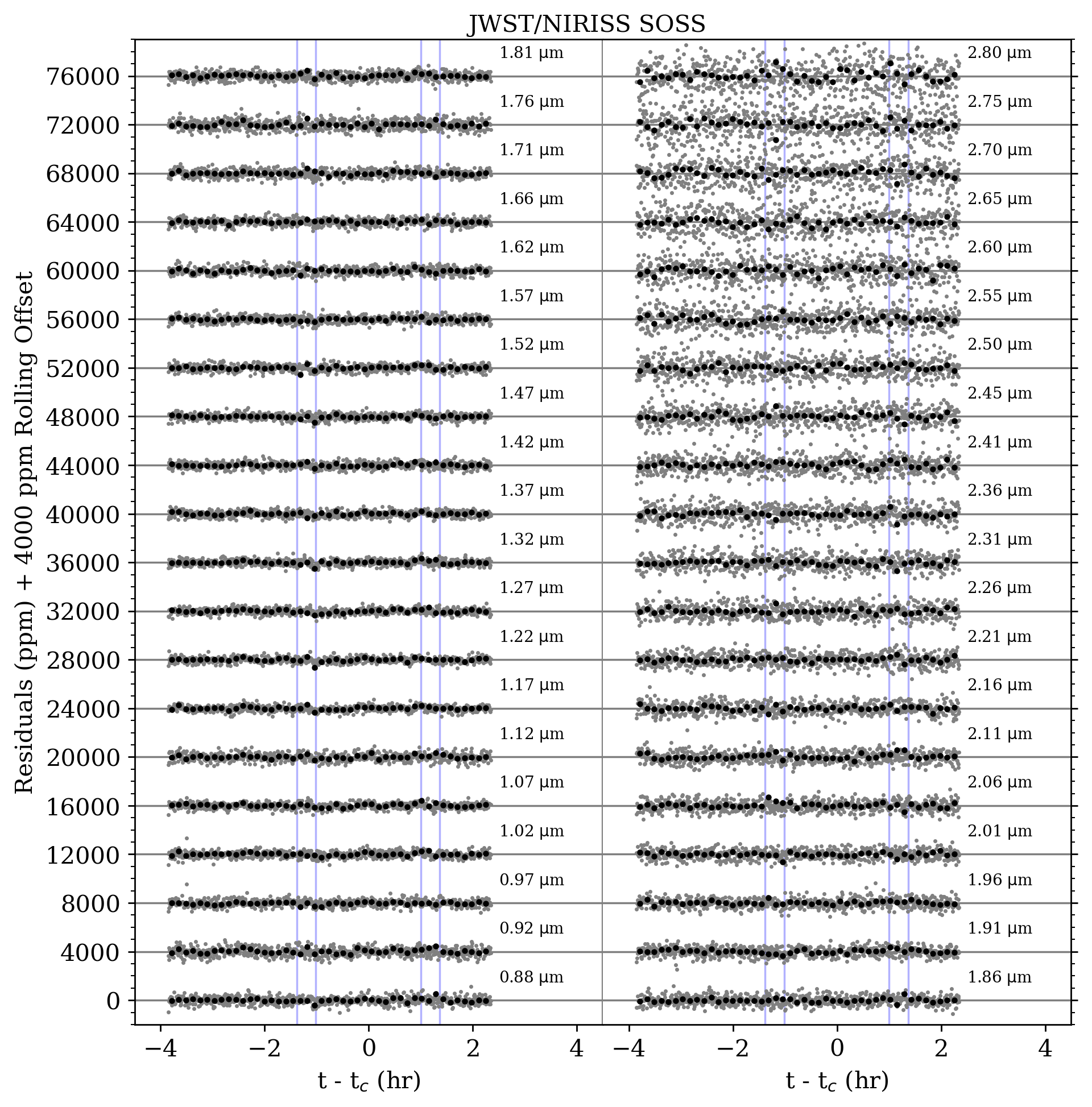}
    \caption{Gallery of light curve residuals from our fits to the JWST/NIRISS SOSS spectroscopic light curves. Gray points are the light curve at native cadence, black points are binned to $\sim$500~s cadence chosen to be consistent across all instruments (Figures~\ref{apxfig:residualgalleryF322W2}--\ref{apxfig:residualgalleryNIRSpec}). Each channel's residual time series is vertically offset for visual separation. The vertical blue lines demarcate the approximate four transit contacts to indicate ingress and egress.}
    \label{apxfig:residualgallerySOSS}
\end{figure*} \clearpage

\begin{figure*}
    \centering
    \includegraphics[width=0.9\textwidth]{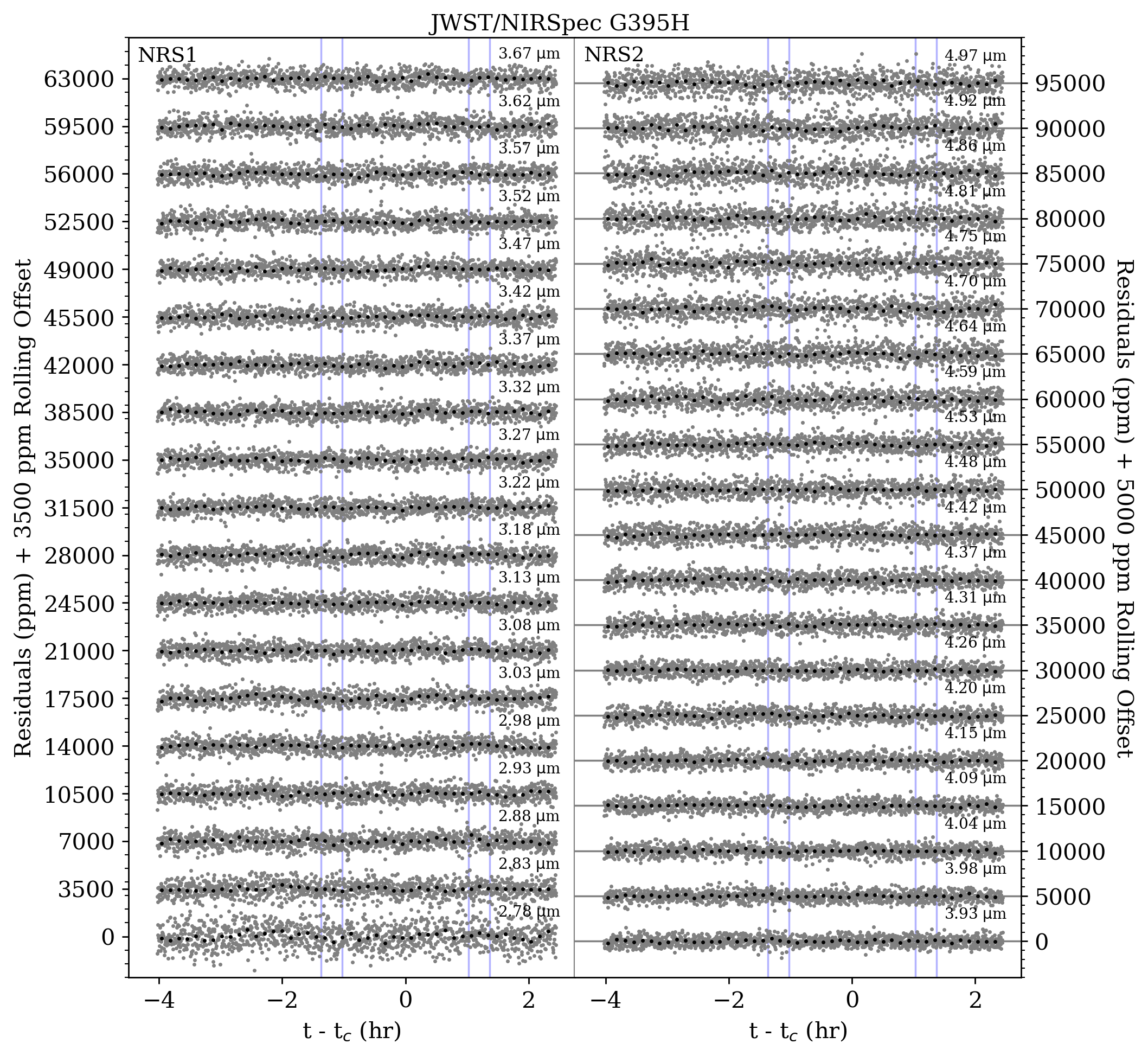}
    \caption{Gallery of light curve residuals from our fits to the JWST/NIRSpec G395H spectroscopic light curves, with NRS1 data on the left and NRS2 on the right. Gray points are the light curve at native cadence, black points are binned to $\sim$500~s cadence chosen to be consistent across all instruments (Figures~\ref{apxfig:residualgalleryF322W2}--\ref{apxfig:residualgalleryNIRSpec}). Each channel's residual time series is vertically offset for visual separation, and note the different offsets in each panel. The vertical blue lines demarcate the approximate four transit contacts to indicate ingress and egress.}
    \label{apxfig:residualgalleryNIRSpec}
\end{figure*} \clearpage

\section{Parameter Variation Tests} \label{apx:paramvartests}

As mentioned in Section~\ref{subsec:ALfits_results}, we tested the effect that the remaining uncertainty on WASP-107b's orbital parameters has on the morning and evening transmission spectra that we derive. Typically, when dealing with the limb-combined transmission spectrum, such uncertainties are of little concern as variations within these parameters' uncertainties would only include an achromatic vertical offset to the entire spectrum. However, these uncertainties may have a stronger affect on the relative limb-resolved transmission spectra. We aim to verify that such uncertainties are not causing any false positive or negatives, biasing our ultimate interpretation of WASP-107b's atmosphere. 

\begin{figure*}[h!]
    \centering
    \includegraphics[width=\columnwidth]{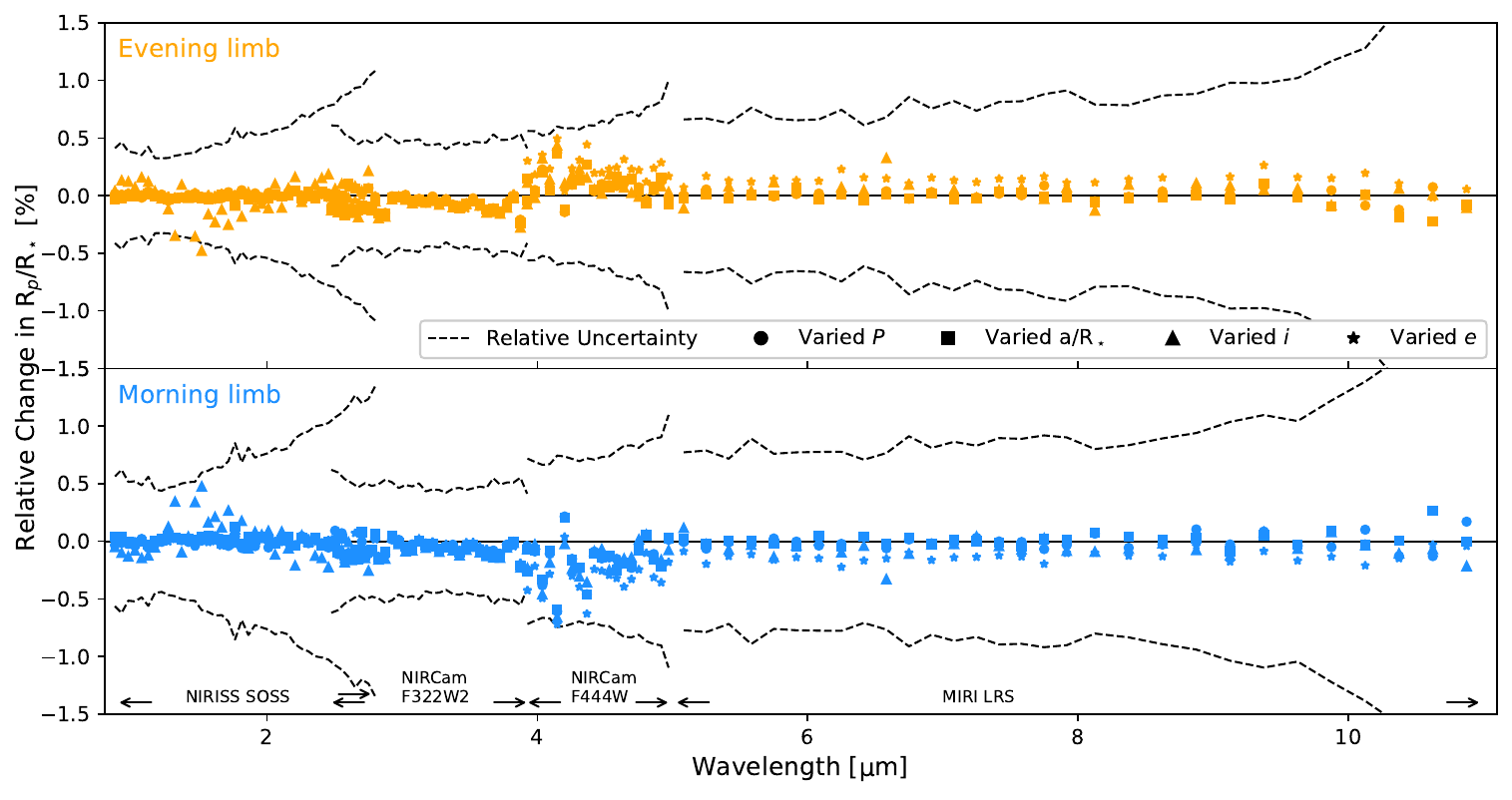}
    \caption{Relative changes in the best-fit evening (top panel) and morning (bottom panel) limb transmission spectra when unfixing certain orbital parameters during the fit. The effects of unfixing the period are shown as circles, semi-major axis as squares, inclination as triangles, and eccentricity as stars. We compare these changes to the relative 1-$\sigma$ uncertainty obtained on the limb depths, shown by the black dashed lines. }
    \label{apxfig:paramvarplot}
\end{figure*}

We repeated the fits with \texttt{catwoman} described in Section~\ref{subsec:datafitting} to the NIRISS, NIRCam, and MIRI data sets but unfixed one orbital parameter at a time. We allowed this unfixed parameter to vary as a sampled parameter during the MCMC, but applied a Gaussian prior on it based on the best-fit value and 1-$\sigma$ uncertainty from \cite{murphy24} whose posterior distributions for the relevant parameters were all approximately Gaussian. We tested the effect of unfixing the orbital period $P$, semi-major axis $a/R_\star$, orbital inclination $i$, and eccentricity $e$. Specifically for the eccentricity, since there are still concerns whether WASP-107b's apparent non-zero eccentricity is real or not \citep{piaulet2021_wasp107b, murphy24}, we did not apply a Gaussian prior. Rather, to allow for a possible circular orbit, we instead applied a uniform prior on $e$ between 0 and 1-$\sigma$ above the best-fit value. For each test, we then computed the relative change in $R_p/R_\star$ compared to the nominal case for each instrument/mode. The results for each instrument are shown in Figure~\ref{apxfig:paramvarplot}, and compared to the relative transit depth uncertainty achieved in the nominal case. 

We find that no single parameter's uncertainty makes a significant difference to the evening and morning limb spectra derived from any of the three datasets. As shown in Figure~\ref{apxfig:paramvarplot}, the relative changes in the transit depths in individual channels are typically very small, around 0.25$\%$, and the largest change in a single channel is only $\sim$1\%. These relative changes are all smaller than the uncertainties on the transit depths, and are negligible compared to the overall difference between the morning and evening spectra. To avoid confusion, as transit depth is often quoted as a percentage, note that these are not \textit{absolute} changes in transit depth, but rather relative changes computed as $\left(\delta_{\rm test} - \delta_{\rm nominal}\right) / \delta_{\rm nominal}$, where $\delta$ is the transit depth. Uncertainties in the inclination and eccentricity seem to have the largest effect on the spectra. This is likely because these parameters are most closely linked to the orientation of the projection of the planet's orbit onto the stellar disk, and thus can effect asymmetries of the transit chord relative to the stellar disk. 

\section{Modeling the Effect of Starspot Crossings} \label{apxsec:starspots}

In Section~\ref{subsec:stellarcontam_niriss_and_nirspec_obs}, we found large discrepancies between the NIRISS and NIRSpec limb spectra and the NIRCam spectra that, as discussed and shown in Section~\ref{subsec:stellarcontam_crossings}, we believe are due to starspot crossings during the NIRISS and NIRSpec observations (Figure~\ref{fig:broadbandresidualsgallery}). To our knowledge, the effect that starspot crossings have when trying to measure limb asymmetry in transit has not been studied in the literature. Therefore, we here conduct an initial investigation of their effect. 

\begin{figure*}[ht!]
    \centering
    \includegraphics[width=\textwidth]{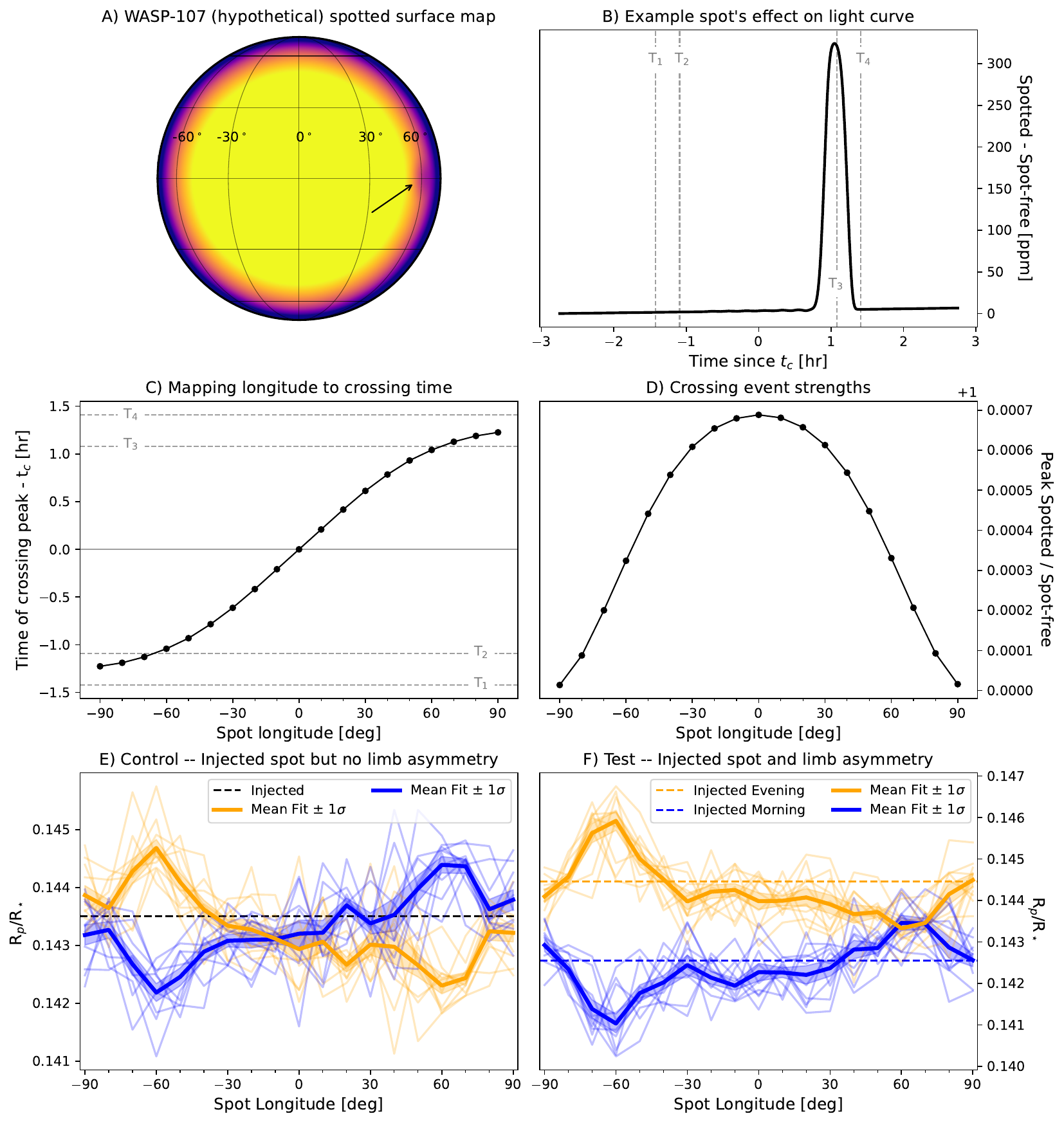}
    \caption{Investigating the impact starspot crossings have on our ability to detect and measure limb asymmetry. We used \texttt{starry} and \texttt{catwoman} to simulate transit observations of WASP-107~b when crossing spots at different locations on the star. Panel~A shows an example of the \texttt{starry} stellar intensity map with a spot of contrast 0.1 and angular radius 5$^\circ$ at a longitude of 60$^\circ$, pointed to by the arrow. Panel~B shows its relative effect on the transit light curve via the difference between the light curve with and without the spot. Panel~C shows how the spot longitude maps to the spot crossing's location within the transit. Panel~D shows the strength of the spot's effect on the light curve as a function of its longitude via the peak value of the ratio of the light curve with and without the spot -- i.e., the factor by which the relative flux increases due to the spot's bump at its peak. In other words, the amplitude of the bump due to the spot is equal to the relative flux of the spot-free light curve at a given time (related to longitude via Panel~C) multiplied by the $y$ value. 
    We then performed injection-recovery tests on these simulated spotted transit observations to test how the spot crossing impacts the best-fit evening and morning limb radii in cases where there is no underlying limb asymmetry (Control case, Panel~E), and where there is underlying limb asymmetry (Test case, Panel~E).
    }
    \label{fig:spoteffects}
\end{figure*}

To measure what effect starspot crossings in the light curve have on the measured limb asymmetry, we perform a series of injection-recovery tests. Specifically, we simulate the transit of a planet, which we model with and without limb asymmetry, across a star with one spot within the transit chord and measure the limb transit depths as a function of spot properties. We intend to make a number of simplifying assumptions that restrict our scope to the situation seen in these observations, and reserve more general modeling of this effect for future work. To model the stellar disk, we used the package \texttt{starry} \citep{luger19_starry_main, luger21_starry_extra1, luger21_starry_extra2}. We assume a stellar mass of 0.683~$M_\odot$ and radius of 0.67~$R_\odot$ \citep{piaulet2021_wasp107b} and quadratic limb darkening with coefficients u$_1$ = 0.15 and u$_2$ = 0.35, which are approximately the median values inferred by our observations. To isolate the effect of the spot itself, we assume stellar rotation is negligible. As mentioned, we assume there is only a single spot on the stellar surface at a given time. We based the model spot's properties on what would reproduce the observed bumps near ingress and egress in the NIRISS and NIRSpec data (Figure~\ref{fig:broadbandresidualsgallery}), which had amplitudes of approximately 300~ppm and durations on the order of 20-30 minutes. We find that a model spot with a contrast of 0.1 and angular radius of 5$^\circ$, when located at a longitude on the star that corresponds to these times in transit ($\pm$60$^\circ$ along the stellar equator), can reproduce such a bump in the light curve. There are likely multiple other combinations of spot properties that could reproduce the observations, but we are only concerned with the net effect on the light curve and not with constraining the properties of the spot itself. Spot contrast is wavelength dependent in reality but, for simplicity in this initial exploration, we neglect any wavelength dependence. We also assume the spots and the planet's transit chord are along the stellar equator. Panel~A of Figure~\ref{fig:spoteffects} shows an example of what this spot looks like on the stellar intensity map. 

We then simulated transit light curves of a hypothetical WASP-107~b across the spotted star, placing a spot of the aforementioned contrast and radius at longitudes spanning -90$^\circ$ to 90$^\circ$, in steps of 10$^\circ$, on the stellar equator. We assumed the orbital parameters for WASP-107~b derived by \cite{murphy24}. Then, we compared these light curves to a reference light curve, calculated with no starspots, to measure the effect of the spot as a function of its location. Panel~B of Figure~\ref{fig:spoteffects} shows an example for the 60$^\circ$ spot. For reference, Panel~C shows the mapping between the spot's longitude and where the peak of the crossing-event would be in time relative to the time of conjunction. To prepare our injection models, we took the ratio of the spotted and spot-free light curves to use as a multiplicative model of the spot crossing's effect. Panel~D shows the value of this ratio at the crossing peak as a function of the spot's location. 

With a measure of the spot crossing's effect in hand, we generated two reference transit models of WASP-107~b using \texttt{catwoman}: a test case with limb asymmetry and a control case with no limb asymmetry. For the test case, we set a true evening limb R$_{\rm p}$/R$_\star$ = 0.144454 and a morning limb R$_{\rm p}$/R$_\star$ = 0.142555. This tests the scenario where the planet intrinsically has limb asymmetry that we are trying to retrieve, but there is a spot crossing during the observation. Then, for the control case, we set equal evening and morning limb radii of R$_{\rm p}$/R$_\star$ = 0.1435, which is the radius of equivalent occulting area as the test case. This control case tests the impact of the spot in the absence of underlying limb asymmetry, and whether the spot crossing can lead to a false detection of limb asymmetry. For both cases, we injected the spot crossings at each longitude by multiplying the model light curve by the corresponding spot crossing model. We then generated simulated observed data of each light curve at 22~second cadence, similar to that used by both the NIRCam and NIRSpec, and injected white Gaussian noise of standard deviation 400~ppm to simulate photon noise. Then, we fit each set of simulated observations following the same method used on the real observations (Section~\ref{subsec:datafitting}). To marginalize over the effect of noise, we repeated each individual fit ten times, generating a new noise realization for each. We assumed that all orbital and limb darkening parameters are known exactly and fixed them in the fit. Figure~\ref{fig:spoteffects} shows the results of these tests, with Panel~E showing the control case results and Panel~F the test case results. Each semi-transparent line shows the best-fit evening (orange) and morning (blue) limb R$_{\rm p}$/R$_\star$ as a function of the injected spot longitude for one iteration, and the thicker solid lines and shaded regions represent the mean and error on the mean of all ten iterations. For reference, the true evening and morning limb R$_{\rm p}$/R$_\star$ values are shown as the horizontal dashed lines. 

From these tests, we find that the spot crossing's effect can vary from negligible to significant depending on where the crossing occurs. This can be readily seen from our control case results, shown in Panel~E of Figure~\ref{fig:spoteffects}. When the spot crossing occurs near the middle of transit, particularly between longitudes of approximately -30$^\circ$ and 10$^\circ$, it has negligible effect on the retrieval of uniform limbs. The best-fit evening and morning limb R$_{\rm p}$/R$_\star$ are equal or consistent within 1-$\sigma$. Interestingly, while the uniformity is correctly retrieved, the retrieved R$_{\rm p}$/R$_\star$ values are slightly smaller than the truth. This is likely because the upward light curve bump induced by the spot crossing -- which is strongest around these longitudes (Figure~\ref{fig:spoteffects}, Panel~D) yields higher relative flux within the transit which, in our setup, can only be compensated by smaller R$_{\rm p}$/R$_\star$. It is unclear why the R$_p$/R$_\star$ values seem to diverge around 20$^\circ$ unlike the uniformity at -20$^\circ$. On the other hand, spots nearer to the limb of the star, around $\pm$60$^\circ$, have a strong impact on the retrieved morning and evening limb radii. Such longitudes correspond to a crossing event at the second and third contact points -- i.e., the end of ingress and the start of egress (Figure~\ref{fig:spoteffects}, Panel~C). A spot near the end of ingress (-60$^\circ$) leads the fit to find a large evening $>$ morning asymmetry while a spot near the beginning of egress (+60$^\circ$) leads the fit to find a large morning $>$ evening asymmetry, both of which are false positives. This is likely because the spot-induced bump, when at these points in the transit, cause higher relative flux values that masquerade as either the ingress appearing late or the egress appearing early relative to uniform-limb case. The fit is then forced to compensate for this through the relative limb radii. For the spot near ingress, the morning limb is made smaller since it is the leading limb, and a smaller leading limb would also explain having a later ingress. Then, to preserve the total occulting area, the evening limb is forced to be larger. The reverse is true for the spot near egress, where the evening limb is made smaller since it is the trailing limb, and a smaller trailing limb would also explain having an earlier egress. All together, this control test shows that a spot crossing during ingress or egress could cause a false positive detection of limb asymmetry, with the polarity depending on whether the crossing is during ingress or egress. 

The test case results shown in Panel~F of Figure \ref{fig:spoteffects} are similar to the control case results, only offset by the injected limb asymmetry. The spot crossings near the middle of transit, between longitudes of -30$^\circ$ and +30$^\circ$, have little effect on the retrieved limb radii. The retrieved evening and morning R$_{\rm p}$/R$_\star$ values are nearly consistent with the truths within 1-$\sigma$ but, similar to in the control case, are both equally biased to slightly lower values by the spot. However, spots nearer the limb of the star, and thus near ingress and egress, again have a significant impact. In this case, spots near ingress (around -60$^\circ$) yield a retrieved limb asymmetry magnitude that is nearly double the truth. Conversely, spots near egress (around 60$^\circ$) have the opposite effect that, in this particular case, combine to yield no limb asymmetry. In some individual tests, the retrieved morning limb R$_{\rm p}$/R$_\star$ was actually equal to the injected evening limb R$_{\rm p}$/R$_\star$ and vice-versa, meaning the spot effectively switched the polarity of the retrieved limb asymmetry relative to the injected. Since the effects of the spot in this test case resemble that of the control case but with the underlying true evening--morning difference providing an additional offset, it seems the effect of spots will generally vary system-to-system. 

It is important to emphasize that the geometry in which we view the real WASP-107 system is not the same as we assumed in these simplified models. In these tests, we assume looking equator-on at WASP-107 and that WASP-107~b transits along the stellar equator because this made for the simplest and most easily interpretable setup. However, the real WASP-107 is seen slightly pole-on and WASP-107~b is likely on a polar orbit \citep{rubenzahl2021_wasp107b}. Therefore, a spot that the planet occults at a certain point during transit would not be at the exact same location on the stellar surface in our models as they would be in real life, so the exact stellar longitudes of each spot should not be directly quoted. What fundamentally matters in this context is where the spot is in relation to the transit itself (e.g., near ingress or egress), as shown in Panel~C of Figure~\ref{fig:spoteffects}, and our results should only be interpreted in that regard. 

\section{Additional GCM Results} \label{apxsec:GCMfigures}

\subsection{Global Temperature and Circulation Structure of WASP-107~b}
\begin{figure*}[ht!]
    \centering
    \includegraphics[width=\textwidth]{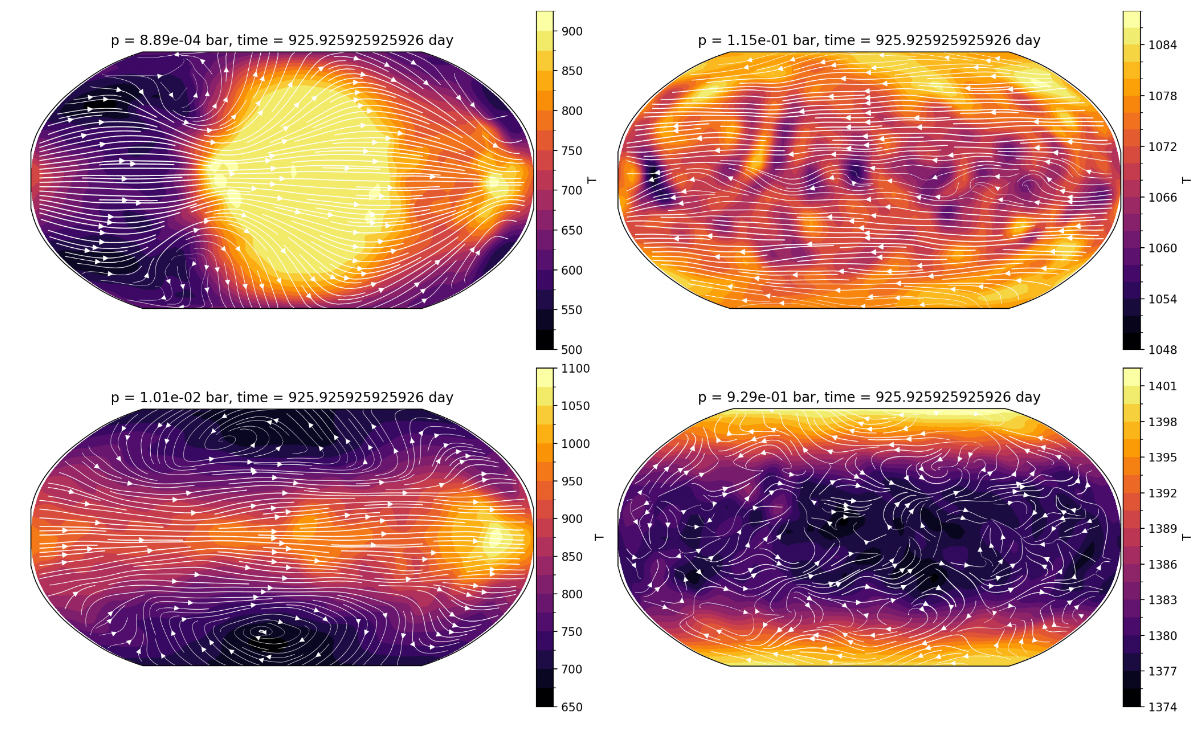}
    \caption{Temperature and circulation maps of WASP-107~b at various pressure levels from our cloudy SPARC/MITgcm model. Each panel represents a slice along a different isobar. The color mapping represents the gas temperature, and we use a different scaling in each panel. The lines with arrows represent the wind direction. }
    \label{apxfig:sparc_temperaturewindmaps}
\end{figure*}


\clearpage

\bibliography{manuscript}{}

\begin{thebibliography}{}
\expandafter\ifx\csname natexlab\endcsname\relax\def\natexlab#1{#1}\fi
\providecommand{\url}[1]{\href{#1}{#1}}
\providecommand{\dodoi}[1]{doi:~\href{http://doi.org/#1}{\nolinkurl{#1}}}
\providecommand{\doeprint}[1]{\href{http://ascl.net/#1}{\nolinkurl{http://ascl.net/#1}}}
\providecommand{\doarXiv}[1]{\href{https://arxiv.org/abs/#1}{\nolinkurl{https://arxiv.org/abs/#1}}}

\bibitem[{{Adcroft} {et~al.}(2004){Adcroft}, {Campin}, {Hill}, \& {Marshall}}]{adcroft_mitgcm}
{Adcroft}, A., {Campin}, J.-M., {Hill}, C., \& {Marshall}, J. 2004, Monthly Weather Review, 132, 2845, \dodoi{10.1175/MWR2823.1}

\bibitem[{{Albert} {et~al.}(2023){Albert}, {Lafreni{\`e}re}, {Doyon}, {Artigau}, {Volk}, {Goudfrooij}, {Martel}, {Radica}, {Rowe}, {Espinoza}, {Roy}, {Filippazzo}, {Darveau-Bernier}, {Talens}, {Sivaramakrishnan}, {Willott}, {Fullerton}, {LaMassa}, {Hutchings}, {Rowlands}, {Vila}, {Zhou}, {Aldridge}, {Maszkiewicz}, {Beaulieu}, {Cook}, {Piaulet}, {Roy}, {Lamontagne}, {Morel}, {Frost}, {Salhi}, {Coulombe}, {Benneke}, {MacDonald}, {Johnstone}, {Turner}, {Fournier-Tondreau}, {Allart}, \& {Kaltenegger}}]{albert_near_2023}
{Albert}, L., {Lafreni{\`e}re}, D., {Doyon}, R., {et~al.} 2023, \pasp, 135, 075001, \dodoi{10.1088/1538-3873/acd7a3}

\bibitem[{{Alderson} {et~al.}(2024){Alderson}, {Batalha}, {Wakeford}, {Wallack}, {Aguichine}, {Teske}, {Adams Redai}, {Alam}, {Batalha}, {Gao}, {Kirk}, {L{\'o}pez-Morales}, {Moran}, {Scarsdale}, {Wogan}, \& {Wolfgang}}]{alderson2024_toi836b_t0systematicerrors}
{Alderson}, L., {Batalha}, N.~E., {Wakeford}, H.~R., {et~al.} 2024, \aj, 167, 216, \dodoi{10.3847/1538-3881/ad32c9}

\bibitem[{{Anderson} {et~al.}(2017){Anderson}, {Collier Cameron}, {Delrez}, {Doyle}, {Gillon}, {Hellier}, {Jehin}, {Lendl}, {Maxted}, {Madhusudhan}, {Pepe}, {Pollacco}, {Queloz}, {S{\'e}gransan}, {Smalley}, {Smith}, {Triaud}, {Turner}, {Udry}, \& {West}}]{anderson2017_wasp107b}
{Anderson}, D.~R., {Collier Cameron}, A., {Delrez}, L., {et~al.} 2017, \aap, 604, A110, \dodoi{10.1051/0004-6361/201730439}

\bibitem[{{Beatty} {et~al.}(2019){Beatty}, {Marley}, {Gaudi}, {Col{\'o}n}, {Fortney}, \& {Showman}}]{beatty2019_phasecurves}
{Beatty}, T.~G., {Marley}, M.~S., {Gaudi}, B.~S., {et~al.} 2019, \aj, 158, 166, \dodoi{10.3847/1538-3881/ab33fc}

\bibitem[{{Beatty} {et~al.}(2024){Beatty}, {Welbanks}, {Schlawin}, {Bell}, {Line}, {Murphy}, {Edelman}, {Greene}, {Fortney}, {Henry}, {Mukherjee}, {Ohno}, {Parmentier}, {Rauscher}, {Wiser}, \& {Arnold}}]{beatty24_gj3470}
{Beatty}, T.~G., {Welbanks}, L., {Schlawin}, E., {et~al.} 2024, \apjl, 970, L10, \dodoi{10.3847/2041-8213/ad55e9}

\bibitem[{{Bell} {et~al.}(2022){Bell}, {Ahrer}, {Brande}, {Carter}, {Feinstein}, {Guzman Caloca}, {Mansfield}, {Zieba}, {Piaulet}, {Benneke}, {Filippazzo}, {May}, {Roy}, {Kreidberg}, \& {Stevenson}}]{EUREKAPIPELINE}
{Bell}, T., {Ahrer}, E.-M., {Brande}, J., {et~al.} 2022, The Journal of Open Source Software, 7, 4503, \dodoi{10.21105/joss.04503}

\bibitem[{{Bell} {et~al.}(2023){Bell}, {Welbanks}, {Schlawin}, {Line}, {Fortney}, {Greene}, {Ohno}, {Parmentier}, {Rauscher}, {Beatty}, {Mukherjee}, {Wiser}, {Boyer}, {Rieke}, \& {Stansberry}}]{bell2023_w80}
{Bell}, T.~J., {Welbanks}, L., {Schlawin}, E., {et~al.} 2023, \nat, 623, 709, \dodoi{10.1038/s41586-023-06687-0}

\bibitem[{Bell {et~al.}(2024)Bell, Crouzet, Cubillos, Kreidberg, Piette, Roman, Barstow, Blecic, Carone, Coulombe, Ducrot, Hammond, Mendonça, Moses, Parmentier, Stevenson, Teinturier, Zhang, Batalha, Bean, Benneke, Charnay, Chubb, Demory, Gao, Lee, López-Morales, Morello, Rauscher, Sing, Tan, Venot, Wakeford, Aggarwal, Ahrer, Alam, Baeyens, Barrado, Caceres, Carter, Casewell, Challener, Crossfield, Decin, Désert, Dobbs-Dixon, Dyrek, Espinoza, Feinstein, Gibson, Harrington, Helling, Hu, Iro, Kempton, Kendrew, Komacek, Krick, Lagage, Leconte, Lendl, Lewis, Lothringer, Malsky, Mancini, Mansfield, Mayne, Mikal-Evans, Molaverdikhani, Nikolov, Nixon, Palle, dit de~la Roche, Piaulet, Powell, Rackham, Schneider, Steinrueck, Taylor, Welbanks, Yurchenko, Zhang, \& Zieba}]{bell2024_wasp43b}
Bell, T.~J., Crouzet, N., Cubillos, P.~E., {et~al.} 2024, Nightside clouds and disequilibrium chemistry on the hot Jupiter WASP-43b.
\newblock \doarXiv{2401.13027}

\bibitem[{{Benneke} {et~al.}(2024){Benneke}, {Roy}, {Coulombe}, {Radica}, {Piaulet}, {Ahrer}, {Pierrehumbert}, {Krissansen-Totton}, {Schlichting}, {Hu}, {Yang}, {Christie}, {Thorngren}, {Young}, {Pelletier}, {Knutson}, {Miguel}, {Evans-Soma}, {Dorn}, {Gagnebin}, {Fortney}, {Komacek}, {MacDonald}, {Raul}, {Cloutier}, {Acuna}, {Lafreni{\`e}re}, {Cadieux}, {Doyon}, {Welbanks}, \& {Allart}}]{benneke_jwst_2024}
{Benneke}, B., {Roy}, P.-A., {Coulombe}, L.-P., {et~al.} 2024, arXiv e-prints, arXiv:2403.03325, \dodoi{10.48550/arXiv.2403.03325}

\bibitem[{Bourque {et~al.}(2021)Bourque, Espinoza, Filippazzo, Fix, King, Martlin, Medina, Batalha, Fox, Fowler, Fraine, Hill, Lewis, Stevenson, Valenti, \& Wakeford}]{exoctk}
Bourque, M., Espinoza, N., Filippazzo, J., {et~al.} 2021, The Exoplanet Characterization Toolkit (ExoCTK), 1.0.0,  Zenodo, \dodoi{10.5281/zenodo.4556063}

\bibitem[{{Bourrier} {et~al.}(2020){Bourrier}, {Ehrenreich}, {Lendl}, {Cretignier}, {Allart}, {Dumusque}, {Cegla}, {Su{\'a}rez-Mascare{\~n}o}, {Wyttenbach}, {Hoeijmakers}, {Melo}, {Kuntzer}, {Astudillo-Defru}, {Giles}, {Heng}, {Kitzmann}, {Lavie}, {Lovis}, {Murgas}, {Nascimbeni}, {Pepe}, {Pino}, {Segransan}, \& {Udry}}]{bourrier2020_wasp121basymmetry}
{Bourrier}, V., {Ehrenreich}, D., {Lendl}, M., {et~al.} 2020, \aap, 635, A205, \dodoi{10.1051/0004-6361/201936640}

\bibitem[{{Bushouse} {et~al.}(2023){Bushouse}, {Eisenhamer}, {Dencheva}, {Davies}, {Greenfield}, {Morrison}, {Hodge}, {Simon}, {Grumm}, {Droettboom}, {Slavich}, {Sosey}, {Pauly}, {Miller}, {Jedrzejewski}, {Hack}, {Davis}, {Crawford}, {Law}, {Gordon}, {Regan}, {Cara}, {MacDonald}, {Bradley}, {Shanahan}, {Jamieson}, {Teodoro}, \& {Williams}}]{bushouse2023jwst}
{Bushouse}, H., {Eisenhamer}, J., {Dencheva}, N., {et~al.} 2023, {JWST Calibration Pipeline}, 1.10.2,  Zenodo, \dodoi{10.5281/zenodo.7829329}

\bibitem[{{Carter} {et~al.}(2024){Carter}, {May}, {Espinoza}, {Welbanks}, {Ahrer}, {Alderson}, {Brahm}, {Feinstein}, {Grant}, {Line}, {Morello}, {O'Steen}, {Radica}, {Rustamkulov}, {Stevenson}, {Turner}, {Alam}, {Anderson}, {Batalha}, {Battley}, {Bayliss}, {Bean}, {Benneke}, {Berta-Thompson}, {Brande}, {Bryant}, {Burleigh}, {Coulombe}, {Crossfield}, {Damiano}, {D{\'e}sert}, {Flagg}, {Gill}, {Inglis}, {Kirk}, {Knutson}, {Kreidberg}, {L{\'o}pez Morales}, {Mansfield}, {Moran}, {Murray}, {Nixon}, {Petit dit de la Roche}, {Rackham}, {Schlawin}, {Sing}, {Wakeford}, {Wallack}, {Wheatley}, {Zieba}, {Aggarwal}, {Barstow}, {Bell}, {Blecic}, {Caceres}, {Crouzet}, {Cubillos}, {Daylan}, {de Val-Borro}, {Decin}, {Fortney}, {Gibson}, {Heng}, {Hu}, {Kempton}, {Lagage}, {Lothringer}, {Lustig-Yaeger}, {Mancini}, {Mayne}, {Mayorga}, {Molaverdikhani}, {Nasedkin}, {Ohno}, {Parmentier}, {Powell}, {Redfield}, {Roy}, {Taylor}, \& {Zhang}}]{cartermay2024_w39datasynth}
{Carter}, A.~L., {May}, E.~M., {Espinoza}, N., {et~al.} 2024, Nature Astronomy, 8, 1008, \dodoi{10.1038/s41550-024-02292-x}

\bibitem[{Castelli \& Kurucz(2004)}]{castelli2004newgridsatlas9model}
Castelli, F., \& Kurucz, R.~L. 2004, New Grids of ATLAS9 Model Atmospheres.
\newblock \doarXiv{astro-ph/0405087}

\bibitem[{{Coulombe} {et~al.}(2023){Coulombe}, {Benneke}, {Challener}, {Piette}, {Wiser}, {Mansfield}, {MacDonald}, {Beltz}, {Feinstein}, {Radica}, {Savel}, {Dos Santos}, {Bean}, {Parmentier}, {Wong}, {Rauscher}, {Komacek}, {Kempton}, {Tan}, {Hammond}, {Lewis}, {Line}, {Lee}, {Shivkumar}, {Crossfield}, {Nixon}, {Rackham}, {Wakeford}, {Welbanks}, {Zhang}, {Batalha}, {Berta-Thompson}, {Changeat}, {D{\'e}sert}, {Espinoza}, {Goyal}, {Harrington}, {Knutson}, {Kreidberg}, {L{\'o}pez-Morales}, {Shporer}, {Sing}, {Stevenson}, {Aggarwal}, {Ahrer}, {Alam}, {Bell}, {Blecic}, {Caceres}, {Carter}, {Casewell}, {Crouzet}, {Cubillos}, {Decin}, {Fortney}, {Gibson}, {Heng}, {Henning}, {Iro}, {Kendrew}, {Lagage}, {Leconte}, {Lendl}, {Lothringer}, {Mancini}, {Mikal-Evans}, {Molaverdikhani}, {Nikolov}, {Ohno}, {Palle}, {Piaulet}, {Redfield}, {Roy}, {Tsai}, {Venot}, \& {Wheatley}}]{coulombe_broadband_2023}
{Coulombe}, L.-P., {Benneke}, B., {Challener}, R., {et~al.} 2023, \nat, 620, 292, \dodoi{10.1038/s41586-023-06230-1}

\bibitem[{{Dai} \& {Winn}(2017)}]{dai2017_wasp107b}
{Dai}, F., \& {Winn}, J.~N. 2017, \aj, 153, 205, \dodoi{10.3847/1538-3881/aa65d1}

\bibitem[{{Darveau-Bernier} {et~al.}(2022){Darveau-Bernier}, {Albert}, {Talens}, {Lafreni{\`e}re}, {Radica}, {Doyon}, {Cook}, {Rowe}, {Allart}, {Artigau}, {Benneke}, {Cowan}, {Dang}, {Espinoza}, {Johnstone}, {Kaltenegger}, {Lim}, {Pauly}, {Pelletier}, {Piaulet}, {Roy}, {Roy}, {Splinter}, {Taylor}, \& {Turner}}]{dareau-bernier_atoca_2022}
{Darveau-Bernier}, A., {Albert}, L., {Talens}, G.~J., {et~al.} 2022, \pasp, 134, 094502, \dodoi{10.1088/1538-3873/ac8a77}

\bibitem[{{Doyon} {et~al.}(2023){Doyon}, {Willott}, {Hutchings}, {Sivaramakrishnan}, {Albert}, {Lafreni{\`e}re}, {Rowlands}, {Bego{\~n}a Vila}, {Martel}, {LaMassa}, {Aldridge}, {Artigau}, {Cameron}, {Chayer}, {Cook}, {Cooper}, {Darveau-Bernier}, {Dupuis}, {Earnshaw}, {Espinoza}, {Filippazzo}, {Fullerton}, {Gaudreau}, {Gawlik}, {Goudfrooij}, {Haley}, {Kammerer}, {Kendall}, {Lambros}, {Ignat}, {Maszkiewicz}, {McColgan}, {Morishita}, {Ouellette}, {Pacifici}, {Philippi}, {Radica}, {Ravindranath}, {Rowe}, {Roy}, {Roy}, {Saad}, {Sohn}, {Talens}, {Touahri}, {Thatte}, {Taylor}, {Vandal}, {Volk}, {Wander}, {Warner}, {Zheng}, {Zhou}, {Abraham}, {Beaulieu}, {Benneke}, {Ferrarese}, {Jayawardhana}, {Johnstone}, {Kaltenegger}, {Meyer}, {Pipher}, {Rameau}, {Rieke}, {Salhi}, \& {Sawicki}}]{doyon_near_2023}
{Doyon}, R., {Willott}, C.~J., {Hutchings}, J.~B., {et~al.} 2023, \pasp, 135, 098001, \dodoi{10.1088/1538-3873/acd41b}

\bibitem[{{Dyrek} {et~al.}(2023){Dyrek}, {Min}, {Decin}, {Bouwman}, {Crouzet}, {Molli{\`e}re}, {Lagage}, {Konings}, {Tremblin}, {G{\"u}del}, {Pye}, {Waters}, {Henning}, {Vandenbussche}, {Ardevol Martinez}, {Argyriou}, {Ducrot}, {Heinke}, {Van Looveren}, {Absil}, {Barrado}, {Baudoz}, {Boccaletti}, {Cossou}, {Coulais}, {Edwards}, {Gastaud}, {Glasse}, {Glauser}, {Greene}, {Kendrew}, {Krause}, {Lahuis}, {Mueller}, {Olofsson}, {Patapis}, {Rouan}, {Royer}, {Scheithauer}, {Waldmann}, {Whiteford}, {Colina}, {van Dishoeck}, {Greve}, {Ostlin}, {Ray}, \& {Wright}}]{dyrek2023_wasp107b}
{Dyrek}, A., {Min}, M., {Decin}, L., {et~al.} 2023, arXiv e-prints, arXiv:2311.12515, \dodoi{10.48550/arXiv.2311.12515}

\bibitem[{{Ehrenreich} {et~al.}(2020){Ehrenreich}, {Lovis}, {Allart}, {Zapatero Osorio}, {Pepe}, {Cristiani}, {Rebolo}, {Santos}, {Borsa}, {Demangeon}, {Dumusque}, {Gonz{\'a}lez Hern{\'a}ndez}, {Casasayas-Barris}, {S{\'e}gransan}, {Sousa}, {Abreu}, {Adibekyan}, {Affolter}, {Allende Prieto}, {Alibert}, {Aliverti}, {Alves}, {Amate}, {Avila}, {Baldini}, {Bandy}, {Benz}, {Bianco}, {Bolmont}, {Bouchy}, {Bourrier}, {Broeg}, {Cabral}, {Calderone}, {Pall{\'e}}, {Cegla}, {Cirami}, {Coelho}, {Conconi}, {Coretti}, {Cumani}, {Cupani}, {Dekker}, {Delabre}, {Deiries}, {D'Odorico}, {Di Marcantonio}, {Figueira}, {Fragoso}, {Genolet}, {Genoni}, {G{\'e}nova Santos}, {Hara}, {Hughes}, {Iwert}, {Kerber}, {Knudstrup}, {Landoni}, {Lavie}, {Lizon}, {Lendl}, {Lo Curto}, {Maire}, {Manescau}, {Martins}, {M{\'e}gevand}, {Mehner}, {Micela}, {Modigliani}, {Molaro}, {Monteiro}, {Monteiro}, {Moschetti}, {M{\"u}ller}, {Nunes}, {Oggioni}, {Oliveira}, {Pariani}, {Pasquini}, {Poretti}, {Rasilla}, {Redaelli}, {Riva}, {Santana Tschudi},
  {Santin}, {Santos}, {Segovia Milla}, {Seidel}, {Sosnowska}, {Sozzetti}, {Span{\`o}}, {Su{\'a}rez Mascare{\~n}o}, {Tabernero}, {Tenegi}, {Udry}, {Zanutta}, \& {Zerbi}}]{ehrenreich2020_wasp76ironasymmetry}
{Ehrenreich}, D., {Lovis}, C., {Allart}, R., {et~al.} 2020, \nat, 580, 597, \dodoi{10.1038/s41586-020-2107-1}

\bibitem[{{Espinoza} \& {Jones}(2021)}]{espinoza2021_catwoman}
{Espinoza}, N., \& {Jones}, K. 2021, \aj, 162, 165, \dodoi{10.3847/1538-3881/ac134d}

\bibitem[{{Espinoza} {et~al.}(2024){Espinoza}, {Steinrueck}, {Kirk}, {MacDonald}, {Savel}, {Arnold}, {Kempton}, {Murphy}, {Carone}, {Zamyatina}, {Lewis}, {Samra}, {Kiefer}, {Rauscher}, {Christie}, {Mayne}, {Helling}, {Rustamkulov}, {Parmentier}, {May}, {Carter}, {Zhang}, {L{\'o}pez-Morales}, {Allen}, {Blecic}, {Decin}, {Mancini}, {Molaverdikhani}, {Rackham}, {Palle}, {Tsai}, {Ahrer}, {Bean}, {Crossfield}, {Haegele}, {H{\'e}brard}, {Kreidberg}, {Powell}, {Schneider}, {Welbanks}, {Wheatley}, {Brahm}, \& {Crouzet}}]{espinoza24_wasp39b}
{Espinoza}, N., {Steinrueck}, M.~E., {Kirk}, J., {et~al.} 2024, arXiv e-prints, arXiv:2407.10294, \dodoi{10.48550/arXiv.2407.10294}

\bibitem[{{Feinstein} {et~al.}(2023){Feinstein}, {Radica}, {Welbanks}, {Murray}, {Ohno}, {Coulombe}, {Espinoza}, {Bean}, {Teske}, {Benneke}, {Line}, {Rustamkulov}, {Saba}, {Tsiaras}, {Barstow}, {Fortney}, {Gao}, {Knutson}, {MacDonald}, {Mikal-Evans}, {Rackham}, {Taylor}, {Parmentier}, {Batalha}, {Berta-Thompson}, {Carter}, {Changeat}, {dos Santos}, {Gibson}, {Goyal}, {Kreidberg}, {L{\'o}pez-Morales}, {Lothringer}, {Miguel}, {Molaverdikhani}, {Moran}, {Morello}, {Mukherjee}, {Sing}, {Stevenson}, {Wakeford}, {Ahrer}, {Alam}, {Alderson}, {Allen}, {Batalha}, {Bell}, {Blecic}, {Brande}, {Caceres}, {Casewell}, {Chubb}, {Crossfield}, {Crouzet}, {Cubillos}, {Decin}, {D{\'e}sert}, {Harrington}, {Heng}, {Henning}, {Iro}, {Kempton}, {Kendrew}, {Kirk}, {Krick}, {Lagage}, {Lendl}, {Mancini}, {Mansfield}, {May}, {Mayne}, {Nikolov}, {Palle}, {Petit dit de la Roche}, {Piaulet}, {Powell}, {Redfield}, {Rogers}, {Roman}, {Roy}, {Nixon}, {Schlawin}, {Tan}, {Tremblin}, {Turner}, {Venot}, {Waalkes}, {Wheatley}, \&
  {Zhang}}]{feinstein_early_2023}
{Feinstein}, A.~D., {Radica}, M., {Welbanks}, L., {et~al.} 2023, \nat, 614, 670, \dodoi{10.1038/s41586-022-05674-1}

\bibitem[{{Foreman-Mackey} {et~al.}(2013){Foreman-Mackey}, {Hogg}, {Lang}, \& {Goodman}}]{emcee}
{Foreman-Mackey}, D., {Hogg}, D.~W., {Lang}, D., \& {Goodman}, J. 2013, \pasp, 125, 306, \dodoi{10.1086/670067}

\bibitem[{{Fortney} {et~al.}(2020){Fortney}, {Visscher}, {Marley}, {Hood}, {Line}, {Thorngren}, {Freedman}, \& {Lupu}}]{fortney2020}
{Fortney}, J.~J., {Visscher}, C., {Marley}, M.~S., {et~al.} 2020, \aj, 160, 288, \dodoi{10.3847/1538-3881/abc5bd}

\bibitem[{{Fournier-Tondreau} {et~al.}(2024){Fournier-Tondreau}, {MacDonald}, {Radica}, {Lafreni{\`e}re}, {Welbanks}, {Piaulet}, {Coulombe}, {Allart}, {Morel}, {Artigau}, {Albert}, {Lim}, {Doyon}, {Benneke}, {Rowe}, {Darveau-Bernier}, {Cowan}, {Lewis}, {Cook}, {Flagg}, {Genest}, {Pelletier}, {Johnstone}, {Dang}, {Kaltenegger}, {Taylor}, \& {Turner}}]{fournier-tondreau_near_2024}
{Fournier-Tondreau}, M., {MacDonald}, R.~J., {Radica}, M., {et~al.} 2024, \mnras, 528, 3354, \dodoi{10.1093/mnras/stad3813}

\bibitem[{{Freedman} {et~al.}(2014){Freedman}, {Lustig-Yaeger}, {Fortney}, {Lupu}, {Marley}, \& {Lodders}}]{freedman2014}
{Freedman}, R.~S., {Lustig-Yaeger}, J., {Fortney}, J.~J., {et~al.} 2014, \apjs, 214, 25, \dodoi{10.1088/0067-0049/214/2/25}

\bibitem[{{Gandhi} {et~al.}(2022){Gandhi}, {Kesseli}, {Snellen}, {Brogi}, {Wardenier}, {Parmentier}, {Welbanks}, \& {Savel}}]{gandhi2022_w76bhighresretrieval}
{Gandhi}, S., {Kesseli}, A., {Snellen}, I., {et~al.} 2022, \mnras, 515, 749, \dodoi{10.1093/mnras/stac1744}

\bibitem[{Grant \& Wakeford(2022)}]{exoticld}
Grant, D., \& Wakeford, H.~R. 2022, Exo-TiC/ExoTiC-LD: ExoTiC-LD v3.0.0, v3.0.0,  Zenodo, \dodoi{10.5281/zenodo.7437681}

\bibitem[{{Grant} \& {Wakeford}(2023)}]{grant2023_transmissionstrings}
{Grant}, D., \& {Wakeford}, H.~R. 2023, \mnras, 519, 5114, \dodoi{10.1093/mnras/stac3632}

\bibitem[{{Grant} {et~al.}(2023){Grant}, {Lewis}, {Wakeford}, {Batalha}, {Glidden}, {Goyal}, {Mullens}, {MacDonald}, {May}, {Seager}, {Stevenson}, {Valenti}, {Visscher}, {Alderson}, {Allen}, {Ca{\~n}as}, {Col{\'o}n}, {Clampin}, {Espinoza}, {Gressier}, {Huang}, {Lin}, {Long}, {Louie}, {Pe{\~n}a-Guerrero}, {Ranjan}, {Sotzen}, {Valentine}, {Anderson}, {Balmer}, {Bellini}, {Hoch}, {Kammerer}, {Libralato}, {Mountain}, {Perrin}, {Pueyo}, {Rickman}, {Rebollido}, {Sohn}, {van der Marel}, \& {Watkins}}]{grant23_wasp17b}
{Grant}, D., {Lewis}, N.~K., {Wakeford}, H.~R., {et~al.} 2023, \apjl, 956, L32, \dodoi{10.3847/2041-8213/acfc3b10.3847/2041-8213/acfdab}

\bibitem[{{Gressier} {et~al.}(2025){Gressier}, {MacDonald}, {Espinoza}, {Wakeford}, {Lewis}, {Goyal}, {Louie}, {Radica}, {Batalha}, {Long}, {May}, {Mullens}, {Seager}, {Stevenson}, {Valenti}, {Alderson}, {Allen}, {Ca{\~n}as}, {Challener}, {Col{\'o}n}, {Glidden}, {Grant}, {Huang}, {Lin}, {Valentine}, {Mountain}, {Pueyo}, {Perrin}, \& {van der Marel}}]{gressier_jwst_2024}
{Gressier}, A., {MacDonald}, R.~J., {Espinoza}, N., {et~al.} 2025, \aj, 169, 57, \dodoi{10.3847/1538-3881/ad97bf}

\bibitem[{Harris {et~al.}(2020)Harris, Millman, van~der Walt, Gommers, Virtanen, Cournapeau, Wieser, Taylor, Berg, Smith, Kern, Picus, Hoyer, van Kerkwijk, Brett, Haldane, del R{\'{i}}o, Wiebe, Peterson, G{\'{e}}rard-Marchant, Sheppard, Reddy, Weckesser, Abbasi, Gohlke, \& Oliphant}]{numpy}
Harris, C.~R., Millman, K.~J., van~der Walt, S.~J., {et~al.} 2020, Nature, 585, 357, \dodoi{10.1038/s41586-020-2649-2}

\bibitem[{{Hoeijmakers} {et~al.}(2020){Hoeijmakers}, {Cabot}, {Zhao}, {Buchhave}, {Tronsgaard}, {Davis}, {Kitzmann}, {Grimm}, {Cegla}, {Bourrier}, {Ehrenreich}, {Heng}, {Lovis}, \& {Fischer}}]{hoeijmakers2020_mascara2basymmetry}
{Hoeijmakers}, H.~J., {Cabot}, S. H.~C., {Zhao}, L., {et~al.} 2020, \aap, 641, A120, \dodoi{10.1051/0004-6361/202037437}

\bibitem[{{Hoskins} \& {Simmons}(1975)}]{hoskins1975}
{Hoskins}, B.~J., \& {Simmons}, A.~J. 1975, Quarterly Journal of the Royal Meteorological Society, 101, 637, \dodoi{10.1002/qj.49710142918}

\bibitem[{Hunter(2007)}]{matplotlib}
Hunter, J.~D. 2007, Computing in Science \& Engineering, 9, 90, \dodoi{10.1109/MCSE.2007.55}

\bibitem[{Jones \& Espinoza(2022)}]{jones2022_catwoman}
Jones, K., \& Espinoza, N. 2022, Journal of Open Source Software, 7, 2382, \dodoi{10.21105/joss.02382}

\bibitem[{{Kataria} {et~al.}(2015){Kataria}, {Showman}, {Fortney}, {Stevenson}, {Line}, {Kreidberg}, {Bean}, \& {D{\'e}sert}}]{kataria2015_w43gcms}
{Kataria}, T., {Showman}, A.~P., {Fortney}, J.~J., {et~al.} 2015, \apj, 801, 86, \dodoi{10.1088/0004-637X/801/2/86}

\bibitem[{{Kataria} {et~al.}(2016){Kataria}, {Sing}, {Lewis}, {Visscher}, {Showman}, {Fortney}, \& {Marley}}]{kataria2016_gcmgrid}
{Kataria}, T., {Sing}, D.~K., {Lewis}, N.~K., {et~al.} 2016, \apj, 821, 9, \dodoi{10.3847/0004-637X/821/1/9}

\bibitem[{Kempton {et~al.}(2023)Kempton, Zhang, Bean, Steinrueck, Piette, Parmentier, Malsky, Roman, Rauscher, Gao, Bell, Xue, Taylor, Savel, Arnold, Nixon, Stevenson, Mansfield, Kendrew, Zieba, Ducrot, Dyrek, Lagage, Stassun, Henry, Barman, Lupu, Malik, Kataria, Ih, Fu, Welbanks, \& McGill}]{kempton2023_gj1214b}
Kempton, E. M.-R., Zhang, M., Bean, J.~L., {et~al.} 2023, Nature, 620, 67–71, \dodoi{10.1038/s41586-023-06159-5}

\bibitem[{{Kesseli} \& {Snellen}(2021)}]{kesseli2021_wasp76ironasymmetry}
{Kesseli}, A.~Y., \& {Snellen}, I.~A.~G. 2021, \apjl, 908, L17, \dodoi{10.3847/2041-8213/abe047}

\bibitem[{{Kiefer} {et~al.}(2024){Kiefer}, {Samra}, {Lewis}, {Schneider}, {Min}, {Carone}, {Decin}, \& {Helling}}]{kiefer2024_mixedclouds}
{Kiefer}, S., {Samra}, D., {Lewis}, D.~A., {et~al.} 2024, \aap, 690, A244, \dodoi{10.1051/0004-6361/202450526}

\bibitem[{Kipping(2025)}]{kipping2025exoplaneteerscallingplotsallan}
Kipping, D. 2025, Exoplaneteers Keep Calling Plots "Allan Variance" Plots When They Aren't.
\newblock \doarXiv{2504.13238}

\bibitem[{{Kirk} {et~al.}(2020){Kirk}, {Alam}, {L{\'o}pez-Morales}, \& {Zeng}}]{kirk2020_wasp107b}
{Kirk}, J., {Alam}, M.~K., {L{\'o}pez-Morales}, M., \& {Zeng}, L. 2020, \aj, 159, 115, \dodoi{10.3847/1538-3881/ab6e66}

\bibitem[{{Knutson} {et~al.}(2012){Knutson}, {Lewis}, {Fortney}, {Burrows}, {Showman}, {Cowan}, {Agol}, {Aigrain}, {Charbonneau}, {Deming}, {D{\'e}sert}, {Henry}, {Langton}, \& {Laughlin}}]{knutson2012_hd189phasecurve}
{Knutson}, H.~A., {Lewis}, N., {Fortney}, J.~J., {et~al.} 2012, \apj, 754, 22, \dodoi{10.1088/0004-637X/754/1/22}

\bibitem[{{Lim} {et~al.}(2023){Lim}, {Benneke}, {Doyon}, {MacDonald}, {Piaulet}, {Artigau}, {Coulombe}, {Radica}, {L'Heureux}, {Albert}, {Rackham}, {de Wit}, {Salhi}, {Roy}, {Flagg}, {Fournier-Tondreau}, {Taylor}, {Cook}, {Lafreni{\`e}re}, {Cowan}, {Kaltenegger}, {Rowe}, {Espinoza}, {Dang}, \& {Darveau-Bernier}}]{lim_atmospheric_2023}
{Lim}, O., {Benneke}, B., {Doyon}, R., {et~al.} 2023, \apjl, 955, L22, \dodoi{10.3847/2041-8213/acf7c4}

\bibitem[{{Line} \& {Parmentier}(2016)}]{line16_nonunifclouds}
{Line}, M.~R., \& {Parmentier}, V. 2016, \apj, 820, 78, \dodoi{10.3847/0004-637X/820/1/78}

\bibitem[{{Luger} {et~al.}(2019){Luger}, {Agol}, {Foreman-Mackey}, {Fleming}, {Lustig-Yaeger}, \& {Deitrick}}]{luger19_starry_main}
{Luger}, R., {Agol}, E., {Foreman-Mackey}, D., {et~al.} 2019, \aj, 157, 64, \dodoi{10.3847/1538-3881/aae8e5}

\bibitem[{{Luger} {et~al.}(2021{\natexlab{a}}){Luger}, {Foreman-Mackey}, \& {Hedges}}]{luger21_starry_extra2}
{Luger}, R., {Foreman-Mackey}, D., \& {Hedges}, C. 2021{\natexlab{a}}, \aj, 162, 124, \dodoi{10.3847/1538-3881/abfdb9}

\bibitem[{{Luger} {et~al.}(2021{\natexlab{b}}){Luger}, {Foreman-Mackey}, {Hedges}, \& {Hogg}}]{luger21_starry_extra1}
{Luger}, R., {Foreman-Mackey}, D., {Hedges}, C., \& {Hogg}, D.~W. 2021{\natexlab{b}}, \aj, 162, 123, \dodoi{10.3847/1538-3881/abfdb8}

\bibitem[{{Maguire} {et~al.}(2024){Maguire}, {Gibson}, {Nugroho}, {Fortune}, {Ramkumar}, {Gandhi}, \& {de Mooij}}]{maguire2024_wasp76limbasymmetries}
{Maguire}, C., {Gibson}, N.~P., {Nugroho}, S.~K., {et~al.} 2024, \aap, 687, A49, \dodoi{10.1051/0004-6361/202449449}

\bibitem[{{Malsky} {et~al.}(2024){Malsky}, {Rauscher}, {Roman}, {Lee}, {Beltz}, {Savel}, {Kempton}, \& {Cinque}}]{malsky2024}
{Malsky}, I., {Rauscher}, E., {Roman}, M.~T., {et~al.} 2024, \apj, 961, 66, \dodoi{10.3847/1538-4357/ad0b70}

\bibitem[{{Mancini} {et~al.}(2018){Mancini}, {Esposito}, {Covino}, {Southworth}, {Biazzo}, {Bruni}, {Ciceri}, {Evans}, {Lanza}, {Poretti}, {Sarkis}, {Smith}, {Brogi}, {Affer}, {Benatti}, {Bignamini}, {Boccato}, {Bonomo}, {Borsa}, {Carleo}, {Claudi}, {Cosentino}, {Damasso}, {Desidera}, {Giacobbe}, {Gonz{\'a}lez-{\'A}lvarez}, {Gratton}, {Harutyunyan}, {Leto}, {Maggio}, {Malavolta}, {Maldonado}, {Martinez-Fiorenzano}, {Masiero}, {Micela}, {Molinari}, {Nascimbeni}, {Pagano}, {Pedani}, {Piotto}, {Rainer}, {Scandariato}, {Smareglia}, {Sozzetti}, {Andreuzzi}, \& {Henning}}]{mancini2018_wasp39b}
{Mancini}, L., {Esposito}, M., {Covino}, E., {et~al.} 2018, \aap, 613, A41, \dodoi{10.1051/0004-6361/201732234}

\bibitem[{{Marley} \& {McKay}(1999)}]{marley1999}
{Marley}, M.~S., \& {McKay}, C.~P. 1999, \icarus, 138, 268, \dodoi{10.1006/icar.1998.6071}

\bibitem[{{May} {et~al.}(2022){May}, {Stevenson}, {Bean}, {Bell}, {Cowan}, {Dang}, {Desert}, {Fortney}, {Keating}, {Kempton}, {Komacek}, {Lewis}, {Mansfield}, {Morley}, {Parmentier}, {Rauscher}, {Swain}, {Zellem}, \& {Showman}}]{may2022_phasecurvestudy}
{May}, E.~M., {Stevenson}, K.~B., {Bean}, J.~L., {et~al.} 2022, \aj, 163, 256, \dodoi{10.3847/1538-3881/ac6261}

\bibitem[{{May} {et~al.}(2023){May}, {MacDonald}, {Bennett}, {Moran}, {Wakeford}, {Peacock}, {Lustig-Yaeger}, {Highland}, {Stevenson}, {Sing}, {Mayorga}, {Batalha}, {Kirk}, {L{\'o}pez-Morales}, {Valenti}, {Alam}, {Alderson}, {Fu}, {Gonzalez-Quiles}, {Lothringer}, {Rustamkulov}, \& {Sotzen}}]{may2023_observedTLSE}
{May}, E.~M., {MacDonald}, R.~J., {Bennett}, K.~A., {et~al.} 2023, \apjl, 959, L9, \dodoi{10.3847/2041-8213/ad054f}

\bibitem[{{Moran} {et~al.}(2023){Moran}, {Stevenson}, {Sing}, {MacDonald}, {Kirk}, {Lustig-Yaeger}, {Peacock}, {Mayorga}, {Bennett}, {L{\'o}pez-Morales}, {May}, {Rustamkulov}, {Valenti}, {Adams Redai}, {Alam}, {Batalha}, {Fu}, {Gonzalez-Quiles}, {Highland}, {Kruse}, {Lothringer}, {Ortiz Ceballos}, {Sotzen}, \& {Wakeford}}]{moran2023_observedTLSE}
{Moran}, S.~E., {Stevenson}, K.~B., {Sing}, D.~K., {et~al.} 2023, \apjl, 948, L11, \dodoi{10.3847/2041-8213/accb9c}

\bibitem[{Morley {et~al.}(2012)Morley, Fortney, Marley, Visscher, Saumon, \& Leggett}]{morley2012cloudcondensationcurves}
Morley, C.~V., Fortney, J.~J., Marley, M.~S., {et~al.} 2012, The Astrophysical Journal, 756, 172, \dodoi{10.1088/0004-637x/756/2/172}

\bibitem[{{Mo{\v{c}}nik} {et~al.}(2017){Mo{\v{c}}nik}, {Hellier}, {Anderson}, {Clark}, \& {Southworth}}]{mocnik2017_wasp107b}
{Mo{\v{c}}nik}, T., {Hellier}, C., {Anderson}, D.~R., {Clark}, B.~J.~M., \& {Southworth}, J. 2017, \mnras, 469, 1622, \dodoi{10.1093/mnras/stx972}

\bibitem[{{Murphy} {et~al.}(2024{\natexlab{a}}){Murphy}, {Beatty}, \& {Apai}}]{murphy24_timingdegen}
{Murphy}, M.~M., {Beatty}, T.~G., \& {Apai}, D. 2024{\natexlab{a}}, arXiv e-prints, arXiv:2407.17564, \dodoi{10.48550/arXiv.2407.17564}

\bibitem[{{Murphy} {et~al.}(2023){Murphy}, {Beatty}, {Roman}, {Malsky}, {Wingate}, {Ochs}, {Cinque}, {Beltz}, {Rauscher}, {Kempton}, \& {Stevenson}}]{murphy2023_wasp43phasecurves}
{Murphy}, M.~M., {Beatty}, T.~G., {Roman}, M.~T., {et~al.} 2023, \aj, 165, 107, \dodoi{10.3847/1538-3881/acaec5}

\bibitem[{{Murphy} {et~al.}(2024{\natexlab{b}}){Murphy}, {Beatty}, {Schlawin}, {Bell}, {Line}, {Greene}, {Parmentier}, {Rauscher}, {Welbanks}, {Fortney}, \& {Rieke}}]{murphy24}
{Murphy}, M.~M., {Beatty}, T.~G., {Schlawin}, E., {et~al.} 2024{\natexlab{b}}, Nature Astronomy, \dodoi{10.1038/s41550-024-02367-9}

\bibitem[{{Parmentier} {et~al.}(2016){Parmentier}, {Fortney}, {Showman}, {Morley}, \& {Marley}}]{parmentier2016_cloudygcms}
{Parmentier}, V., {Fortney}, J.~J., {Showman}, A.~P., {Morley}, C., \& {Marley}, M.~S. 2016, \apj, 828, 22, \dodoi{10.3847/0004-637X/828/1/22}

\bibitem[{{Parmentier} \& {Guillot}(2014)}]{parmentier2014}
{Parmentier}, V., \& {Guillot}, T. 2014, \aap, 562, A133, \dodoi{10.1051/0004-6361/201322342}

\bibitem[{{Parmentier} {et~al.}(2015){Parmentier}, {Guillot}, {Fortney}, \& {Marley}}]{parmentier2015}
{Parmentier}, V., {Guillot}, T., {Fortney}, J.~J., \& {Marley}, M.~S. 2015, \aap, 574, A35, \dodoi{10.1051/0004-6361/201323127}

\bibitem[{{Parmentier} {et~al.}(2021){Parmentier}, {Showman}, \& {Fortney}}]{parmentier2021_hJphasecurves}
{Parmentier}, V., {Showman}, A.~P., \& {Fortney}, J.~J. 2021, \mnras, 501, 78, \dodoi{10.1093/mnras/staa3418}

\bibitem[{{Parmentier} {et~al.}(2013{\natexlab{a}}){Parmentier}, {Showman}, \& {Lian}}]{parmentier2013_gcms}
{Parmentier}, V., {Showman}, A.~P., \& {Lian}, Y. 2013{\natexlab{a}}, \aap, 558, A91, \dodoi{10.1051/0004-6361/201321132}

\bibitem[{{Parmentier} {et~al.}(2013{\natexlab{b}}){Parmentier}, {Showman}, \& {Lian}}]{parmentier2013_gcmtracers}
---. 2013{\natexlab{b}}, \aap, 558, A91, \dodoi{10.1051/0004-6361/201321132}

\bibitem[{{Piaulet} {et~al.}(2021){Piaulet}, {Benneke}, {Rubenzahl}, {Howard}, {Lee}, {Thorngren}, {Angus}, {Peterson}, {Schlieder}, {Werner}, {Kreidberg}, {Jaouni}, {Crossfield}, {Ciardi}, {Petigura}, {Livingston}, {Dressing}, {Fulton}, {Beichman}, {Christiansen}, {Gorjian}, {Hardegree-Ullman}, {Krick}, \& {Sinukoff}}]{piaulet2021_wasp107b}
{Piaulet}, C., {Benneke}, B., {Rubenzahl}, R.~A., {et~al.} 2021, \aj, 161, 70, \dodoi{10.3847/1538-3881/abcd3c}

\bibitem[{{Piaulet-Ghorayeb} {et~al.}(2024){Piaulet-Ghorayeb}, {Benneke}, {Radica}, {Raul}, {Coulombe}, {Ahrer}, {Kubyshkina}, {Howard}, {Krissansen-Totton}, {MacDonald}, {Roy}, {Louca}, {Christie}, {Fournier-Tondreau}, {Allart}, {Miguel}, {Schlichting}, {Welbanks}, {Cadieux}, {Dorn}, {Evans-Soma}, {Fortney}, {Pierrehumbert}, {Lafreni{\`e}re}, {Acu{\~n}a}, {Komacek}, {Innes}, {Beatty}, {Cloutier}, {Doyon}, {Gagnebin}, {Gapp}, \& {Knutson}}]{piaulet_jwst_2024}
{Piaulet-Ghorayeb}, C., {Benneke}, B., {Radica}, M., {et~al.} 2024, \apjl, 974, L10, \dodoi{10.3847/2041-8213/ad6f00}

\bibitem[{{Pont} {et~al.}(2006){Pont}, {Zucker}, \& {Queloz}}]{pont2006}
{Pont}, F., {Zucker}, S., \& {Queloz}, D. 2006, \mnras, 373, 231, \dodoi{10.1111/j.1365-2966.2006.11012.x}

\bibitem[{{Powell} {et~al.}(2019){Powell}, {Louden}, {Kreidberg}, {Zhang}, {Gao}, \& {Parmentier}}]{powell2019_HJlimbasymmetries}
{Powell}, D., {Louden}, T., {Kreidberg}, L., {et~al.} 2019, \apj, 887, 170, \dodoi{10.3847/1538-4357/ab55d9}

\bibitem[{{Powell} \& {Zhang}(2024)}]{powell2024twoDcloudmodels}
{Powell}, D., \& {Zhang}, X. 2024, \apj, 969, 5, \dodoi{10.3847/1538-4357/ad3de4}

\bibitem[{{Rackham} {et~al.}(2018){Rackham}, {Apai}, \& {Giampapa}}]{rackham18_TLSEmstars}
{Rackham}, B.~V., {Apai}, D., \& {Giampapa}, M.~S. 2018, \apj, 853, 122, \dodoi{10.3847/1538-4357/aaa08c}

\bibitem[{{Rackham} {et~al.}(2019){Rackham}, {Apai}, \& {Giampapa}}]{rackham19_TLSEfgkstars}
---. 2019, \aj, 157, 96, \dodoi{10.3847/1538-3881/aaf892}

\bibitem[{Radica(2024)}]{exotedrfJOSS}
Radica, M. 2024, Journal of Open Source Software, 9, 6898, \dodoi{10.21105/joss.06898}

\bibitem[{{Radica} {et~al.}(2022){Radica}, {Albert}, {Taylor}, {Lafreni{\`e}re}, {Coulombe}, {Darveau-Bernier}, {Doyon}, {Cook}, {Cowan}, {Espinoza}, {Johnstone}, {Kaltenegger}, {Piaulet}, {Roy}, \& {Talens}}]{radica_applesoss_2022}
{Radica}, M., {Albert}, L., {Taylor}, J., {et~al.} 2022, \pasp, 134, 104502, \dodoi{10.1088/1538-3873/ac9430}

\bibitem[{{Radica} {et~al.}(2023){Radica}, {Welbanks}, {Espinoza}, {Taylor}, {Coulombe}, {Feinstein}, {Goyal}, {Scarsdale}, {Albert}, {Baghel}, {Bean}, {Blecic}, {Lafreni{\`e}re}, {MacDonald}, {Zamyatina}, {Allart1}, {Artigau}, {Batalha}, {Cook}, {Cowan}, {Dang}, {Doyon}, {Fournier-Tondreau}, {Johnstone}, {Line}, {Moran}, {Mukherjee}, {Pelletier}, {Roy}, {Talens}, {Filippazzo}, {Pontoppidan}, \& {Volk}}]{radica_awesome_2023}
{Radica}, M., {Welbanks}, L., {Espinoza}, N., {et~al.} 2023, \mnras, 524, 835, \dodoi{10.1093/mnras/stad1762}

\bibitem[{{Radica} {et~al.}(2024){Radica}, {Coulombe}, {Taylor}, {Albert}, {Allart}, {Benneke}, {Cowan}, {Dang}, {Lafreni{\`e}re}, {Thorngren}, {Artigau}, {Doyon}, {Flagg}, {Johnstone}, {Pelletier}, \& {Roy}}]{radica_muted_2024}
{Radica}, M., {Coulombe}, L.-P., {Taylor}, J., {et~al.} 2024, \apjl, 962, L20, \dodoi{10.3847/2041-8213/ad20e4}

\bibitem[{{Radica} {et~al.}(2025){Radica}, {Piaulet-Ghorayeb}, {Taylor}, {Coulombe}, {Benneke}, {Albert}, {Artigau}, {Cowan}, {Doyon}, {Lafreni{\`e}re}, {L'Heureux}, \& {Lim}}]{radica_promise_2024}
{Radica}, M., {Piaulet-Ghorayeb}, C., {Taylor}, J., {et~al.} 2025, \apjl, 979, L5, \dodoi{10.3847/2041-8213/ada381}

\bibitem[{Rauscher(2023)}]{RM-GCM}
Rauscher, E. 2023, {RM-GCM code}, Version 5.0,  Zenodo, \dodoi{10.5281/zenodo.7962911}

\bibitem[{{Rauscher} \& {Menou}(2010)}]{rauscher2010}
{Rauscher}, E., \& {Menou}, K. 2010, \apj, 714, 1334, \dodoi{10.1088/0004-637X/714/2/1334}

\bibitem[{{Rauscher} \& {Menou}(2012)}]{rauscher2012}
---. 2012, \apj, 750, 96, \dodoi{10.1088/0004-637X/750/2/96}

\bibitem[{{Rigby} {et~al.}(2023){Rigby}, {Perrin}, {McElwain}, {Kimble}, {Friedman}, {Lallo}, {Doyon}, {Feinberg}, {Ferruit}, {Glasse}, {Rieke}, {Rieke}, {Wright}, {Willott}, {Colon}, {Milam}, {Neff}, {Stark}, {Valenti}, {Abell}, {Abney}, {Abul-Huda}, {Acton}, {Adams}, {Adler}, {Aguilar}, {Ahmed}, {Albert}, {Alberts}, {Aldridge}, {Allen}, {Altenburg}, {{\'A}lvarez-M{\'a}rquez}, {Alves de Oliveira}, {Andersen}, {Anderson}, {Anderson}, {Argyriou}, {Armstrong}, {Arribas}, {Artigau}, {Arvai}, {Atkinson}, {Bacon}, {Bair}, {Banks}, {Barrientes}, {Barringer}, {Bartosik}, {Bast}, {Baudoz}, {Beatty}, {Bechtold}, {Beck}, {Bergeron}, {Bergkoetter}, {Bhatawdekar}, {Birkmann}, {Blazek}, {Blome}, {Boccaletti}, {B{\"o}ker}, {Boia}, {Bonaventura}, {Bond}, {Bosley}, {Boucarut}, {Bourque}, {Bouwman}, {Bower}, {Bowers}, {Boyer}, {Bradley}, {Brady}, {Braun}, {Breda}, {Bresnahan}, {Bright}, {Britt}, {Bromenschenkel}, {Brooks}, {Brooks}, {Brown}, {Brown}, {Brown}, {Bunker}, {Burger}, {Bushouse}, {Cale}, {Cameron}, {Cameron},
  {Canipe}, {Caplinger}, {Caputo}, {Cara}, {Carey}, {Carniani}, {Carrasquilla}, {Carruthers}, {Case}, {Catherine}, {Chance}, {Chapman}, {Charlot}, {Charlow}, {Chayer}, {Chen}, {Cherinka}, {Chichester}, {Chilton}, {Chonis}, {Clampin}, {Clark}, {Clark}, {Coe}, {Coleman}, {Comber}, {Comeau}, {Connolly}, {Cooper}, {Cooper}, {Coppock}, {Correnti}, {Cossou}, {Coulais}, {Coyle}, {Cracraft}, {Curti}, {Cuturic}, {Davis}, {Davis}, {Dean}, {DeLisa}, {deMeester}, {Dencheva}, {Dencheva}, {DePasquale}, {Deschenes}, {Hunor Detre}, {Diaz}, {Dicken}, {DiFelice}, {Dillman}, {Dixon}, {Doggett}, {Donaldson}, {Douglas}, {DuPrie}, {Dupuis}, {Durning}, {Easmin}, {Eck}, {Edeani}, {Egami}, {Ehrenwinkler}, {Eisenhamer}, {Eisenhower}, {Elie}, {Elliott}, {Elliott}, {Ellis}, {Engesser}, {Espinoza}, {Etienne}, {Etxaluze}, {Falini}, {Feeney}, {Ferry}, {Filippazzo}, {Fincham}, {Fix}, {Flagey}, {Florian}, {Flynn}, {Fontanella}, {Ford}, {Forshay}, {Fox}, {Franz}, {Fu}, {Fullerton}, {Galkin}, {Galyer}, {Garc{\'\i}a Mar{\'\i}n}, {Gardner},
  {Gardner}, {Garland}, {Garrett}, {Gasman}, {Gaspar}, {Gaudreau}, {Gauthier}, {Geers}, {Geithner}, {Gennaro}, {Giardino}, {Girard}, {Giuliano}, {Glassmire}, \& {Glauser}}]{Rigby2023}
{Rigby}, J., {Perrin}, M., {McElwain}, M., {et~al.} 2023, \pasp, 135, 048001, \dodoi{10.1088/1538-3873/acb293}

\bibitem[{{Roman} \& {Rauscher}(2019)}]{roman2019}
{Roman}, M., \& {Rauscher}, E. 2019, \apj, 872, 1, \dodoi{10.3847/1538-4357/aafdb5}

\bibitem[{{Roman} {et~al.}(2021){Roman}, {Kempton}, {Rauscher}, {Harada}, {Bean}, \& {Stevenson}}]{roman2021}
{Roman}, M.~T., {Kempton}, E. M.~R., {Rauscher}, E., {et~al.} 2021, \apj, 908, 101, \dodoi{10.3847/1538-4357/abd549}

\bibitem[{{Roth} {et~al.}(2024){Roth}, {Parmentier}, \& {Hammond}}]{roth2024_HJmodels}
{Roth}, A., {Parmentier}, V., \& {Hammond}, M. 2024, \mnras, 531, 1056, \dodoi{10.1093/mnras/stae984}

\bibitem[{{Rubenzahl} {et~al.}(2021){Rubenzahl}, {Dai}, {Howard}, {Chontos}, {Giacalone}, {Lubin}, {Rosenthal}, {Isaacson}, {Batalha}, {Crossfield}, {Dressing}, {Fulton}, {Huber}, {Kane}, {Petigura}, {Robertson}, {Roy}, {Weiss}, {Beard}, {Hill}, {Mayo}, {Mocnik}, {Murphy}, \& {Scarsdale}}]{rubenzahl2021_wasp107b}
{Rubenzahl}, R.~A., {Dai}, F., {Howard}, A.~W., {et~al.} 2021, \aj, 161, 119, \dodoi{10.3847/1538-3881/abd177}

\bibitem[{{Schlawin} \& {Glidic}(2022)}]{TSHIRTPIPELINE}
{Schlawin}, E., \& {Glidic}, K. 2022, {tshirt},  GitHub.
\newblock \url{https://github.com/eas342/tshirt}

\bibitem[{{Schlawin} {et~al.}(2018){Schlawin}, {Greene}, {Line}, {Fortney}, \& {Rieke}}]{schlawin2018_manateepaper}
{Schlawin}, E., {Greene}, T.~P., {Line}, M., {Fortney}, J.~J., \& {Rieke}, M. 2018, \aj, 156, 40, \dodoi{10.3847/1538-3881/aac774}

\bibitem[{{Schlawin} {et~al.}(2020){Schlawin}, {Leisenring}, {Misselt}, {Greene}, {McElwain}, {Beatty}, \& {Rieke}}]{schlawin20_1fnoise}
{Schlawin}, E., {Leisenring}, J., {Misselt}, K., {et~al.} 2020, \aj, 160, 231, \dodoi{10.3847/1538-3881/abb811}

\bibitem[{{Schlawin} {et~al.}(2024{\natexlab{a}}){Schlawin}, {Mukherjee}, {Ohno}, {Bell}, {Beatty}, {Greene}, {Line}, {Challener}, {Parmentier}, {Fortney}, {Rauscher}, {Wiser}, {Welbanks}, {Murphy}, {Edelman}, {Batalha}, {Moran}, {Mehta}, \& {Rieke}}]{schlawin24_wasp69b}
{Schlawin}, E., {Mukherjee}, S., {Ohno}, K., {et~al.} 2024{\natexlab{a}}, \aj, 168, 104, \dodoi{10.3847/1538-3881/ad58e0}

\bibitem[{{Schlawin} {et~al.}(2024{\natexlab{b}}){Schlawin}, {Ohno}, {Bell}, {Murphy}, {Welbanks}, {Beatty}, {Greene}, {Fortney}, {Parmentier}, {Edelman}, {Gill}, {Anderson}, {Wheatley}, {Henry}, {Mehta}, {Kreidberg}, \& {Rieke}}]{schlawin24_gj1214b}
{Schlawin}, E., {Ohno}, K., {Bell}, T.~J., {et~al.} 2024{\natexlab{b}}, \apjl, 974, L33, \dodoi{10.3847/2041-8213/ad7fef}

\bibitem[{{Schmidt} {et~al.}(2025){Schmidt}, {MacDonald}, {Tsai}, {Radica}, {Wang}, {Ahrer}, {Bell}, {Fisher}, {Thorngren}, {Wogan}, {May}, {Ferrari}, {Bennett}, {Rustamkulov}, {L{\'o}pez-Morales}, \& {Sing}}]{schmidt_comprehensive_2025}
{Schmidt}, S.~P., {MacDonald}, R.~J., {Tsai}, S.-M., {et~al.} 2025, arXiv e-prints, arXiv:2501.18477, \dodoi{10.48550/arXiv.2501.18477}

\bibitem[{{Showman} {et~al.}(2009){Showman}, {Fortney}, {Lian}, {Marley}, {Freedman}, {Knutson}, \& {Charbonneau}}]{showman2009_gcms}
{Showman}, A.~P., {Fortney}, J.~J., {Lian}, Y., {et~al.} 2009, \apj, 699, 564, \dodoi{10.1088/0004-637X/699/1/564}

\bibitem[{{Showman} {et~al.}(2019){Showman}, {Tan}, \& {Zhang}}]{showman19_gcm}
{Showman}, A.~P., {Tan}, X., \& {Zhang}, X. 2019, \apj, 883, 4, \dodoi{10.3847/1538-4357/ab384a}

\bibitem[{Sing {et~al.}(2024)Sing, Rustamkulov, Thorngren, Barstow, Tremblin, Alves~de Oliveira, Beck, Birkmann, Challener, Crouzet, Espinoza, Ferruit, Giardino, Gressier, Lee, Lewis, Maiolino, Manjavacas, Rauscher, Sirianni, \& Valenti}]{sing2024_wasp107b}
Sing, D.~K., Rustamkulov, Z., Thorngren, D.~P., {et~al.} 2024, Nature, 630, 831–835, \dodoi{10.1038/s41586-024-07395-z}

\bibitem[{{Spake} {et~al.}(2021){Spake}, {Oklop{\v{c}}i{\'c}}, \& {Hillenbrand}}]{spake2021_wasp107b}
{Spake}, J.~J., {Oklop{\v{c}}i{\'c}}, A., \& {Hillenbrand}, L.~A. 2021, \aj, 162, 284, \dodoi{10.3847/1538-3881/ac178a}

\bibitem[{{Spake} {et~al.}(2018){Spake}, {Sing}, {Evans}, {Oklop{\v{c}}i{\'c}}, {}, {Bourrier}, {Kreidberg}, {Rackham}, {Irwin}, {Ehrenreich}, {Wyttenbach}, {Wakeford}, {Zhou}, {Chubb}, {Nikolov}, {Goyal}, {Henry}, {Williamson}, {Blumenthal}, {Anderson}, {Hellier}, {Charbonneau}, {Udry}, \& {Madhusudhan}}]{spake2018_wasp107b}
{Spake}, J.~J., {Sing}, D.~K., {Evans}, T.~M., {et~al.} 2018, \nat, 557, 68, \dodoi{10.1038/s41586-018-0067-5}

\bibitem[{{Steinrueck} {et~al.}(2019){Steinrueck}, {Parmentier}, {Showman}, {Lothringer}, \& {Lupu}}]{steinrueck2019_gcms}
{Steinrueck}, M.~E., {Parmentier}, V., {Showman}, A.~P., {Lothringer}, J.~D., \& {Lupu}, R.~E. 2019, \apj, 880, 14, \dodoi{10.3847/1538-4357/ab2598}

\bibitem[{{Tan} \& {Showman}(2021)}]{tanshowman21b}
{Tan}, X., \& {Showman}, A.~P. 2021, \mnras, 502, 2198, \dodoi{10.1093/mnras/stab097}

\bibitem[{{Tsai} {et~al.}(2023{\natexlab{a}}){Tsai}, {Moses}, {Powell}, \& {Lee}}]{tsai2023_so2transport}
{Tsai}, S.-M., {Moses}, J.~I., {Powell}, D., \& {Lee}, E. K.~H. 2023{\natexlab{a}}, \apjl, 959, L30, \dodoi{10.3847/2041-8213/ad1405}

\bibitem[{{Tsai} {et~al.}(2023{\natexlab{b}}){Tsai}, {Lee}, {Powell}, {Gao}, {Zhang}, {Moses}, {H{\'e}brard}, {Venot}, {Parmentier}, {Jordan}, {Hu}, {Alam}, {Alderson}, {Batalha}, {Bean}, {Benneke}, {Bierson}, {Brady}, {Carone}, {Carter}, {Chubb}, {Inglis}, {Leconte}, {Line}, {L{\'o}pez-Morales}, {Miguel}, {Molaverdikhani}, {Rustamkulov}, {Sing}, {Stevenson}, {Wakeford}, {Yang}, {Aggarwal}, {Baeyens}, {Barat}, {de Val-Borro}, {Daylan}, {Fortney}, {France}, {Goyal}, {Grant}, {Kirk}, {Kreidberg}, {Louca}, {Moran}, {Mukherjee}, {Nasedkin}, {Ohno}, {Rackham}, {Redfield}, {Taylor}, {Tremblin}, {Visscher}, {Wallack}, {Welbanks}, {Youngblood}, {Ahrer}, {Batalha}, {Behr}, {Berta-Thompson}, {Blecic}, {Casewell}, {Crossfield}, {Crouzet}, {Cubillos}, {Decin}, {D{\'e}sert}, {Feinstein}, {Gibson}, {Harrington}, {Heng}, {Henning}, {Kempton}, {Krick}, {Lagage}, {Lendl}, {Lothringer}, {Mansfield}, {Mayne}, {Mikal-Evans}, {Palle}, {Schlawin}, {Shorttle}, {Wheatley}, \& {Yurchenko}}]{tsai2023_so2production}
{Tsai}, S.-M., {Lee}, E. K.~H., {Powell}, D., {et~al.} 2023{\natexlab{b}}, \nat, 617, 483, \dodoi{10.1038/s41586-023-05902-2}

\bibitem[{{von Paris} {et~al.}(2016){von Paris}, {Gratier}, {Bord{\'e}}, {Leconte}, \& {Selsis}}]{vonparis16_LA}
{von Paris}, P., {Gratier}, P., {Bord{\'e}}, P., {Leconte}, J., \& {Selsis}, F. 2016, \aap, 589, A52, \dodoi{10.1051/0004-6361/201527894}

\bibitem[{{Wall} \& {Jenkins}(2012)}]{practicalstatsforastrobook}
{Wall}, J.~V., \& {Jenkins}, C.~R. 2012, {Practical Statistics for Astronomers} (Cambridge University Press)

\bibitem[{{Wardenier} {et~al.}(2023){Wardenier}, {Parmentier}, {Line}, \& {Lee}}]{wardenier2023_w76bGCM}
{Wardenier}, J.~P., {Parmentier}, V., {Line}, M.~R., \& {Lee}, E. K.~H. 2023, \mnras, 525, 4942, \dodoi{10.1093/mnras/stad2586}

\bibitem[{{Welbanks} {et~al.}(2024){Welbanks}, {Bell}, {Beatty}, {Line}, {Ohno}, {Fortney}, {Schlawin}, {Greene}, {Rauscher}, {McGill}, {Murphy}, {Parmentier}, {Tang}, {Edelman}, {Mukherjee}, {Wiser}, {Lagage}, {Dyrek}, \& {Arnold}}]{welbanks24}
{Welbanks}, L., {Bell}, T.~J., {Beatty}, T.~G., {et~al.} 2024, arXiv e-prints, arXiv:2405.11018, \dodoi{10.48550/arXiv.2405.11018}

\bibitem[{{Zamyatina} {et~al.}(2024){Zamyatina}, {Christie}, {H{\'e}brard}, {Mayne}, {Radica}, {Taylor}, {Baskett}, {Moore}, {Lils}, {Sergeev}, {Ahrer}, {Manners}, {Kohary}, \& {Feinstein}}]{zamyatina24_wasp96bGCMs_limbasym}
{Zamyatina}, M., {Christie}, D.~A., {H{\'e}brard}, E., {et~al.} 2024, \mnras, 529, 1776, \dodoi{10.1093/mnras/stae600}

\end{thebibliography}
\bibliographystyle{aasjournal}



\end{document}